\documentclass[aps,nofootinbib,english,prx,floatfix,amsmath,superscriptaddress,tightenlines,twocolumn]{revtex4-2}
\usepackage[pdftex,colorlinks=true,linkcolor=darkblue,citecolor=blue,urlcolor=darkred]{hyperref}
\usepackage[english]{babel}
\usepackage{braket}
\usepackage{amsmath,amsfonts, amssymb, amsthm, dsfont}
\usepackage{yfonts,bbm}
\usepackage{mathrsfs}
\usepackage{graphicx}
\usepackage{diagbox} 
\usepackage[export]{adjustbox}
\usepackage[normalem]{ulem}
\usepackage{verbatim}
\usepackage{xcolor}
\usepackage{float}
\usepackage{enumitem}
\usepackage{mathtools}
\usepackage{physics}
\usepackage{makecell}
\usepackage[inkscapelatex=false]{svg}

\DeclareMathSymbol{:}{\mathord}{operators}{"3A}

\makeatletter
\def\maketitle{
\@author@finish
\title@column\titleblock@produce
\suppressfloats[t]}
\makeatother

\definecolor{darkblue}{rgb}{0.,0.,0.4}
\definecolor{darkred}{rgb}{0.5,0.,0.}

\theoremstyle{plain}

\usepackage{verbatim}
\usepackage{stackengine}

\usepackage{multirow}

\begin{document}
\title{Anomalous spin transport in integrable random quantum circuits}

\author{Songlei Wang}
\affiliation{School of Physics, Peking University, Beijing 100871, China}

\author{Chenguang Liang}
\thanks{cgliang@pku.edu.cn}
\affiliation{School of Physics, Peking University, Beijing 100871, China}

\author{Hongzheng Zhao}
\affiliation{State Key Laboratory of Artificial Microstructure and Mesoscopic
Physics,School of Physics, Peking University, Beijing 100871, China}

\author{Zhi-Cheng Yang}
\thanks{zcyang19@pku.edu.cn}
\affiliation{School of Physics, Peking University, Beijing 100871, China}
\affiliation{Center for High Energy Physics, Peking University, Beijing 100871, China}

\begin{abstract} 
High-temperature spin transport in integrable quantum spin chains exhibits a rich dynamical phase diagram, including ballistic, superdiffusive, and diffusive regimes. While integrability is known to survive in static and periodically driven systems, its fate in the complete absence of time-translation symmetry, particularly in interacting random quantum circuits, has remained unclear. Here we construct integrable random quantum circuits built from inhomogeneous XXZ $R$-matrices. Remarkably, integrability is preserved for arbitrary sequences of gate layers, ranging from quasiperiodic to fully random, thereby explicitly breaking both continuous and discrete time-translation symmetry. Using large-scale time-dependent density-matrix renormalization group simulations at infinite temperature and half filling, we map out the resulting spin-transport phase diagram and identify ballistic, superdiffusive, and diffusive regimes controlled by the spectral parameters of the $R$-matrices. The spatiotemporal structure of spin correlations within each regime depends sensitively on the inhomogeneity, exhibiting spatial asymmetry and sharp peak structures tied to near-degenerate quasiparticle velocities. To account for these findings, we develop a generalized hydrodynamics framework adapted to time-dependent integrable circuits, yielding Euler-scale predictions for correlation functions, Drude weights, and diffusion bounds. This approach identifies the quasiparticles governing transport and quantitatively captures both the scaling exponents and fine structures of the correlation profiles observed numerically. Our results demonstrate that exact Yang-Baxter integrability is compatible with stochastic quantum dynamics and establish generalized hydrodynamics as a predictive framework for transport in time-dependent integrable systems.
\end{abstract}

\maketitle 

\tableofcontents

\section{Introduction}
The non-equilibrium dynamics of closed quantum many-body systems has garnered significant attention in recent years, driven by rapid advances in quantum simulation platforms such as ultracold atoms, trapped ions~\cite{ions1,ions4,ions3}, superconducting qubits, and neutral-atom arrays~\cite{endres2016,Rydberg}. A paradigmatic setting for exploring such dynamics is the transport of conserved quantities~\cite{prosenRMPtransport}, which serves as a canonical probe of many-body dynamics in both quantum simulation and condensed matter experiments. A widely held expectation is that, at sufficiently long timescales, the transport of conserved quantities is governed by an emergent coarse-grained hydrodynamic description~\cite{kadanoff1963,spohnbook,forsterbook,chaikinbook}. In generic chaotic quantum many-body systems, transport is typically diffusive, with the mean square displacement growing linearly in time. However, deviations from diffusion in non-integrable systems have attracted sustained interest, particularly in the context of anomalous transport, whose microscopic mechanisms remain an open frontier. By introducing structured disorder~\cite{disordertransport1,disordertransport2,disordertransport3,disordertransport4}, kinetic constraints~\cite{PhysRevB.100.214301,PhysRevE.103.022142,constraintsubdiff,fractontransport}, or long-range interactions~\cite{longrangeconstraints,LRSYKtransport}, a wide variety of quantum many-body models exhibiting subdiffusive, superdiffusive, or even localized dynamics have been constructed and studied extensively.

Many-body integrable systems \cite{classicalint,arutyunov2019elements} play an important role in the study of quantum dynamics \cite{intoutreview2016}. Owing to an extensive set of exactly conserved charges, their spectral and dynamical properties differ qualitatively from those of quantum-chaotic systems. Under unitary evolution they do not thermalize to canonical ensembles; instead, they relax to maximum-entropy stationary states described by generalized Gibbs ensembles (GGEs)~\cite{Essler_2016,Vidmar_2016,Ilievski_2016,GGE17ilievski}. In integrable models, coherent quasiparticle excitations with effectively infinite lifetimes—even at high temperatures—underlie distinctive mechanisms for transporting conserved charges and quantum information, typically leading to ballistic rather than diffusive behavior.

Many integrable models admit exact solutions for their spectra; nevertheless, characterizing their out-of-equilibrium dynamics remains a significant challenge. In recent years, a powerful framework, which is now known as generalized hydrodynamics (GHD)~\cite{GHD2016doyon,GHD2016prl,doyonGHDnote,Alba_2021,DeNardis_2022,EsslerGHDnote}, has been developed to describe the large space–time limit dynamics of many body integrable systems. In contrast to conventional hydrodynamics, where one obtains a closed set of
macroscopic evolution equations by keeping only a finite number of conserved densities
(and their associated currents), integrable models possess an infinite hierarchy of conserved charges, so attempting to build hydrodynamics by tracking all of them is not practical. GHD resolves this by working in the basis of stable quasiparticle excitations: the state is encoded in quasiparticle densities, and the dynamics at the Euler scale (the joint limit
$x,t\to\infty$ with $\xi=x/t$ fixed) are governed by continuity equations whose state-dependent effective
velocities are renormalized by interactions through a set of self-consistent dressing
relations.  In this way, the evolution of physical observables and correlation functions can be expressed directly in terms of quasiparticle distributions, bypassing full form-factor resummations \cite{DeNardis_2022}. The GHD approach has thus yielded quantitative insights into diverse phenomena, including inhomogeneous quantum quenches \cite{gapghd2017,ghdnewton,GHDatomchip}, linear-response transport (e.g., Drude weights)~\cite{doyonspohnDrude,DenardisIlievskiDrude,GHDmoore}, and entanglement dynamics~\cite{qppentanglement,ghdentanglement}. It has also established a unified language that bridges integrability, kinetic transport, and experimentally relevant non-equilibrium protocols, thereby elevating GHD to a cornerstone of the theoretical description for dynamics in integrable quantum matter~\cite{GHDrevintro,GHDperspective}.

The GHD framework has become a key tool for understanding anomalous transport in integrable
spin chains. A canonical example is the spin-$\tfrac12$ XXZ model,
\begin{equation}
    H_{\text{XXZ}}=\sum_{j=1}^N\!\left(
    s_j^x s_{j+1}^x+s_j^y s_{j+1}^y+\Delta\, s_j^z s_{j+1}^z
    \right)-h\sum_{j=1}^N s_j^z, \label{eq:xxzH}
\end{equation}
with $s_j^\alpha=\sigma_j^\alpha/2$. It is Yang--Baxter integrable \cite{CNYang1967,BAXTER1972,baxterexactly,korepin1997,faddeev1996ABA,Vsamaj—book}: a transfer matrix
generates an infinite set of commuting charges, and the spectrum is obtained by Bethe
ansatz. At $h=0$, the anisotropy $\Delta$ organizes the well-known ground-state
phase diagram into a gapless easy-plane regime $|\Delta|<1$, a gapped easy-axis regime
$|\Delta|>1$, and $\mathrm{SU}(2)$-symmetric isotropic points $\Delta=\pm1$. This equilibrium structure foreshadows a finite-temperature dynamical phase diagram, which GHD captures within a unified description.

At half filling and high temperature, the three XXZ regimes exhibit distinct spin-transport
behaviors. Spin-flip symmetry forces the spin current to have zero overlap with the
standard family of local Yang--Baxter charges, but in the easy-plane regime one can
construct \emph{quasi-local}, spin-flip odd charges with finite current overlap~\cite{xxzqlc2,Pereira_2014,PROSENnpb,Ilievski_2016}, which
rigorously yields ballistic transport and a nonzero Drude weight computable within GHD~\cite{xxzqlc1,xxzqlc3,affleckxxz1,affleckxxz2,doyonspohnDrude,DenardisIlievskiDrude}. In
the easy-axis regime, dressed quasiparticles still propagate ballistically, yet their spin
content is effectively demagnetized on Euler scales by inter-species scattering~\cite{sachdevdiffusion,sarangromainkinetic}; the
residual spin dynamics is then controlled by rare, heavy magnon strings with exponentially
small velocities $v_s\sim e^{-s}$(with \(s\) the string length), whose collision-induced wandering produces diffusion
with diffusion constants and hydrodynamic equations accessible to GHD~\cite{diffusionGHD,otocint}. At the isotropic
point $\Delta=1$, the same mechanism persists but with algebraically slow strings
$v_s\sim 1/s$, leading to a divergent diffusion constant~\cite{znidaricxxx2011,prosen17NC,xxxsuperdif18ilievskiprosen,sarangromainkinetic} and superdiffusion with dynamical
exponent $z=3/2$~\cite{XXXmagicformula,super-superdiffusion,Bulchandani_superdiffusion}. In this regime, dynamical spin correlations fall into the KPZ universality
class~\cite{KPZ1986,KPZ2019,KPZgaugemode,KPZsoliton,KPZdmt,coupleKPZ,KPZfullcheck2025}, with signatures reported in condensed-matter experiments~\cite{cmekpz} and
quantum-simulation platforms \cite{coldatomkpz,googlekpz}. While a coherent phenomenology
has emerged, the microscopic origin and the precise domain of universality remain active
topics \cite{sarangromain_rev2023,sarangromain_rev2024}. Closely analogous dynamical phases and mechanisms have also been identified in integrable Floquet settings~\cite{intfloquet2017,XXXcircuit,XXZcircuitprosen,ALEINERcircuit,XXZCircuitGGE,XXZCircuitGHD,integrablecircuitratchets,znidaricinthaar,vznidarivcnonlocalsu2}.

Recently, quantum many-body dynamics without time-translation symmetry—encompassing quasiperiodic drives~\cite{prx7aperiodic,else,vasseurqpdr,mori,CHSE} and random quantum circuits~\cite{nahumprx7,nahumprx8,vonrcnocharge,vonrcwithcharge,circuitreivewnahum}—has emerged as a powerful paradigm for exploring thermalization, information scrambling, and transport. Despite this progress, it has remained unclear whether interacting quantum integrability can persist beyond static Hamiltonian and time-periodic Floquet settings~\cite{intfloquet2017,XXXcircuit,XXZcircuitprosen,ALEINERcircuit}. Closely related is the open question of whether strictly ballistic transport, protected by integrability, can exist in \textit{interacting} quantum systems that lack any form of time-translation symmetry. The central challenge is that generic aperiodic driving in interacting systems tends to wash out the fine-tuned constraints underlying integrability, driving the dynamics toward generic chaotic behavior—typically diffusive at high temperature. Existing attempts to break time-translation symmetry while preserving integrability have so far been largely confined to classical spin chains~\cite{randintegrablekrajnik2025}, time-dependent free-fermion~\cite{fibdrivenfree} and impurity Kondo~\cite{tdkondo} systems, as well as classical integrable cellular automata~\cite{inteautomata2017,Klobas_2018,Krajnik_2025,PhysRevE.111.024141,yoshimura2506}. There also exist GHD treatments of quantum integrable models subject to weak spatial inhomogeneities and slowly varying drives \cite{GHDstinh}; however, such perturbations typically spoil exact integrability and only admit an effective hydrodynamic description.

These developments motivate two questions:
(i) can one construct a genuinely \emph{interacting} quantum circuit that lacks strict time-translation symmetry, yet remains integrable \emph{throughout the entire evolution};
(ii) if so, how does this symmetry breaking affect spin transport and the associated scaling laws under such integrable dynamics?
In this work, we construct such an integrable, time-translation–breaking interacting circuit and systematically characterize its consequences for spin transport and scaling. Inspired by Refs.~\cite{floquetbaxter,Paletta2025IntegrabilityAC,randominteEssler,randomintsr}, we use the XXZ (six-vertex) $R$-matrix to define a three-site inhomogeneous transfer matrix, from which we construct two distinct, unitarily inequivalent and yet \emph{commuting}, genuinely interacting circuit layers, $U_0$ and $U_1$. Both layers are generated from the same Yang--Baxter structure and, crucially, they commute with (and hence preserve) the same inhomogeneous transfer matrix for all spectral parameters.
As a consequence, $U_0$ and $U_1$ share an identical infinite family of conserved charges, and \emph{any} layer sequence formed as a word in $\{U_0,U_1\}$---including quasiperiodic Fibonacci and Thue--Morse sequences \cite{prx7aperiodic,else,vasseurqpdr,mori,CHSE}, as well as fully random sequences with a fixed bias---remains integrable at every step of the evolution, despite the absence of discrete or continuous time-translation symmetry.

We then investigate infinite-temperature spin transport at half filling under these
circuits, characterized by the dynamical two-point function
$C(x,t)=\langle S^z(x,t)\,S^z(0,0)\rangle$. Using large-scale time-dependent
density-matrix renormalization group (tDMRG) simulations, we map out the resulting
spin-transport phase diagram and identify ballistic, superdiffusive, and diffusive
regimes. A central finding is that the transport universality class---and hence the
dynamical exponent $z$---is a robust macroscopic property: it is insensitive to the
choice of circuit sequence (quasiperiodic or random) and is fixed solely by the $R$-matrix
type defining the circuit layers. Concretely, the trigonometric (XXZ easy-plane) case is
ballistic, the hyperbolic (easy-axis) case is diffusive, and the rational (XXX) limit is
superdiffusive.

In sharp contrast, the full spatiotemporal lineshape of $C(x,t)$ encodes a set of highly tunable, genuinely microscopic dynamical fingerprints. The dominant
sensitivity is to the inhomogeneity parameters, while the choice of sequence (Fibonacci,
Thue--Morse, or stochastic driving) provides a secondary but clearly visible modulation.
As we show below, even within a fixed transport regime the correlation profile can be drastically
reshaped. In the ballistic (gapless) regime, for example, one encounters parameter choices yielding a smooth, continuum-like profile, as well as regimes in which multiple sharp ballistic peaks emerge. These features have a direct quasiparticle origin: the corresponding thermodynamic Bethe ansatz (TBA) data reveal that
the dominant root densities can reorganize from a single broad support into effectively
bimodal or trimodal distributions in rapidity. Whenever a substantial fraction of the spectral weight accumulates in regions where the effective velocity $v^{\rm eff}(\lambda)$ is nearly
$\lambda$-independent, spectral weight concentrates onto narrow rays $\xi=x/t$, producing extremely sharp peaks in the correlation function.

At the isotropic point, where the asymptotic transport universality is superdiffusive, we further observe a nontrivial finite-time interplay between the intrinsic $z=3/2$ background and emergent ballistic components arising from proximity to dual unitarity. In particular, sharp ballistic peaks can transiently develop atop the broad superdiffusive background. The spectral weight carried by these ballistic contributions decays in time, indicating that they do not correspond to stable quasiparticles but rather to finite-lifetime modes. Nonetheless, within the time window accessible to our simulations, these ballistic features remain clearly visible and strongly influence the observed correlation profiles. 

To provide a unified physical interpretation of both the robust transport universality classes and the sequence- and parameter-dependent correlation lineshapes, we develop a TBA and
GHD framework tailored to arbitrary evolution protocols built from $\{U_0,U_1\}$. The inhomogeneous transfer
matrix fixes the underlying quasiparticle content (root densities and dressed charges),
while the specific driving protocol enters solely through the dynamical dressing data, most importantly the effective velocities. We show how to consistently combine the quasiparticle information associated with the two circuit layers into protocol-dependent, \emph{weight-averaged} effective velocities, and
use the resulting TBA/GHD data to reconstruct $C(x,t)$ at Euler (ballistic) scale. Across all protocols and inhomogeneity parameters studied, these hydrodynamic predictions quantitatively reproduce the bulk tDMRG correlation profiles. In particular, the GHD calculations capture all prominent fine structures of the correlation functions, with systematic deviations arising only from beyond-Euler effects—most notably front broadening—and from finite-time limitations in resolving extremely sharp features. To the best of our knowledge, this constitutes the first systematic benchmark of GHD predictions for linear-response correlation functions against large-scale tDMRG simulations. Beyond providing a transparent physical interpretation of the observed transport behavior, our results offer a stringent test of the applicability and predictive power of GHD in random, time-aperiodic integrable quantum dynamics.

The remainder of this paper is organized as follows.
In Sec.~\ref{sec:ABA and models} we review the Yang--Baxter structure of the XXZ model, construct the inhomogeneous transfer matrix, derive the integrable circuit building blocks $U_0$ and $U_1$,and specify the classes of circuits
considered in this work. In Sec.~\ref{sec:tba and ghd} we solve the Bethe equations associated to our circuit in various parameter regimes, develop the thermodynamic Bethe ansatz, and set up the GHD description that provides dressed quasiparticle data and hydrodynamic observables for our circuits.
Section~\ref{sec:Numerical method} summarizes our numerical methods, including tDMRG simulations of spin transport and the practical implementation of the TBA/GHD equations for quasiperiodic and random circuits. In Secs.~\ref{sec:Fibresult} and \ref{sec:random result} we present our numerical results for quasiperiodic (Fibonacci sequence in particular) and random quantum circuits, respectively,and compare tDMRG simulations with GHD predictions. We conclude with a summary and outlook in Sec.~\ref{sec:Outlook}, while additional numerical details and further data, including transport under the Thue--Morse sequence, are presented in the Appendix.

\section{Yang--Baxter integrability and circuit models}
\label{sec:ABA and models}
In this section, we present our construction of integrable quantum circuits without time-translation symmetry. We begin by reviewing the standard Yang–Baxter framework for XXZ-type spin chains and the associated transfer matrix formalism. As a warm-up, we demonstrate how a simple two-site inhomogeneous transfer matrix gives rise to an integrable Floquet circuit. Building on this, we construct a three-site inhomogeneous transfer matrix and derive a pair of commuting circuit unitaries, $U_0$ and $U_1$. These unitaries form the fundamental building blocks for realizing integrable dynamics under quasiperiodic and completely random protocols, which will be explored in detail throughout the rest of the paper.

\subsection{Yang--Baxter integrability and XXZ-type quantum circuits}
Consider an integrable spin-\(\tfrac12\) chain. Although a universal
definition of quantum integrability is still lacking, a central and widely
accepted feature is the presence of an infinite family of mutually commuting
conserved quantities with local densities,
\begin{equation}
    [Q_m,Q_n] = 0 \quad \forall\, m,n,
\end{equation}
where \(Q_n\) is an \(n\)-local (or \(n\)-site local) operator. Within the Yang--Baxter framework, these conserved charges are generated by the derivatives of the logarithm of the transfer matrix \(T(u)\)\cite{korepin1997,faddeev1996ABA,Vsamaj—book},
\begin{equation}
    \begin{aligned}
        \frac{\mathrm{d}}{\mathrm{d}u} \ln T(u)
        &= \sum_{n=1}^{\infty} Q_n\, u^{n-1}, \\
        [T(u),T(v)]&=0 \qquad \forall\,u,v,
    \end{aligned}
\end{equation}
so that the fundamental commutation relation of transfer matrices for arbitrary
spectral parameters \(u\) and \(v\) guarantees that all charges are mutually commuting. For example, the familiar XXZ Hamiltonian~(\ref{eq:xxzH}) corresponds to $Q_1$ of the six-vertex model transfer matrix. The commutativity of the transfer matrices is ensured by the Yang--Baxter equation (YBE) obeyed by the
\(R\)-matrix~\cite{korepin1997,faddeev1996ABA,Vsamaj—book},
\begin{equation}
  \begin{aligned}
    &R_{\chi_1,\chi_2}(u-v)\,
      R_{\chi_1,\chi_3}(u-w)\,
      R_{\chi_2,\chi_3}(v-w)
      \\[2pt]
    &= R_{\chi_2,\chi_3}(v-w)\,
       R_{\chi_1,\chi_3}(u-w)\,
       R_{\chi_1,\chi_2}(u-v)\, ,
  \end{aligned}
  \label{YBE Equation}
\end{equation}
where \(\chi_n\) denotes the Hilbert space associated with the spin at lattice site \(n\) and \(\chi_0\) is an auxiliary space isomorphic to the local spin-\(\tfrac12\) space, \(\chi_n \simeq \mathbb{C}^2\). The transfer matrix $T(u)$ is then constructed from the above $R$-matrices, as illustrated in Fig.~\ref{fig:example of Circuits}(a),
\begin{equation}
    T(u) = \operatorname{Tr}_{\chi_0}
    \left( \prod_{n=1}^N R_{\chi_0,\chi_n}(u) \right).
    \label{transfer operator}
\end{equation}
In the Yang-Baxter framework, different solutions to the YBE correspond to different integrable models.

In this work we focus on XXZ-type integrability, characterized by
$R$-matrix of the form~\cite{faddeev1996ABA,Vsamaj—book} 
\begin{equation}
    R(u;\eta) =
    \begin{pmatrix}
        1 & 0 & 0 & 0 \\
        0 & \dfrac{\sin u}{\sin(u+\eta)} &
            \dfrac{\sin\eta}{\sin(u+\eta)} & 0 \\
        0 & \dfrac{\sin\eta}{\sin(u+\eta)} &
            \dfrac{\sin u}{\sin(u+\eta)} & 0 \\
        0 & 0 & 0 & 1
    \end{pmatrix},
    \label{Rmatrix}
\end{equation}
where \(u\in\mathbb{C}\) is the spectral parameter and \(\eta\in\mathbb{C}\) is related to the XXZ anisotropy \(\Delta\) in Hamiltonian~(\ref{eq:xxzH}). Note that the \(R\)-matrix reduces to the permutation operator at \(u=0\),
\begin{equation}
    R(0;\eta) = P =
    \begin{pmatrix}
        1 & 0 & 0 & 0 \\
        0 & 0 & 1 & 0 \\
        0 & 1 & 0 & 0 \\
        0 & 0 & 0 & 1
    \end{pmatrix}.
\end{equation}

As discussed in the introduction, spin transport at half filling depends qualitatively on the anisotropy \(\Delta\): the isotropic point, gapless regime and gapped regime exhibit distinct dynamical behaviors. It is therefore convenient to classify the corresponding \(R\)-matrices into three functional forms: trigonometric (gapless phase), hyperbolic (gapped phase) and rational (isotropic limit). A detailed account of this classification and our convention is given in Appendix~\ref{Appendix A}.

\subsection{A warm-up: integrable Floquet circuits from two-site inhomogeneous transfer matrix}
\label{subsec:2site-circuit}

We now turn to the construction of integrable quantum circuits and review how a simple two-site inhomogeneous transfer matrix gives rise to a Floquet circuit that explicitly breaks continuous time-translation symmetry while remaining integrable. This example also motivates our use of a three-site periodic transfer matrix to construct integrable models that completely break time-translation symmetry.

\begin{figure}[!t]
    \centering
    \includegraphics[width=0.9\linewidth]{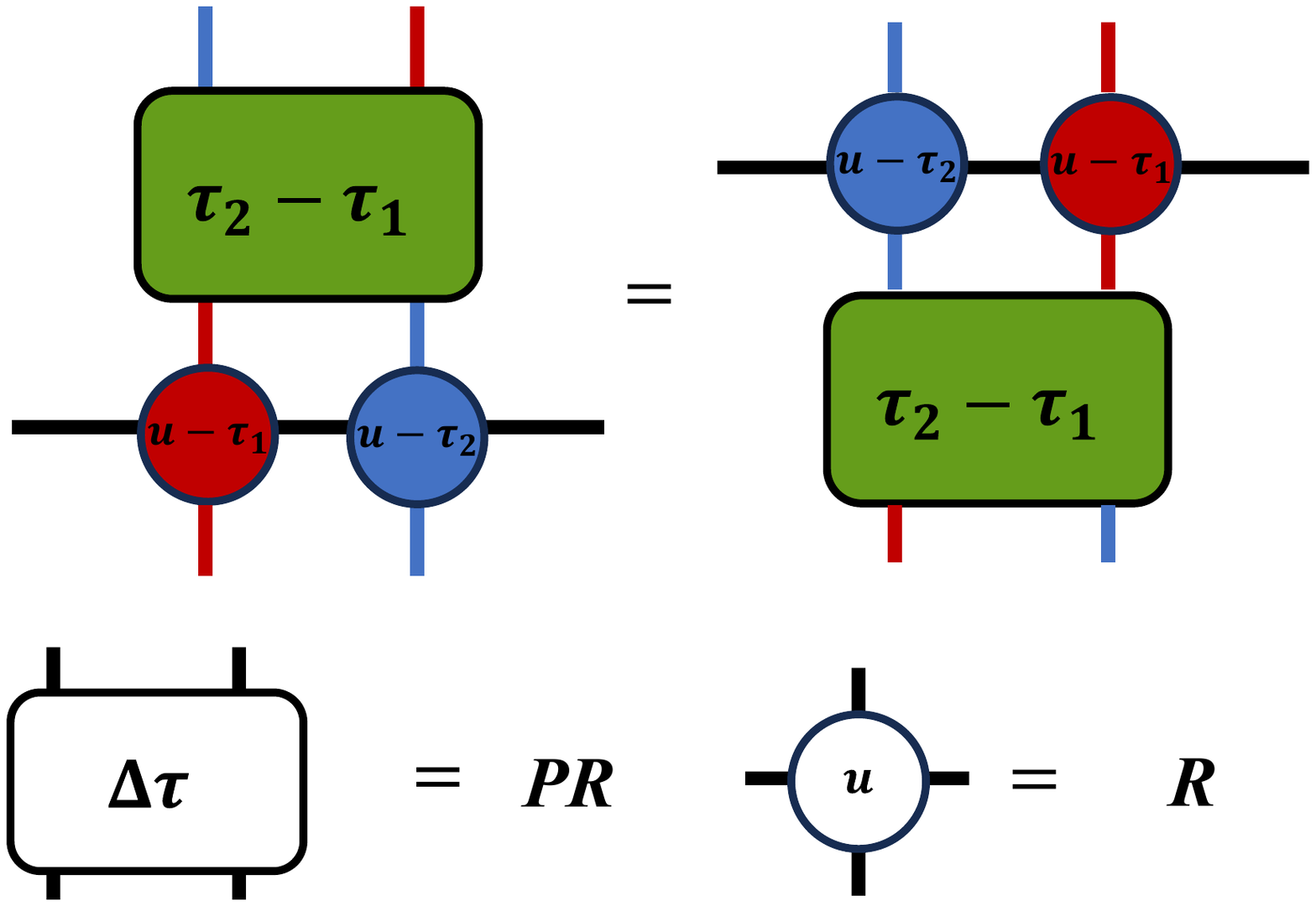}
    \caption{Graphical interpretation of Eq.~\eqref{YBE}. The red (blue)
    circle denotes \(R(u-\tau_1)\) [\(R(u-\tau_2)\)], while the rounded
    rectangle denotes the local gate \((PR)(\tau_2-\tau_1)\), which swaps the
    two \(R\)-matrices.}
    \label{fig:Picture of YBE}
\end{figure}

We first introduce a convenient modified Yang--Baxter relation
\cite{integrablecircuitratchets,Paletta2025IntegrabilityAC},
\begin{equation}
    \begin{aligned}
        &\bigl(PR\bigr)_{\chi_1\chi_2}(\tau_{2}-\tau_{1})\,
        R_{\chi_1\chi_0}(u-\tau_1)\,
        R_{\chi_2\chi_0}(u-\tau_2) \\
        &\qquad = R_{\chi_1\chi_0}(u-\tau_2)\,
           R_{\chi_2\chi_0}(u-\tau_1)\,
           \bigl(PR\bigr)_{\chi_1\chi_2}(\tau_{2}-\tau_{1}) ,
    \end{aligned}
    \label{YBE}
\end{equation}
where \(P_{\chi_1\chi_2}\) is the permutation operator and
\((PR)_{\chi_1\chi_2} \equiv P_{\chi_1\chi_2}R_{\chi_1\chi_2}\).
Figure~\ref{fig:Picture of YBE} provides a graphical representation:
the rounded rectangle labelled \(\tau_2-\tau_1\) represents the local gate \((PR)(\tau_2-\tau_1)\), which exchanges two \(R\)-matrices with spectral parameters \(u-\tau_1\) and \(u-\tau_2\). The Floquet circuit is built from the $PR$ local gate above,  and its integrability follows from demonstrating its commutativity with a two-site inhomogeneous transfer matrix constructed from the red and blue $R$-matrices in Fig.~\ref{fig:Picture of YBE}, as discussed in more detail below~\cite{faddeev1996ABA,XXXcircuit,XXZcircuitprosen,XXZCircuitGGE,randominteEssler,randomintsr}.

\begin{figure}[!t]
    \centering
    \includegraphics[width=0.9\linewidth]{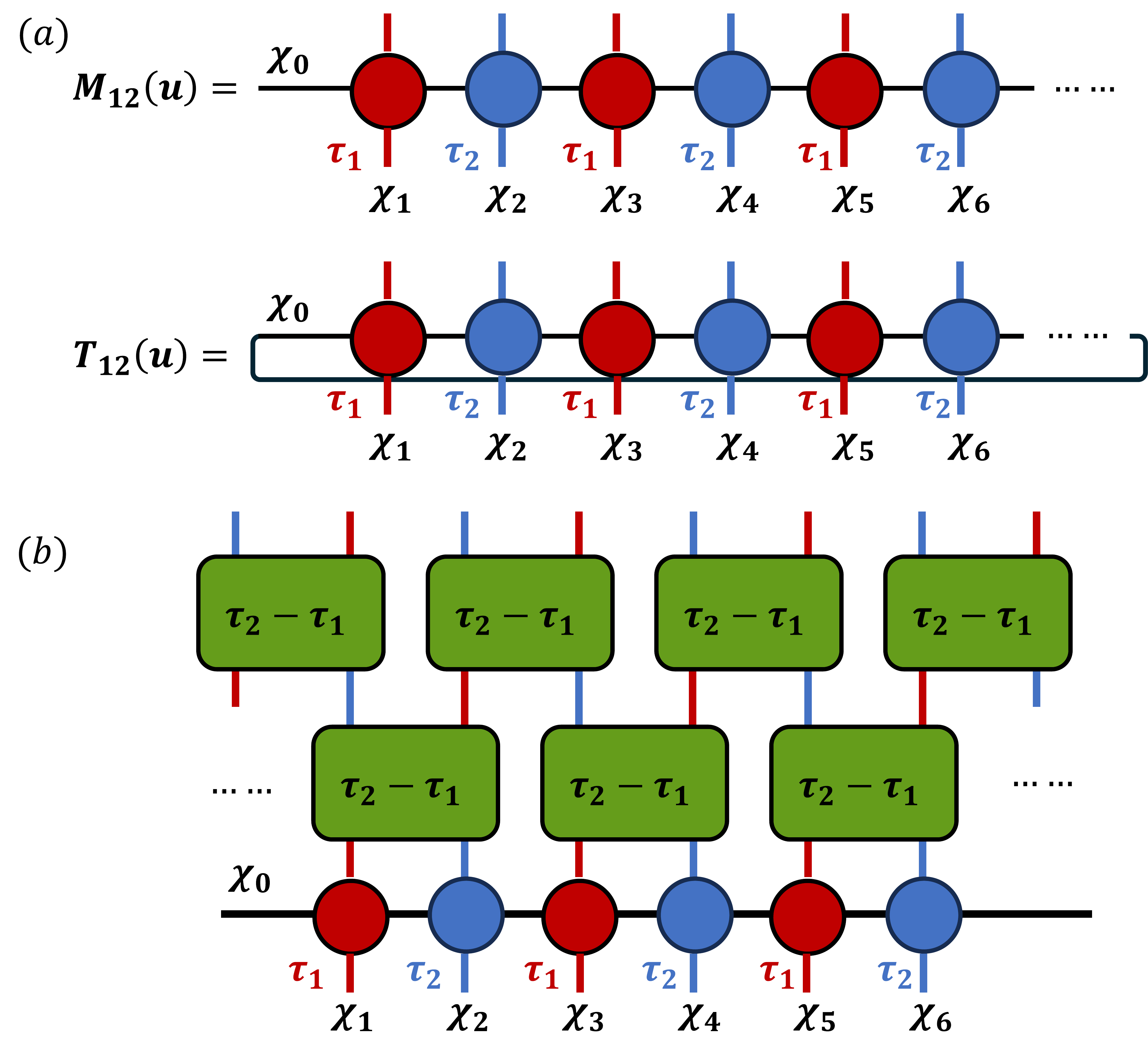}
    \caption{(a) Schematic illustration of the monodromy and transfer matrices with two-site translational symmetry.  (b) Action of the circuit $U$ on the staggered monodromy matrix. A single layer of $U$ exchanges the $R$ - matrices on odd and even bonds. In the thermodynamic limit this corresponds to a one-site translation of the inhomogeneity pattern depicted by the blue and red lines. Two successive layers restore the original configuration. As a result, $U$ commutes with $T_{12}(u)$ for all $u$, confirming the integrability of the circuit.}
    \label{fig:example of Circuits}
\end{figure}

Consider an infinite chain with a staggered pattern of \(R\)-matrices, \(R(u-\tau_1)\) on odd sites and \(R(u-\tau_2)\) on even sites. The corresponding inhomogeneous monodromy matrix is
\begin{equation}
      M_{12}(u)
      = \prod_{n=-\infty}^{\infty}
        R_{\chi_0,\chi_{2n-1}}(u-\tau_1)\,
        R_{\chi_0,\chi_{2n}}(u-\tau_2),
        \label{eq:monodromy}
\end{equation}
where \(\chi_0\) labels the auxiliary space and \(\chi_n\) the physical space on the chain. Fig.~\ref{fig:example of Circuits}(a) shows a sketch of the monodromy matrix~(\ref{eq:monodromy}) and the corresponding transfer matrix $T_{12}(u) = \mathrm{Tr}_{\chi_0}\bigl[M_{12}(u)\bigr]$. The two-layer Floquet unitary $U$ is constructed by tiling the chain with the local gate \((PR)(\tau_2-\tau_1)\) applied alternately on even and odd bonds,
\begin{equation}
      \begin{aligned}
        U_{\mathrm{even}} &=
        \prod_{n\in\mathrm{even}}
        (PR)_{\chi_{n},\chi_{n+1}}(\tau_2-\tau_1), \\
        U_{\mathrm{odd}} &=
        \prod_{n\in\mathrm{odd}}
        (PR)_{\chi_{n},\chi_{n+1}}(\tau_2-\tau_1), \\
        U &= U_{\mathrm{even}} U_{\mathrm{odd}},
      \end{aligned}
\end{equation}
as depicted in Fig.~\ref{fig:example of Circuits}(b). 
We begin by demonstrating the integrability of the Floquet unitary $U$ through a pictorial proof, showing that it commutes with the family of two-site inhomogeneous transfer matrices obtained from Eq.~(\ref{eq:monodromy}), using the modified YBE depicted in Fig.~\ref{fig:Picture of YBE}.  Consider the action on the transfer matrix $T_{12}(u)$ with the Floquet unitary $U$ as shown in Fig.~\ref{fig:example of Circuits}(b). Using the YBE in Fig.~\ref{fig:Picture of YBE}, when passing $T_{12}(u)$ through a single layer \(U_{\mathrm{odd}}\) or \(U_{\mathrm{even}}\), neighboring $R$-matrices on alternating bonds are exchanged, which is equivalent to shifting the staggered pattern of inhomogeneities by one lattice site in the thermodynamic limit. Two unitary layers thus restore the original inhomogeneity configuration and hence $T_{12}(u)$ commutes with the full Floquet unitary.

To formalize the above pictorial argument, we now introduce the ``flipped" monodromy matrix \(M_{21}(u)\), obtained by interchanging the inhomogeneous parameters \(\tau_1\) and \(\tau_2\),
\begin{equation}
    M_{21}(u)
    = \prod_{n=-\infty}^{\infty}
      R_{\chi_0,\chi_{2n-1}}(u-\tau_2)\,
      R_{\chi_0,\chi_{2n}}(u-\tau_1).
\end{equation}
Using Eq.~\eqref{YBE} locally on each bond one finds the intertwining
relations
\begin{equation}
    \begin{aligned}
        U_{\mathrm{odd}}\, M_{12}(u) &= M_{21}(u)\, U_{\mathrm{odd}}, \\
        U_{\mathrm{even}}\, M_{21}(u) &= M_{12}(u)\, U_{\mathrm{even}}.
    \end{aligned}
\end{equation}
Multiplying these two relations we obtain
\begin{equation}
    U\,M_{12}(u)=M_{12}(u)\,U,
\end{equation}
so the two-layer circuit \(U\) commutes with the inhomogeneous monodromy matrix \(M_{12}(u)\). Tracing over the auxiliary space \(\chi_0\) yields the corresponding inhomogeneous transfer matrix and the commutation relation between the ciruit and the tranfer matrix:
\begin{equation}
    %\begin{aligned}
        %T_{12}(u) = \mathrm{Tr}_{\chi_0}\bigl[M_{12}(u)\bigr], \\
        [U,T_{12}(u)] = 0 \qquad \forall\,u.
        \label{Commutation}
    %\end{aligned}
\end{equation}

Although \(T_{12}(u)\) is invariant under translation by two lattice sites as opposed to the homogeneous transfer matrix in Eq.~\eqref{transfer operator}, the Yang--Baxter equation still guarantees
\begin{equation}
    [T_{12}(u),T_{12}(v)] = 0 \qquad \forall\,u,v,
\end{equation}
thus the Floquet circuit \(U\) preserves the integrable structure encoded in the inhomogeneous transfer matrix. 

Moreover, one can show that the Floquet unitary $U$ can be directly expressed using the transfer matrix itself as~\cite{pozsgaymedint,Paletta2025IntegrabilityAC,floquetbaxter,integrablecircuitratchets}
\begin{equation} 
    U = T_{12}^{-1}(\tau_1)\, T_{12}(\tau_2), 
\end{equation} 
which provides a more refined characterization of the integrable structure and
immediately implies Eq.~\eqref{Commutation}.

\subsection{Elementary circuit layers from three-site inhomogeneous transfer matrix}
\begin{figure}[!t]
    \centering
    \includegraphics[width=1.0\linewidth]{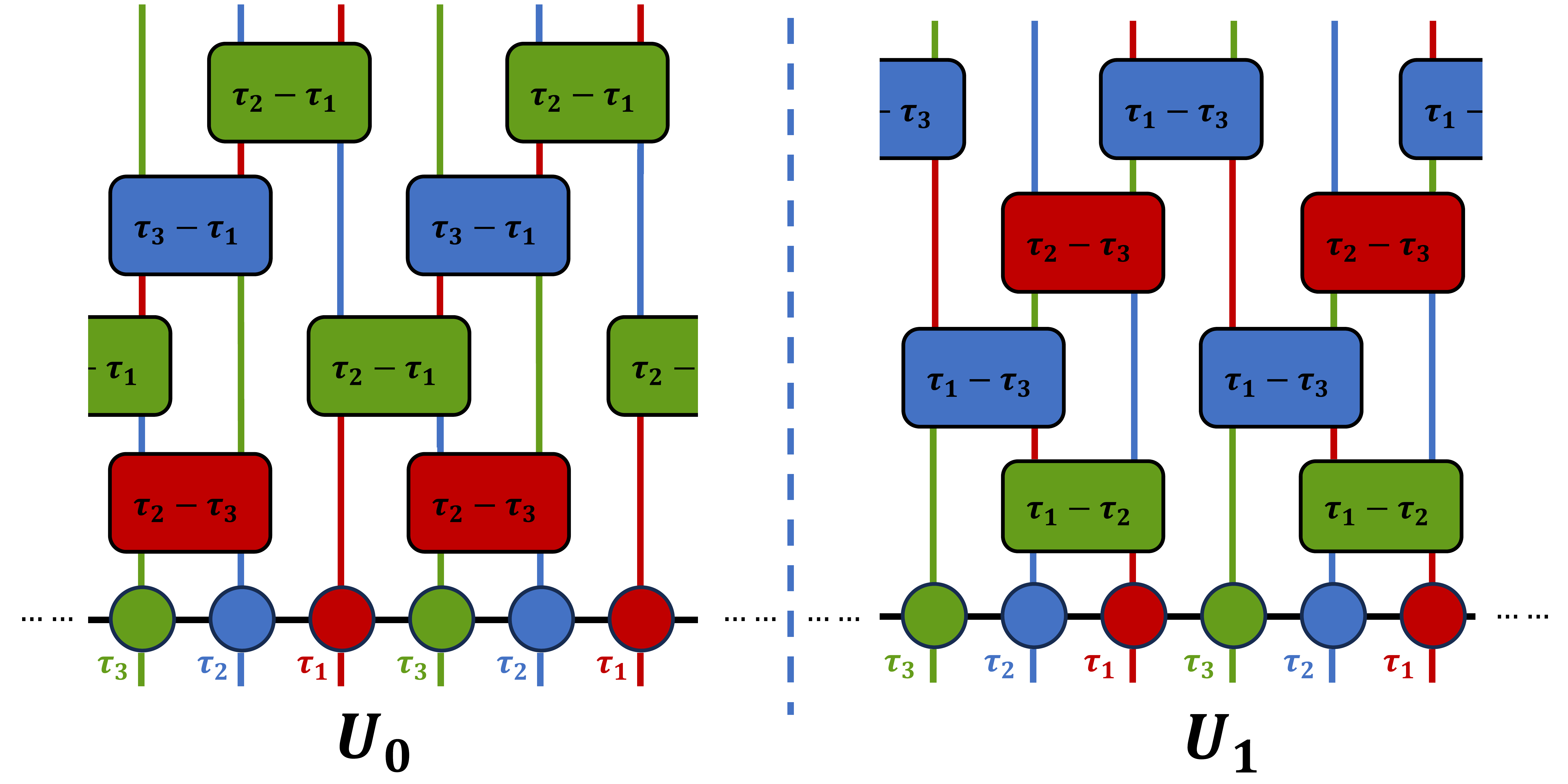}
    \caption{Schematic circuit construction from a three-site inhomogeneous transfer matrix. Red, blue and green circles represent \(R(u-\tau_1)\), \(R(u-\tau_2)\), and \(R(u-\tau_3)\), respectively. Local gates \((PR)(\tau_i-\tau_j)\) are color-coded using the same RGB convention: when a gate exchanges two \(R\)-matrices associated with \(\tau_i\) and \(\tau_j\), the resulting block is assigned the remaining third color. We construct two unitarily inequivalent circuit elements $U_0$ and $U_1$ that both commute with the transfer matrix $T(u)$.}
    \label{fig:Our Circuits}
\end{figure}

The above construction can be naturally generalized to three inhomogeneity parameters \(\tau_1,\tau_2,\tau_3\)\cite{Paletta2025IntegrabilityAC}, which is central to our integrable random circuit models. We define a three-site inhomogeneous transfer matrix:
\begin{equation}
  \begin{aligned}
    T(u)
    &= \operatorname{Tr}_{\chi_0}\Biggl(
       \prod_{n=-\infty}^{\infty}
       R_{\chi_0,3n-2}(u-\tau_3)\,
       R_{\chi_0,3n-1}(u-\tau_2)\,
       \\
    &\hphantom{=\operatorname{Tr}_{\chi_0}\Biggl(
       \prod_{n=-\infty}^{\infty}}
       {}\times
       R_{\chi_0,3n}(u-\tau_1)
       \Biggr).
  \end{aligned}
  \label{eq:TMour}
\end{equation}
Correspondingly, we contruct unitary circuit elements built from local gates \((PR)(\tau_i - \tau_j)\), with \((i,j) \in \{(2,3), (2,1), (1,3)\}\). Crucially, to guarantee integrability, we require that the resulting circuit element commutes with the above transfer matrix~(\ref{eq:TMour}).  Two inequivalent unitary elements $U_0$ and $U_1$ are constructed, as illustrated in Fig.~\ref{fig:Our Circuits}. By following the YBE in Fig.~\ref{fig:Picture of YBE}, one can verify that the inhomogeneous parameter returns to the original pattern after passing through four layers of unitary gates, thereby ensuring that the transfer matrix commutes with both $U_0$ and $U_1$.

In direct analogy with the two-site case, one can similary show that the two circuit elements $U_0$ and $U_1$ can be written explicitly in terms of the transfer matrix itself:
\begin{equation}
    \begin{aligned}
        U_0 &= T^{-1}(\tau_1)\, T(\tau_2), \\
        U_1 &= T^{-1}(\tau_3)\, T(\tau_1). 
    \end{aligned} \label{eq:utoTMour}
\end{equation}
Since both circuit elements commute with the transfer matrix, \textit{any} sequence of $\{U_0, U_1\}$ preserves the underlying integrability structure, including those that lack any time-translation symmetry. From the above expressions, it is obvious that \([U_0,U_1]=0\). Nevertheless, since they are not unitarily equivalent to one another, the resulting dynamics from different sequences of $\{U_0, U_1\}$ are still nontrivial.

\subsection{Random and quasiperiodic circuits considered in this work}
\label{subsec:protocols}
Having constructed the circuit elements \(U_0\) and \(U_1\), we now specify the classes of circuit sequences considered in this work. In all cases, the discrete time evolution is generated by products of \(U_0\) and \(U_1\) according to a binary word \(\{w_t\}_{t=1}^L\) with letters \(w_t\in\{0,1\}\), where \(w_t=0\) (\(1\)) denotes the application of \(U_0\) (\(U_1\)) at time step
\(t\).

\begin{figure}[!t]
    \centering
    \includegraphics[width=0.95\linewidth]{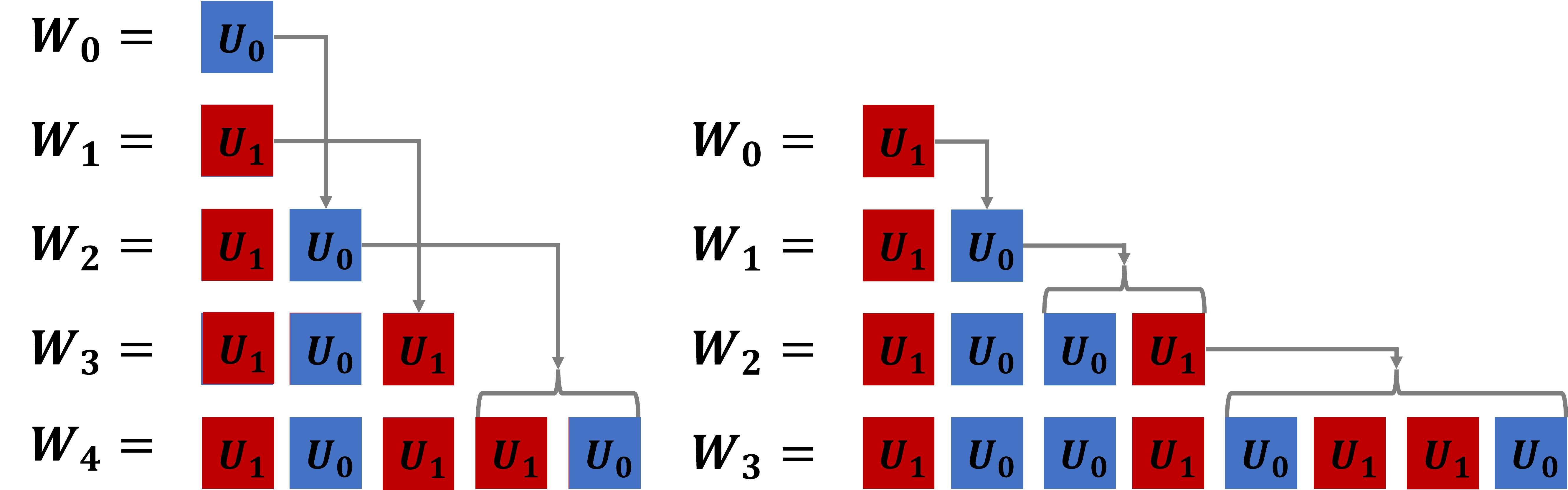}
    \caption{ Quasiperiodic circuits generated from Thue-Morse (left) and Fibonacci (right) words. Blue (red) blocks represent applications of \(U_0\) (\(U_1\)). Both sequences are aperiodic but have well-defined densities of \(U_0\) and \(U_1\) in the long time limit.}
    \label{fig:TM_FB_sequences}
\end{figure}

We consider two classes of protocols:

\begin{itemize} 
    \item \textbf{Stochastic sequence.}  
    For each of $N$ independent realizations we generate a random binary word by choosing $w_t=1$ with probability $p$ and $w_t=0$ with probability $1-p$, independently at each time step. The resulting evolution operator is $U^{(r)}(L) = \prod_{t=1}^L U_{w_t}$ for realization $r$. Observables are computed by averaging over all realizations. \label{stochastic protocol}
    
    \item \textbf{Deterministic quasiperiodic sequence.}
    We consider two prototypical quasiperiodic sequences, namely, the Fibonacci and Thue-Morse sequence.

    The Thue--Morse sequence is generated by the substitution rule \(0 \to 01\), \(1 \to 10\). Starting from the seed word \(1\), the first few generations are
    \begin{equation}
        1 \;\xrightarrow{}\; 10 \;\xrightarrow{}\; 1001
        \;\xrightarrow{}\; 10010110 \;\xrightarrow{}\; \cdots.
    \end{equation}
    In the infinite-length limit the densities of \(0\) and \(1\) are exactly equal, but the sequence is strictly aperiodic.

    The Fibonacci sequence is generated by the concatenation rule
    \(w_{n+1} = w_n w_{n-1}\) with initial words \(w_0=0\) and \(w_1=1\). The first few words are
    \(1,\ 10,\ 101,\ 10110,\ 10110101,\dots\).
    In the infinite-length limit the density of \(1\) approaches the inverse golden ratio, \((\sqrt{5}-1)/2 \approx 0.618\).

    A schematic representation of the two quasiperiodic sequences is shown in Fig.~\ref{fig:TM_FB_sequences}, where blue (red) blocks correspond to \(U_0\) (\(U_1\)).
    
\end{itemize}

\section{Thermodynamic Bethe ansatz and generalized hydrodynamics for quantum circuits}
\label{sec:tba and ghd}
In the previous section we introduced the three-site inhomogeneous transfer matrix,
parametrized by $(\tau_1,\tau_2,\tau_3)$, as the basic building block of our integrable
circuit construction. The circuit layers $U_0$ and $U_1$ are defined as suitable combinations of this transfer matrix and its inverse, so that their quasiparticle content can be read off directly from
the transfer-matrix spectrum. In this section we determine the spectrum of the
inhomogeneous transfer matrix in various parameter regimes using the algebraic Bethe
ansatz (ABA), and, invoking the string hypothesis, derive the corresponding
TBA description.

To avoid repeating essentially identical manipulations across regimes, we present the
isotropic (XXX) limit as a worked example, where we spell out the full chain of steps:
ABA eigenvalues $\rightarrow$ Bethe equations $\rightarrow$ TBA root densities $\rightarrow$
the quasiparticle data (quasi-energies, dressed charges, and effective velocities) entering
the circuit unitaries $U_0$ and $U_1$. In the gapped and gapless XXZ regimes we therefore focus on
the regime-specific inputs---namely the string content, and kernels
$a_n(\lambda)$ and $A_{nm}(\lambda)$---since once these are specified, the subsequent
dressing relations and the mapping from transfer-matrix data to the quasiparticles of
$\{U_0,U_1\}$ follow verbatim from the XXX construction. Finally, we summarize the GHD
ingredients that will be used for Euler-scale predictions of
spin transport and correlation functions, which will be compared against tDMRG simulations.

\subsection{Thermodynamic Bethe ansatz for inhomogeneous transfer matrix} \label{sub:TBAintro}
\subsubsection{Isotropic (XXX) point}
At the isotropic point the $R$--matrix reduces to
\begin{equation}
    \begin{aligned}
        R_{\mathrm{XXX}}(\lambda) &=
        \begin{pmatrix}
            1 & 0 & 0 & 0 \\
            0 & \dfrac{-\mathrm{i}\lambda}{-\mathrm{i}\lambda+1 } &
                \dfrac{1}{-\mathrm{i}\lambda+1} & 0 \\
            0 & \dfrac{1}{-\mathrm{i}\lambda+1} &
                \dfrac{-\mathrm{i}\lambda}{-\mathrm{i}\lambda+1} & 0 \\
            0 & 0 & 0 & 1
        \end{pmatrix}.
    \end{aligned}
    \label{RXXX}
\end{equation}
Using the ABA, the $N$--particle
eigenvalues of the transfer matrix in Eq.~\eqref{eq:TMour},
$T(\lambda)\ket{\{\lambda_j\}} = \Lambda(\lambda;\{\lambda_j\})\ket{\{\lambda_j\}}$ are given by~\cite{faddeev1996ABA}
\begin{align}
    \Lambda(\lambda;\{\lambda_j\})
    &= \prod_{j=1}^{N}
       \frac{\lambda-\lambda_{j}-\frac{\mathrm{i}}{2}}
            {\lambda-\lambda_{j}+\frac{\mathrm{i}}{2}} \nonumber
    \\[-2pt]
    &\hspace{-1.4em}+ % pull the second line a bit to the left
    \Bigl[\prod_{a=1}^{3}\frac{\lambda-\tau_a}{\lambda-\tau_a+\mathrm{i}}\Bigr]^{\!L/3}
    \prod_{j=1}^{N}
       \frac{\lambda-\lambda_{j}+\frac{3\mathrm{i}}{2}}
            {\lambda-\lambda_{j}+\frac{\mathrm{i}}{2}} \, .
    \label{eq:xxxtmev}
\end{align}
Here $\{\lambda_j\}_{j=1}^N$ are the rapidities of the quasiparticle excitations, $\lambda$ is the spectral parameter of the transfer matrix, and $\{\tau_a\}_{a=1,2,3}$ denotes the inhomogeneity parameters. The corresponding Bethe equations for the transfer matrix read
\begin{equation}
    \begin{aligned}
        \Biggl[\prod_{a=1}^{3}
        \Bigl(\frac{\lambda_j-\tau_{a}+\frac{\mathrm{i}}{2}}
        {\lambda_j-\tau_{a}-\frac{\mathrm{i}}{2}}\Bigr)
        \Biggr]^{\tfrac{L}{3}}
        &=
        \prod_{\substack{l=1\\ l\neq j}}^{N}
        \frac{\lambda_{j}-\lambda_{l}+\mathrm{i}}
        {\lambda_{j}-\lambda_{l}-\mathrm{i}}, \\
        &\qquad j = 1,2,\dots,N.
    \end{aligned}
    \label{eq:bexxx}
\end{equation}
In the thermodynamic limit, the solutions of Eq.~\eqref{eq:bexxx} are assumed to organize into string-like patterns, as is standard in both Hamiltonian and Floquet-circuit settings. This assumption is known as the string hypothesis~\cite{takahashi1972,takahashiTBA}, and takes the form:
\begin{equation}
    \lambda^{n,k}_{\alpha}
    = \lambda^{n}_{\alpha}
      + \mathrm{i}\Bigl(\frac{n+1}{2}-k\Bigr),
    \qquad k = 1,2,\dots,n.
    \label{eq:xxxrootstring}
\end{equation}
Here $\lambda^{n}_{\alpha}\in\mathbb{R}$ is the real string center, $n$ is the string length, and $\alpha$ labels different string species of the same length. An $n$--string is interpreted as a bound state of $n$ magnons.

Based on the string hypothesis, the TBA~\cite{takahashiTBA} focuses on the thermodynamic limit $L\to\infty$ at fixed quasiparticle density $N/L$. It postulates that a thermodynamic macrostate of the integrable system is fully characterized by the distribution of the string centers. We therefore introduce density functions $\rho_n(\lambda)$ such that $L\rho_n(\lambda)\,\mathrm{d}\lambda$
gives the number of occupied $n$--string states with centers in the interval $[\lambda,\lambda+\mathrm{d}\lambda)$, and total densities $\rho_n^t(\lambda)$ such that $L\rho_n^t(\lambda)\,\mathrm{d}\lambda$ is the total number of available $n$--string states in the same interval. The relation between $\rho_n(\lambda)$ and $\rho^t_n(\lambda)$ are defined by the filling factor $\theta_n(\lambda)$: 
\begin{equation}
    \rho_n(\lambda)=\theta_n(\lambda) \rho_n^{t}(\lambda).
\end{equation}

The TBA equations for $\rho_n(\lambda)$ and $\rho^t_n(\lambda)$ follow
from the Bethe equation ~\eqref{eq:bexxx} in the usual way. Following the standard
procedure~\cite{takahashiTBA,Vsamaj—book}, we obtain a closed integral equation of $\rho_n^{t}(\lambda)$:
\begin{equation}
    \begin{aligned}
        \rho_n^{t}(\lambda)= {}&
        \tfrac{1}{3}\Bigl(a_n(\lambda-\tau_1)
                        + a_n(\lambda-\tau_2)
                        + a_n(\lambda-\tau_3)\Bigr) \\
        &\;-\sum_{m=1}^{\infty}\int_{-\infty}^{\infty} \mathrm{d}\lambda'\,
           A_{nm}(\lambda-\lambda')\,
           \theta_m(\lambda')\,\rho^t_m(\lambda') \,.
    \end{aligned}
    \label{eq:TBArhoxxx}
\end{equation}
Here the kernels entering Eq.~\eqref{eq:TBArhoxxx} are
\begin{align}
    a_n(\lambda)
      &= \frac{1}{2\pi}\,\frac{n}{\lambda^2 + \tfrac{n^2}{4}}\ ,\label{eq:anxxx}
    \\[2mm]
    A_{nm}(\lambda)
     &= (1-\delta_{nm})\,a_{|n-m|}(\lambda)
        + 2\,a_{|n-m|+2}(\lambda) + \cdots
        \notag\\[-2pt]
     &\qquad {}+ 2\,a_{n+m-2}(\lambda) + a_{n+m}(\lambda).
     \label{eq:Anm}
\end{align}
When the inhomogeneities vanish, $\tau_1=\tau_2=\tau_3$, the TBA equations
\eqref{eq:TBArhoxxx} reduce to those of the homogeneous XXX Hamiltonian
chain~\cite{takahashiTBA}.

The filling functions $\theta_n(\lambda)$ are determined by the Yang--Yang
equations~\cite{yangyangtba,takahashiTBA}. In this paper we focus on transport
around the infinite-temperature, half-filled equilibrium state. In this case
the Yang--Yang equations are not affected by the inhomogeneities
$\tau_a$~\cite{randominteEssler}, and the filling functions take a simple,
rapidity-independent form 
\begin{equation}
    \theta_n(\lambda) = \frac{1}{(1+n)^2},
    \label{eq:theta}
\end{equation}
see Refs.~\cite{takahashiTBA,xxxsuperdif18ilievskiprosen}. Plugging this into
Eq.~\eqref{eq:TBArhoxxx} one can solve the resulting linear integral equations
numerically to obtain $\rho_n^t(\lambda)$ and $\rho_n(\lambda)$. Equation~\eqref{eq:TBArhoxxx} is a closed integral equation for the total
densities $\rho^t_n(\lambda)$. Physically, it expresses the fact that the bare
distributions generated by the inhomogeneities are dressed by the two-body
scattering encoded in the kernel $A_{nm}$.

Recall from Eq.~(\ref{eq:utoTMour}) that our circuit elements $U_0$ and $U_1$ can be
written in terms of transfer matrices. For example,
\begin{equation}
    U_0=\bigl[T(\tau_1)\bigr]^{-1}T(\tau_2),
\end{equation}
using Eq.~\eqref{eq:xxxtmev}, the quasi-energy spectrum of $U_0$,
\mbox{$U_0\ket{\{\lambda\}}= \mathrm{e}^{\mathrm{i} \sum_{j=1}^N
\varepsilon(\lambda_j)}\ket{\{\lambda\}}$}, is readily found to be
\begin{equation}
    U_0\ket{\{\lambda\}}
    = \prod_{j=1}^{N}
    \left(
      \frac{\lambda_j - \tau_1 -\tfrac{\mathrm{i}}{2}}
           {\lambda_j - \tau_1 + \tfrac{\mathrm{i}}{2}}
    \right)
    \left(
      \frac{\lambda_j - \tau_2 + \tfrac{\mathrm{i}}{2}}
           {\lambda_j - \tau_2 - \tfrac{\mathrm{i}}{2}}
    \right)
    \ket{\{\lambda\}}.
\end{equation}
Thus the single-particle quasi-energy is a simple function of the Bethe root
$\lambda_j$,
\begin{equation}
    \varepsilon(\lambda_j)
    = k(\lambda_j - \tau_{1})
      - k(\lambda_j - \tau_{2})\,,
\end{equation}
where
\begin{equation}
    k(\lambda)=\mathrm{i}\log\!\left(
    \frac{\lambda+\frac{\mathrm{i}}{2}}{\lambda-\frac{\mathrm{i}}{2}}
    \right).
\end{equation}
The quasi-energy is not an extensive quantity and is defined only modulo
multiples of $2\pi$, but in the GHD description used below only its derivative
with respect to $\lambda$ is relevant. We have
\begin{equation}
    k'(\lambda) = 2\pi a_1(\lambda),
\end{equation}
so that
\begin{equation}
    \varepsilon'(\lambda)
    = 2\pi\bigl(a_1(\lambda-\tau_1)-a_1(\lambda-\tau_2)\bigr).
\end{equation}
Because of the string structure of the Bethe roots, the derivative of the quasi-energy for an $n$--string magnon $\varepsilon_n(\lambda^n_\alpha)=\sum_{j=1}^n \varepsilon(\lambda^{n,j}_\alpha)$ is
\begin{equation}
    \varepsilon'_n(\lambda)
    = 2\pi\bigl(a_n(\lambda-\tau_1)-a_n(\lambda-\tau_2)\bigr),
    \label{eq:quaen_xxxstring}
\end{equation}
which differs from the corresponding Hamiltonian dispersion, see
Refs.~\cite{takahashiTBA,XXZCircuitGHD,integrablecircuitratchets,Paletta2025IntegrabilityAC}.

\subsubsection{Gapped regime}
For the gapped regime we consider the transfer matrix constructed
from the hyperbolic $R$--matrix
\begin{equation}
    R(\lambda;\gamma) =
    \begin{pmatrix}
        1 & 0 & 0 & 0 \\
        0 & \dfrac{\sinh(\lambda)}{\sinh(\lambda+\gamma)} &
            \dfrac{\sinh(\gamma)}{\sinh(\lambda+\gamma)} & 0 \\
        0 & \dfrac{\sinh(\gamma)}{\sinh(\lambda+\gamma)} &
            \dfrac{\sinh(\lambda)}{\sinh(\lambda+\gamma)} & 0 \\
        0 & 0 & 0 & 1
    \end{pmatrix}.
    \label{RGapped}
\end{equation}
Following the same ABA steps as in the isotropic case, one obtains the Bethe
equations
\begin{equation}
   \begin{aligned}
    \Biggl[\prod_{a=1}^{3}
    \frac{\sin(\lambda_j-\tau_a + \mathrm{i}\tfrac{\gamma}{2})}
         {\sin(\lambda_j-\tau_a - \mathrm{i}\tfrac{\gamma}{2})}
    \Biggr]^{\frac{L}{3}}
    &=
    \prod_{\substack{l=1\\l\neq j}}^N
    \frac{\sin(\lambda_j - \lambda_l + \mathrm{i}\gamma)}
         {\sin(\lambda_j - \lambda_l - \mathrm{i}\gamma)},\\
    &\qquad j = 1,2,\dots,N.
   \end{aligned}
    \label{eq:begapped}
\end{equation}
In this regime the Bethe roots $\{\lambda_j\}$ are generally complex, and in the thermodynamic limit they also form strings with the same pattern as in the isotropic case, cf.~Eq.~\eqref{eq:xxxrootstring}. The TBA equations obtained from Eq.~\eqref{eq:begapped} take the form
\begin{equation}
     \begin{aligned}
        \rho_n^{t}(\lambda)= {}&
        \tfrac{1}{3}\Bigl(a_n(\lambda-\tau_1)
                        + a_n(\lambda-\tau_2)
                        + a_n(\lambda-\tau_3)\Bigr) \\
        &\;-\sum_{m=1}^{\infty}\int_{-\frac{\pi}{2}}^{\frac{\pi}{2}}
           \mathrm{d}\lambda'\,
           A_{nm}(\lambda-\lambda')\,
           \theta_m(\lambda')\,\rho^t_m(\lambda') \,,
    \end{aligned}
    \label{eq:TBArhoGAP}
\end{equation}
where the rapidity range is now $\lambda\in[-\pi/2,\pi/2]$. The kernels are
\begin{align}
    a_n(\lambda)
    &= \frac{1}{\pi}\,
       \frac{\sinh(n\gamma)}
            {\cosh(n\gamma)-\cos(2\lambda)} , \label{eq:anGAP}
    \\[2mm]
    A_{nm}(\lambda)
    &= (1-\delta_{nm})\,a_{|n-m|}(\lambda)
       + 2\,a_{|n-m|+2}(\lambda) + \cdots
       \notag\\[-2pt]
    &\qquad {}+ 2\,a_{n+m-2}(\lambda) + a_{n+m}(\lambda).
    \label{eq:AnmGAP}
\end{align}
At half filling and infinite temperature the filling functions $\theta_n(\lambda)$
coincide with those of the isotropic chain [see Eq.~\eqref{eq:theta}],
so that the only difference with respect to the isotropic case lies in the domain of $\lambda$ and the scattering kernels~\eqref{eq:anGAP}--\eqref{eq:AnmGAP}. The
quasi-energies of the $n$--string magnons of the circuit unitary $U$ are
expressed in terms of $a_n(\lambda)$ in exactly the same way as in
Eq.~\eqref{eq:quaen_xxxstring}.

\subsubsection{Gapless regime}
In the gapless regime the $R$--matrix takes the trigonometric form
\begin{equation}
    R(\lambda;\gamma) =
    \begin{pmatrix}
        1 & 0 & 0 & 0 \\
        0 & \dfrac{\sin(\lambda)}{\sin(\lambda+\gamma)} &
            \dfrac{\sin(\gamma)}{\sin(\lambda+\gamma)} & 0 \\
        0 & \dfrac{\sin(\gamma)}{\sin(\lambda+\gamma)} &
            \dfrac{\sin(\lambda)}{\sin(\lambda+\gamma)} & 0 \\
        0 & 0 & 0 & 1
    \end{pmatrix}.
    \label{RGapless}
\end{equation}
The Bethe equations for the transfer matrix~\eqref{eq:TMour} in this regime
take the form
\begin{equation}
  \begin{aligned}
      \Biggl[\prod_{a=1}^{3}
    \frac{\sinh(\lambda_j-\tau_a + \mathrm{i}\tfrac{\gamma}{2})}
         {\sinh(\lambda_j-\tau_a - \mathrm{i}\tfrac{\gamma}{2})}
    \Biggr]^{\frac{L}{3}}
    &=
    \prod_{\substack{l=1\\l\neq j}}^N
    \frac{\sinh(\lambda_j - \lambda_l + \mathrm{i}\gamma)}
         {\sinh(\lambda_j - \lambda_l - \mathrm{i}\gamma)},\\
    &\qquad j = 1,2,\dots,N.
  \end{aligned}
    \label{eq:begapless}
\end{equation}
In the thermodynamic limit the solutions of Eq.~\eqref{eq:begapless} again form
strings, but their structure is more involved than in the previous cases
\cite{takahashiTBA,takahashi1972}. In this work we restrict ourselves to the
anisotropy
\begin{equation}
    \gamma = \frac{\pi}{P}, \qquad P\in\mathbb{Z},\ P\geq 3.
\end{equation}
In this case there is a finite number $P$ of different string types, labelled
by $j=1,\dots,P$. Each type is characterized by a string length $n_j$ and a
parity $v_j=\pm 1$:
\begin{equation}
    \begin{aligned}
    n_j &= j, \qquad  v_j = 1, \qquad  j=1,2,\ldots,P-1,\\
    n_P &= 1, \qquad  v_P = -1 .
    \end{aligned}
\end{equation}
The string structure in the gapless phase is
\begin{equation}
    \lambda^{n,k}_{\alpha}
    = \lambda^{n}_{\alpha}
    + \mathrm{i}\gamma\Bigl(\frac{n+1}{2}-k\Bigr)
    + \mathrm{i}\frac{\pi(1-v_n)}{4},
    \qquad k = 1,2,\dots,n.
\end{equation}
With this notation, the TBA equations derived from Eq.~\eqref{eq:begapless}
can be written as
\begin{equation}
     \begin{aligned}
        \sigma_n\,\rho_n^{t}(\lambda)= {}&
        \tfrac{1}{3}\Bigl(a_n(\lambda-\tau_1)
                        + a_n(\lambda-\tau_2)
                        + a_n(\lambda-\tau_3)\Bigr) \\
        &\;-\sum_{m=1}^{P}\int_{-\infty}^{\infty} \mathrm{d}\lambda'\,
           A_{nm}(\lambda-\lambda')\,
           \theta_m(\lambda')\,\rho^t_m(\lambda') \,,
    \end{aligned}
    \label{eq:TBArhogapless}
\end{equation}
with kernels
\begin{align}
    a_j(\lambda)
    &= a^{v_j}_{n_j}(\lambda)
      \equiv \frac{v_j}{\pi}\,
    \frac{\sin(\gamma n_j)}
         {\cosh(2\lambda)-v_j\cos(\gamma n_j)} ,
    \label{eq:angapless}
    \\[2mm]
    A_{jk}(\lambda)&=
    (1-\delta_{n_j n_k})\,
    a^{v_j v_k}_{|n_j-n_k|}(\lambda)
    + 2\,a^{v_j v_k}_{|n_j-n_k|+2}(\lambda) + \cdots
    \notag\\[-2pt]
    &\qquad {}+ a^{v_j v_k}_{n_j+n_k}(\lambda) .
    \label{eq:Anmgapless}
\end{align}
Here the parity factor $\sigma_n$ on the left-hand side of
Eq.~\eqref{eq:TBArhogapless} is defined as
\begin{eqnarray}
    && \sigma_j = 1, \qquad j=1,2,\ldots,P-1, \nonumber \\
    && \sigma_P = -1, \label{eq:stringparitygapless}
\end{eqnarray}
and ensures the positivity of the total densities $\rho^t_n(\lambda)$. In
Eq.~\eqref{eq:Anmgapless} we used the shorthand
\begin{equation}
    a^{v_j v_k}_{n_j+n_k}(\lambda)
    = \frac{v_j v_k}{\pi}\,
      \frac{\sin\bigl(\gamma (n_j+n_k)\bigr)}
           {\cosh(2\lambda)-v_j v_k
            \cos\bigl(\gamma (n_j+n_k)\bigr)}.
\end{equation}
Thus, in the TBA equations of the gapless XXZ chain there are two parity
indices, $v_j$ and $\sigma_j$, which can change the sign of the kernels
$a_n(\lambda)$ and of the densities.

Finally, the filling factors in the gapless regime at infinite temperature are
modified with respect to the isotropic/gapped cases~\cite{takahashiTBA,XXZgaplessdwGHD}. For $j\leq P-2$ they retain the
simple form
\begin{equation}
    \theta_j(\lambda)=\frac{1}{(j+1)^2},
    \label{eq:theta_gapless1}
\end{equation}
while for the last two string types one finds
\begin{equation}
    \theta_{P-1}(\lambda)=\frac{1}{P}, \qquad
    \theta_{P}(\lambda)=\frac{P-1}{P}.
    \label{eq:theta_gapless2}
\end{equation}
These values are independent of $\lambda$ and encode the finite string
spectrum characteristic of the root-of-unity anisotropy $\gamma=\pi/P$.

\subsection{Generalized hydrodynamics}
\label{GHD formula}
Using the Bethe-ansatz solution of the transfer matrix, we have obtained the quasiparticle content of the circuit unitary $\{U_0, U_1\}$, in particular the rapidity distributions $\rho_n(\lambda)$ and the derivatives of the quasienergy $\varepsilon'_n(\lambda)$ in all parameter regimes. Based on these ingredients, we now briefly summarize the GHD framework as far as it is needed in the following sections.

As discussed in the introduction, GHD is a hydrodynamic theory formulated
directly in terms of the quasiparticle distribution functions
$\rho_n(\lambda)$, and the dynamics of local observables and correlation
functions can be expressed in terms of these distributions. For spin transport
the central object is the infinite-temperature spin–spin correlation function
\begin{equation}
    C(x,t) = \frac{1}{2^L}{\rm tr} \left[S^z(x,t)\,S^z(0,0)\right],
\end{equation}
where the trace is taken within the subspace of half-filling.
In the Euler scaling limit, GHD predicts the ballistic contribution to
$C(x,t)$ in the form~\cite{doyonspohnDrude,DenardisIlievskiDrude}
\begin{equation}
\begin{aligned}
C_{\rm ghd}(x,t) &\;\simeq\; \sum_{n}\int \mathrm{d}\lambda\;
\delta \bigl(x - v_n^{\mathrm{eff}}(\lambda)\,t \bigr) \\
&\quad\times \bigl(1-\theta_n(\lambda)\bigr)\,
\rho_n(\lambda)\,\bigl(m_n^{\mathrm{dr}}(\lambda)\bigr)^2 .
\end{aligned}
\label{eq:GHDformula}
\end{equation}
where $v^{\mathrm{eff}}_n(\lambda)$ is the dressed (effective) group velocity
of quasiparticles of type $n$, and $m_n^{\rm dr}(\lambda)$ is their dressed
magnetization.

The associated Drude weight can be defined from the ballistic growth of the
second moment of $C(x,t)$ as
\begin{equation}
    \mathcal{D}
    = \frac{\mathrm{d}^2}{\mathrm{d}t^2}\int \mathrm{d}x\,x^2\,C(x,t),
\end{equation}
which within GHD leads to the expression~\cite{doyonspohnDrude,DenardisIlievskiDrude}
\begin{equation}
    \mathcal{D}=
    \sum_{n}\int \mathrm{d}\lambda\;
        \bigl[v_n^{\mathrm{eff}}(\lambda)\bigr]^2\,
        \bigl(1-\theta_n(\lambda)\bigr)\,
        \rho_n(\lambda)\,\bigl(m_n^{\rm dr}(\lambda)\bigr)^2 .
    \label{eq:DWghd}
\end{equation}

Just as the densities $\rho_n(\lambda)$ are dressed by interactions through
the TBA integral equations~\eqref{eq:TBArhoxxx}, \eqref{eq:TBArhoGAP} and
\eqref{eq:TBArhogapless}, the effective velocities $v^{\rm eff}_n(\lambda)$
and dressed magnetizations $m_n^{\rm dr}(\lambda)$ are also dressed
quantities. The effective velocity is defined as~\cite{GHD2016doyon,GHD2016prl}
\begin{equation}
    v^{\rm eff}_n(\lambda)
    =\frac{\varepsilon'^{\rm dr}_n(\lambda)}{p'^{\rm dr}_n(\lambda)},
\end{equation}
where $p'^{\rm dr}_n(\lambda)={\rm d}p^{\rm dr}_n/{\rm d}\lambda$ and $\varepsilon'^{\rm dr}_n(\lambda)={\rm d}\varepsilon^{\rm dr}_n/{\rm d}\lambda$ are the
dressed derivatives of the quasiparticle momentum and energy, respectively.
They are obtained from the linear integral equations
\begin{align}
    p'^{\rm dr}_n(\lambda)
    &= p'_n(\lambda)
    -\sum_{m=1}\sigma_m\int_{I}\mathrm{d}\lambda'\,
      A_{nm}(\lambda-\lambda')\,
      \theta_m(\lambda')\,p'^{\rm dr}_m(\lambda'), 
      \label{eq:Linear source in rhoTBA} \\
    \varepsilon'^{\rm dr}_n(\lambda)
    &= \varepsilon'_n(\lambda)
    -\sum_{m=1}\sigma_m\int_{I}\mathrm{d}\lambda'\,
      A_{nm}(\lambda-\lambda')\,
      \theta_m(\lambda')\,\varepsilon'^{\rm dr}_m(\lambda'),
    \label{eq:Linear source in EdrTBA}
\end{align}
with the same scattering kernels $A_{nm}$ and filling functions $\theta_m(\lambda)$ appearing in the TBA equations. The parity index $\sigma_m$ take valuses as equations (\ref{eq:stringparitygapless}) in the gapless regime, and all $\sigma_m$ equal to 1 in the isotropic and gapped regimes. The rapidity integration domain $I$ is $(- \infty,\infty)$ in the isotropic and gapless regimes and $[-\pi/2,\pi/2]$ in the gapped regime, in accordance with Eqs.~\eqref{eq:TBArhoxxx} and \eqref{eq:TBArhogapless}.

The source terms \(p'_n(\lambda)\) and \(\varepsilon'_n(\lambda)\) in
Eq.~\eqref{eq:Linear source in EdrTBA} are the bare derivatives of the quasiparticle momentum and energy. For the inhomogeneous transfer matrix considered in this work, the bare momentum derivatives are
\begin{equation}
    p'_n(\lambda)
    = \frac{2\pi}{3}\Bigl(
    a_n(\lambda-\tau_1)
    +a_n(\lambda-\tau_2)
    +a_n(\lambda-\tau_3)\Bigr),
    \label{eq:pdot}
\end{equation}
which follows from the general relation \(p'^{\rm dr}_n(\lambda)=2\pi\sigma_n
\rho^t_n(\lambda)\)~\cite{GHD2016prl,doyonGHDnote}. By contrast, the bare energy derivatives depend on which circuit layer \(U\) is applied. For later convenience we denote them by
\(\varepsilon'_{0n}\) and \(\varepsilon'_{1n}\) for the layers \(U_0\) and
\(U_1\), respectively:
\begin{align}
   \varepsilon'_{0n}(\lambda)
    &= 2\pi\bigl(
       a_n(\lambda-\tau_1)
      -a_n(\lambda-\tau_2)\bigr),
      \label{eq:edu0} \\
    \varepsilon'_{1n}(\lambda)
    &= 2\pi\bigl(
       a_n(\lambda-\tau_3)
      -a_n(\lambda-\tau_1)\bigr).
    \label{eq:edu1}
\end{align}
Equation~\eqref{eq:edu0} coincides with the expression
\eqref{eq:quaen_xxxstring} obtained previously for \(U_0\), while
Eq.~\eqref{eq:edu1} follows from the same construction with the replacement
\((\tau_1,\tau_2)\to(\tau_3,\tau_1)\).

Finally, the dressed magnetizations $m_n^{\rm dr}(\lambda)$ satisfy a closed
integral equation of the same structure:
\begin{equation}
    m^{\rm dr}_n(\lambda)
    = m_n
    -\sum_{k=1}\sigma_k\int_{I}\mathrm{d}\lambda'\,
      A_{nk}(\lambda-\lambda')\,
      \theta_k(\lambda')\,m_k^{\rm dr}(\lambda') ,
      \label{eq:Linear source in mdrTBA}
\end{equation}
where $m_n$ is the bare magnetization carried by an $n$--string. In the
case of interest one has simply $m_n = n$, i.e.~the bare magnetization is
proportional to the string length.

\section{Numerical methods} \label{sec:Numerical method}
In this section, we present two complementary numerical approaches for computing the spin–spin correlation function to investigate spin transport in the infinite-temperature, half-filling regime. (1) We numerically simulate the real-time evolution of the correlation function $C(x,t)$ via tDMRG calculations. (2) We self-consistently solve the set of linear integral equations in TBA and GHD, which leads to the GHD prediction for the Euler scale correlation function. The second approach not only elucidates the dominating quasiparticle contributions to the observed lineshapes of $C(x,t)$, but also serves as a benchmark for the validity of GHD beyond Floquet integrable dynamics.

\subsection{Spin correlation function via tDMRG}
We employ the tDMRG method to compute the real-time evolution of spin chains under quasiperiodic Fibonacci, Thue-Morse sequences, as well as completely random sequences. We consider evolution from a weakly polarized domain wall initial state~\cite{prosen17NC,KPZ2019}
\begin{equation}
    \rho_{\mu} \propto (\mathrm{I}+\mu S_z)^{\bigotimes L/2} \otimes (\mathrm{I}-\mu S_z)^{\bigotimes L/2}.
\end{equation}
The infinite-temperature correlation function is then extracted from the leading-order coefficient in the Taylor expansion of $S^z(t)$ with respect to a small bias parameter $\mu$,

\begin{equation}
    \begin{aligned}
        \expval{ S_{i}^{z}(t)}_\mu = \mu \sum_{j=1}^L \mathrm{sign}\left(\frac{L}{2}-j\right)  \expval{S_i^z(t) S_j^z}+o(\mu). \\
    \end{aligned}
\end{equation}
We set $\mu=0.002$ in all simulations. See Appendix~\ref{appendix:numerical-check} for details on the choice of parameters.

Due to a discrete $k$-site translational symmetry in our systems, we consider the difference in magnetization over $k$-sites, 
\begin{equation}
    \begin{aligned}
        &\expval{ S_{i-k}^{z}(t)}_\mu - \expval{ S_{i}^{z}(t)}_\mu \\
        &=\mu \sum_{j=1}^L \mathrm{sign}\left(\frac{L}{2}-j\right)  \expval{(S_{i-k}^z(t) - S_{i}^z(t)) S_j^z} \\
        &=\mu \sum_{j=1}^L \mathrm{sign}\left(\frac{L}{2}-j\right)  \expval{S_{i}^z(t) (S_{j+k}^z - S_j^z)} \\
        &=2\mu \expval{S_i^z(t) \left(\sum_{m=1}^k S_{\frac{L}{2}+m}^z -  \sum_{m=1}^k S_{m}^z \right)} \\
        &\xrightarrow[i \sim L/2]{L \to \infty}  \ 2\mu \sum_{m=1}^k \expval{S_i^z(t) S_{\frac{L}{2}+m}^z}.\\
    \end{aligned}
    \label{correlation averaged}
\end{equation}
For the case we study $k=3$. The last line indicates that, in the thermodynamic limit, the magnetization difference probes a sum of $k$-distinct two-point correlations. Direct use of the $k$-site difference can exhibit strong oscillations due to the underlying k-site translational symmetry of the circuit elements. To obtain a spatially smoothed observable that faithfully reflects the hydrodynamic correlation function in the long-wavelength limit, we define the $k$-site averaged correlation function $C(x,t)$ by performing a moving average within the translational unit cell:
\begin{equation}
    C(x,t) = \frac{1}{18 \mu} \sum_{m=0}^{k-1} \left( \expval{ S_{x-k+m}^{z}(t)}_\mu - \expval{ S_{x+m}^{z}(t)}_\mu \right). 
\end{equation}
The normalization factor $18 \mu$ is chosen such that the spatial integral of $C(x,t)$ yields a total weight of $0.25$. Throughout this work, the term “correlation function $C(x,t)$ ” refers exclusively to this spatially averaged quantity.

\subsection{Spin correlation function via TBA and GHD} \label{Constructing Correlations via TBA and GHD}

\begin{table}[!t]
    \centering
    \renewcommand{\cellalign}{cc}
    \begin{tabular}{|c|c|c|c|c|}
        \hline
        \textbf{Quantity} & \textbf{Equations} &  \multicolumn{3}{c|}{\textbf{Inputs}}\\
        \hline
        \textbf{$\rho_n(\lambda)$} & 
        \makecell{
            Eq.~\eqref{eq:Linear source in rhoTBA} \\
            Eq.~\eqref{eq:pdot}
        } & 
        \makecell{
            \textbf{Isotropic} \\
            \\
            \\
            \\
            Eq.~\eqref{eq:anxxx} \\
            Eq.~\eqref{eq:Anm} \\
            Eq.~\eqref{eq:theta}
        } & 
        \makecell{
            \textbf{Gapped} \\
            \\
            \\
            \\
            Eq.~\eqref{eq:anGAP} \\
            Eq.~\eqref{eq:AnmGAP} \\
            Eq.~\eqref{eq:theta}
        } & 
        \makecell{
            \textbf{Gapless} \\
            \\
            \\
            Eq.~\eqref{eq:angapless} \\
            Eq.~\eqref{eq:Anmgapless} \\
            Eq.~\eqref{eq:theta_gapless1} \\
            Eq.~\eqref{eq:theta_gapless2}
        } \\
        \cline{1-2}
        \textbf{$\varepsilon'^{\rm dr}_n$} & 
        \makecell{
            Eq.~\eqref{eq:Linear source in EdrTBA} \\
            Eq.~\eqref{eq:edu0} \\
            Eq.~\eqref{eq:edu1}
        } & & & \\
        \cline{1-2}
        \textbf{$m^{\rm dr}_n$} & 
        Eq.~\eqref{eq:Linear source in mdrTBA} & & & \\
        \hline
    \end{tabular}
    \caption{Lookup table for implementing the GHD calculations based on TBA. The middle column points to the closed integral equations (and their source terms) that determine each quantity, while the right columns list the regime-specific inputs $a_n(\lambda)$, $A_{mn}(\lambda)$, and $\theta_n$ (isotropic / gapped / gapless) entering those equations.} \label{tab:integration}
\end{table}

As discussed in Sec.~\ref{GHD formula}, the computation of correlation
functions within the GHD framework requires three essential ingredients: the
quasiparticle densities \(\rho_n(\lambda)\), the effective velocities
\(v^{\text{eff}}_n(\lambda)\), and the dressed magnetizations
\(m^{\mathrm{dr}}_n(\lambda)\). In our setting, the corresponding equations,
Eqs.~\eqref{eq:Linear source in rhoTBA}, \eqref{eq:Linear source in EdrTBA},
and \eqref{eq:Linear source in mdrTBA}, are linear integral equations that we
solve numerically using a fast Fourier transform (FFT) method. 

As emphasized above, the bare momentum derivatives \(p'_n(\lambda)\) depend only on the inhomogeneous transfer matrix and are independent of the specific circuit layer $\{U_0,U_1\}$. The layer dependence enters solely through the bare energy derivatives $\varepsilon'_{\alpha n}(\lambda)$, which are different for $U_0$ and $U_1$, cf. Eqs.~\eqref{eq:edu0}-\eqref{eq:edu1}. In our circuit models, the evolution operator over \(t\) steps is a sequence in \(\{U_0,U_1\}\). In the long-time limit, all evolution protocols we focus on in this work have well-defined asymptotic letter frequencies: $U_1$ appears with fraction $p$ and $U_0$ with fraction $1-p$. For example, the Fibonacci sequence has $p\approx 0.618$ asymptotically, and Thue-Morse sequence has $p=0.5$. An ensemble of stochastic sequences is also defined by the probability $p$ and $1-p$ for $U_1$ and $U_0$, respectively. Because the dressing equations are linear in the bare source terms, the time dependence can be incorporated through a probability-weighted average of the corresponding energy sources. In practice, we replace \(\varepsilon'_n(\lambda)\) in Eq.~\eqref{eq:Linear source in EdrTBA} by
\begin{equation}
    \bar{\varepsilon}'_n(\lambda)
    = p\,\varepsilon'_{1n}(\lambda)
      + (1-p)\,\varepsilon'_{0n}(\lambda),
\end{equation}
which corresponds to an effective Floquet generator with quasi-energy
\(\bar{\varepsilon}=p\,\varepsilon_1 + (1-p)\,\varepsilon_0\).
All subsequent TBA and GHD quantities (dressed energies, effective velocities, and correlation functions) are then computed using the common \(p'_n(\lambda)\) and the averaged source \(\bar{\varepsilon}'_n(\lambda)\), providing an effective hydrodynamic description for circuits whose gate sequences are non-repetitive in time.

We summarize the set of equations one needs to solve numerically for GHD calculations in Table~\ref{tab:integration}. In practice, one first picks the appropriate regime (isotropic / gapped / gapless) and reads off the corresponding inputs $a_n(\lambda)$, $A_{mn}(\lambda)$, and $\theta_n$. Then, one computes $\varepsilon'^{ \rm dr}_n$, $p'^{\rm dr}_n$, $v_n^{\rm eff}$, $m_n^{\rm dr}$ and $\rho_n$ by solving the corresponding integral equations numerically. Finally, one obtains $C_{\rm ghd}(x,t)$ using Eq.~(\ref{eq:GHDformula}) based on the data above.

\section{Numerical results for quasiperiodic quantum circuits}  
\label{sec:Fibresult}
In this section, we present numerical results for quasiperiodic quantum circuits built from $\{U_0, U_1\}$. In the main text, we focus on the Fibonacci sequence, while results on Thue-Morse sequence are shown in Appendix~\ref{Appendix B}. Following the TBA classification in Sec.\ref{sub:TBAintro} (see also App.~\ref{Appendix A}), we organize the results into three regimes distinguished by the underlying $R$-matrix/transfer-matrix structure (the analogs of the easy-plane, isotropic, and easy-axis regimes of the XXZ chain), which allows for a direct comparison with Hamiltonian and Floquet resutls. Our numerical data $C(x,t)$, obtained from large-scale tDMRG simulations at infinite temperature and half filling, are compared with the correlation function $C_{\text{ghd}}(x,t)$ constructed via GHD with weight-averaged effective velocities introduced in Sec.~\ref{Constructing Correlations via TBA and GHD}. As outlined in Fig.~\ref{fig:Our Circuits}, the circuit depends only on spectral-parameter differences, so we fix $\tau_2=0$ without loss of generality. 

\subsection{Gapless phase}
In the gapless phase, we construct the circuit using the local gate $PR$ with anisotropy parameter fixed at $\gamma = \frac{\pi}{3}$ and sample spectral parameters $\tau_1$, $\tau_3$ as follows:
\begin{equation*}
    \begin{aligned}
        \tau_1 \in \{-3.9\mathrm{i}, 0.1\mathrm{i}, 4.1\mathrm{i}\}, \\
        \tau_3 \in \{-2.9\mathrm{i}, 1.1\mathrm{i}, 5.1\mathrm{i}\}.
    \end{aligned}
\end{equation*}

We first compute the dynamic exponent $z$ across all sampled parameter combinations by fitting the asymptotic slope of $\log\left( \sigma^2(t) \right)$ versus $\log(t)$,
\begin{equation}
    \begin{aligned}
        \log\left( \sigma^2(t) \right) &\approx \frac{2}{z} \log(t) + C.
    \end{aligned}
\end{equation}
The variance $\sigma^2(t)$ is calculated directly from the tDMRG data as:
\begin{equation}
    \begin{aligned}
        \sigma^2(t) & = \sum_m m^2 C(m,t),\\
    \end{aligned}
\end{equation}

\begin{figure}[!t]
    \centering
    \includegraphics[width=0.8\linewidth]{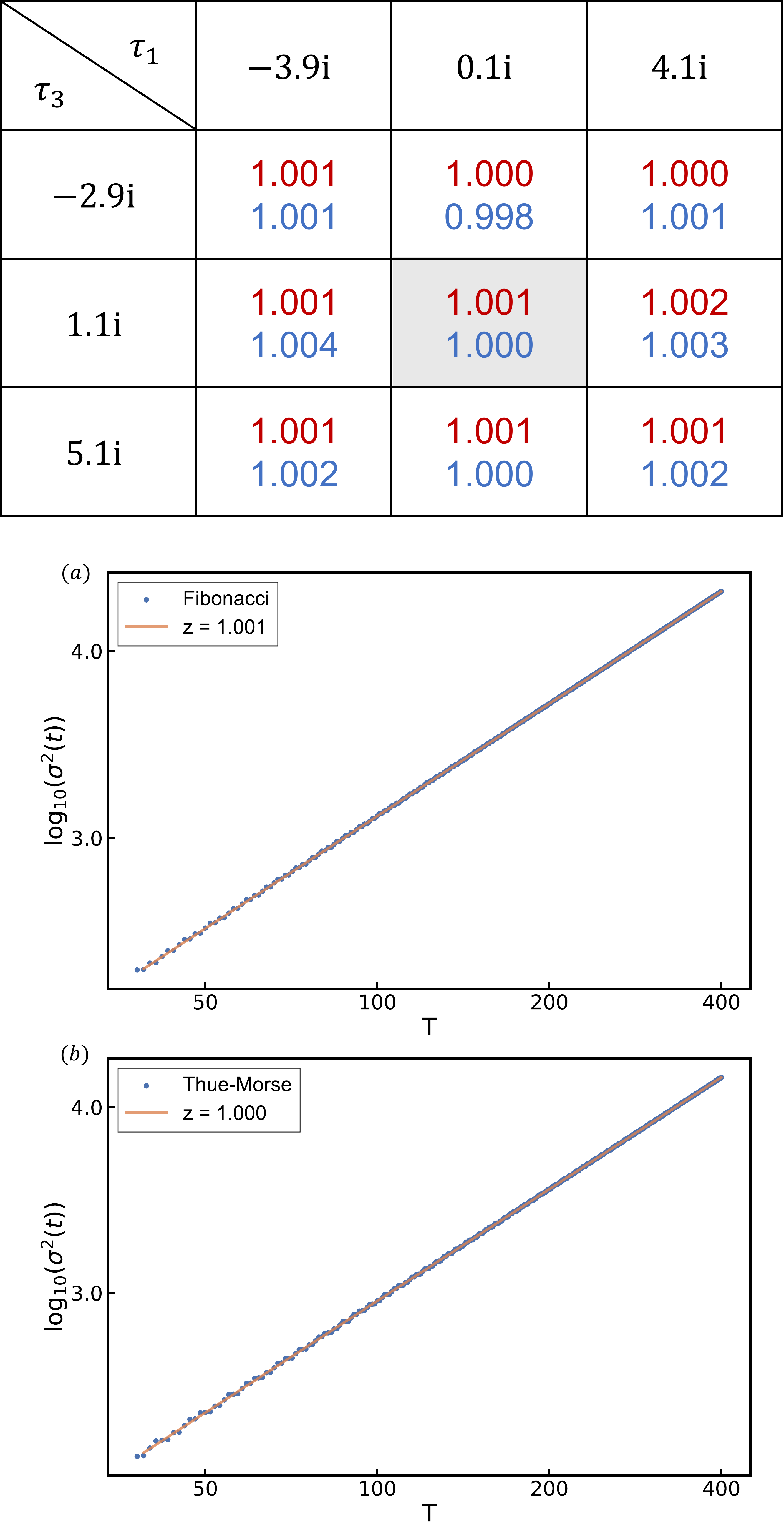}
    \caption{Estimated values of the dynamical exponent \( z \) in the gapless phase. Results for Fibonacci (Thue--Morse) sequence are shown in red (blue). All dynamical exponents converge to $1$, independent of both the sequence and spectral parameters. These results are obtained from simulations of a system of size $L= 1500$ with bond dimension $\chi= 256$, evolved for $400$ time steps (each step consists of applying either $U_0$ or $U_1$).}
    \label{tab:Gapless z(t)}
\end{figure}

\begin{figure*}
    \centering
    \includegraphics[width=0.95\linewidth]{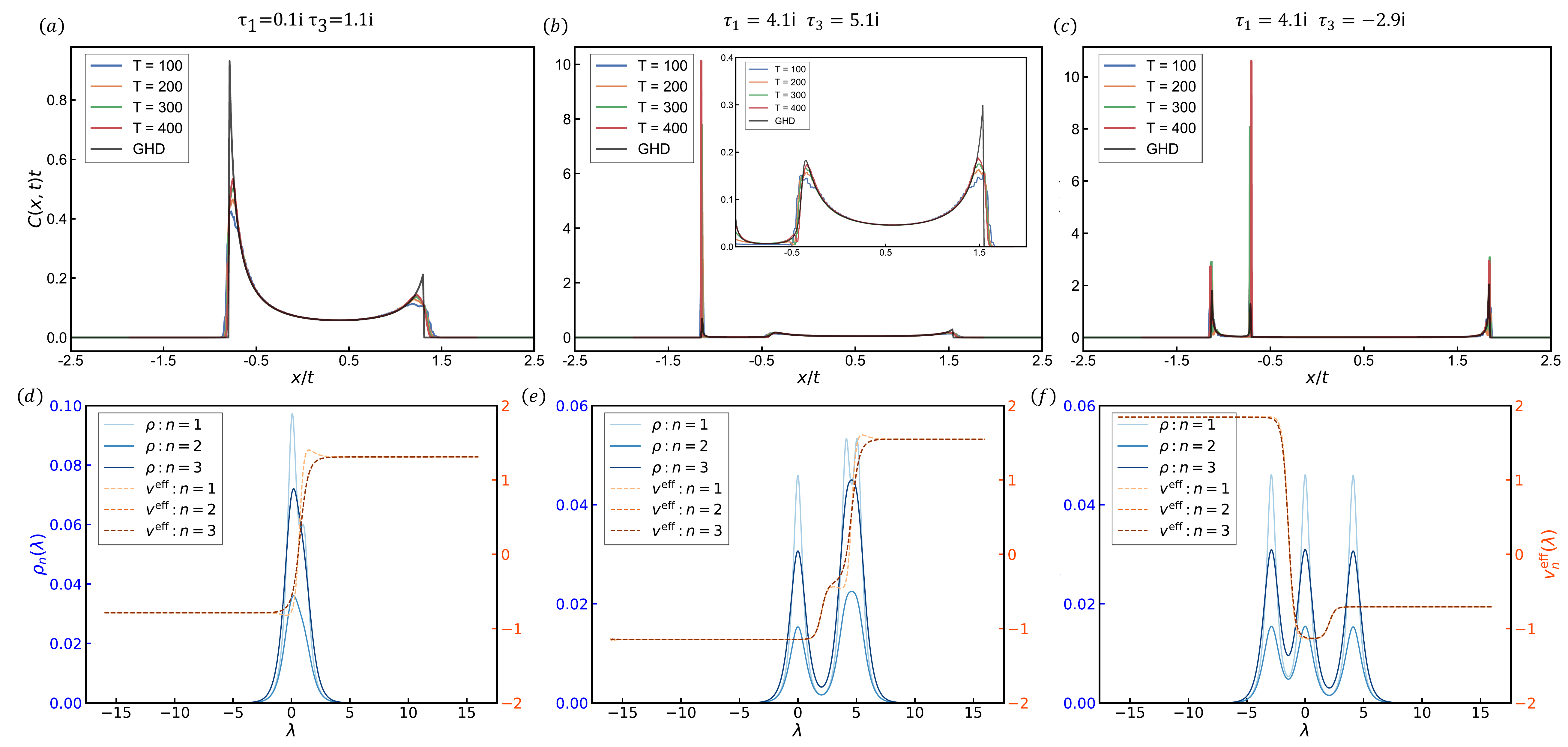}
    \caption{{\bf Gapless phase.} (a), (b), and (c): Scaling collapse of the correlation function $C(x,t)$ obtained from tDMRG simulations compared with the GHD prediction $C_{\text{ghd}}(x,t)$ under Fibonacci sequence. Simulations are performed with system size $L = 1500$, bond dimension $\chi = 256$ and $\mu=0.002$. The inset of (b) is a zoom-in plot of the low-amplitude region of the profile, which shows excellent agreement between tDMRG and GHD there.
    (d), (e), and (f): Corresponding TBA results for the effective velocity $v_{\mathrm{eff}}(\lambda)$ and quasi-particle density $\rho_n(\lambda)$ used as inputs for GHD calculation.
    The tDMRG data are well captured by GHD and the only notable deviations can be attributed to finite-size effects.}
    \label{fig:FB TEBD GHD TBA}
\end{figure*}

As summarized in Fig.~\ref{tab:Gapless z(t)}, the dynamical exponent remains $z=1$ across all tested parameter sets, confirming the persistence of ballistic spin transport throughout the entire gapless phase. This universality is independent of both the spectral parameter $\tau$ and the choice of sequence, and can be attributed to the presence of quasi-local charges that are odd under spin flip and have finite overlap with the spin current \cite{xxzqlc2, Pereira_2014, PROSENnpb, Ilievski_2016}.

The variance $\sigma^2(t)$, despite providing an estimate for the dynamical exponent $z$, gives only crude information on the finer structure of the spreading of correlations. To further validate ballistic spin transport and elucidate the impact of the spectral parameters on the transport behavior, we now turn to the full spatiotemporal profile of $C(x,t)$, obtained via tDMRG simulations and GHD calculations.
In this case, Eq.~\eqref{eq:GHDformula} simplifies to
\begin{equation}
    \begin{aligned}
        &C_{\rm ghd}(x,t) \simeq \\
        &\sum_{n=P-1,P}\Big(\frac{P}{2}\Big)^2(1-\theta_n)\int d\lambda\,
        \delta \!\left(x - v_n^{\mathrm{eff}}(\lambda)\,t \right)\,
        \rho_n(\lambda),
    \end{aligned}
    \label{eq:xxzGHDformula}
\end{equation}
with $P=3$ for our current choice of $\gamma=\frac{\pi}{3}$. Here $v_n^{\mathrm{eff}}(\lambda)$ is obtained from the weight-averaged dressing procedure described in Sec.~\ref{Constructing Correlations via TBA and GHD}. The sum is restricted to the last two string species $n=P-1,P$ because, in our setting, the filling functions $\theta_n$ are rapidity-independent and only these two species carry a non-vanishing dressed magnetization,  $m_n^{\rm dr}=P/2$ ~\cite{lfepXXZ,prlide2017} (which can be verified numerically). 
Equation~\eqref{eq:xxzGHDformula} has a transparent quasiparticle interpretation: each mode labelled by $(n,\lambda)$ propagates ballistically with velocity $v_n^{\mathrm{eff}}(\lambda)$. Modes that share the same effective velocity contribute along the same ray $\xi=x/t$ selected by the $\delta$-function, while $\rho_n(\lambda)$ provides the spectral weight, determining which modes dominate the overall signal. As a direct consequence, we expect that the correlation function exhibits Euler scaling,
\begin{equation}
    C(x,t)=\frac{1}{t}\,f\!\left(\frac{x}{t}\right).
    \label{eq:seuler}
\end{equation}

Figure~\ref{fig:FB TEBD GHD TBA} shows three representative scaled profiles for different choices of $(\tau_1,\tau_3)$. In the first row we compare the tDMRG simulation $C(x,t)$ with GHD results $C_{\rm ghd}(x,t)$. In the second row we plot the corresponding TBA inputs $\rho_n(\lambda)$ and the averaged effective velocities $v_n^{\rm eff}(\lambda)$ that enter Eq.~\eqref{eq:xxzGHDformula}. We find that the full correlation function at different times indeed collapse into a universal form, confirming the scaling hypothesis~(\ref{eq:seuler}) with $z=1$. In all cases studied, the profiles exhibit an inversion asymmetry due to the lack of spatial inversion symmetry in the underlying circuit. Moreover, we find that our modified GHD scheme quantitatively captures all fine structures in the full profile of the correlation function, including the location and shape of various peaks. As we now demonstrate, the structure of $\rho_n(\lambda)$ together with $v_n^{\rm eff}(\lambda)$ provides a direct, intuitive explanation of the observed profiles.

For $(\tau_1,\tau_3)=(0.1\mathrm{i},\,1.1\mathrm{i})$, the
scaled correlation profile exhibits two broad peaks, and the tDMRG data agree well with
$C_{\rm ghd}(x,t)$ throughout an extended bulk region [Fig.~\ref{fig:FB TEBD GHD TBA}(a)]. The remaining discrepancies are
confined to the outer edges and can be attributed to finite-time effect in tDMRG together with front broadening beyond the Euler scale. The origin of the broad, smooth bulk profile is clarified by the corresponding TBA data in Fig.~\ref{fig:FB TEBD GHD TBA}(d): the effective velocity $v_n^{\rm eff}(\lambda)$ varies appreciably over the rapidity range where $\rho_n(\lambda)$ carries most of its weight. Through Eq.~\eqref{eq:xxzGHDformula}, this spread of velocities naturally produces an
extended continuum contribution to the correlation profile, rather than a narrowly
localized ballistic peak.

For $(\tau_1,\tau_3)=(4.1\mathrm{i},\,5.1\mathrm{i})$, a pronounced sharp peak emerges on one side of the profile, while the remaining part stays smooth and extended [Fig.~\ref{fig:FB TEBD GHD TBA}(b)]. Although a deviation is visible in the \emph{amplitude} of the ballistic peak, GHD still
predicts the peak positions and accurately reproduces the smooth background; the latter
agreement is highlighted in the inset of Fig.~\ref{fig:FB TEBD GHD TBA}(b), which zooms
into the low-amplitude region of the main plot.
 The mixed ``peak + continuum" structure is clearly explained by the TBA data in Fig.~\ref{fig:FB TEBD GHD TBA}(e): the root density becomes effectively bimodal in $\lambda$, and one of the dominant components sits in a region where $v_n^{\rm eff}(\lambda)$ is nearly $\lambda$-independent (velocity-degenerate). Through Eq.~\eqref{eq:xxzGHDformula}, such a near-degeneracy concentrates spectral weight into a narrow range of rays $\xi=x/t$, producing a sharp peak. The other component, where $v_n^{\rm eff}(\lambda)$ varies with $\lambda$, instead yields the broad off-peak background.

Finally, for $(\tau_1,\tau_3)=(4.1\mathrm{i},\,-2.9\mathrm{i})$, the scaled correlation function is dominated by three sharp peaks [Fig.~\ref{fig:FB TEBD GHD TBA}(c)]. The peak \emph{positions} from tDMRG and GHD remain aligned, whereas the amplitude of the central peak shows a noticeable discrepancy. Consistently, the corresponding TBA data indicate that $\rho_n(\lambda)$ develops an approximately trimodal structure, and within each dominant lobe the effective velocity is nearly degenerate in $\lambda$; as a result, each of which generates a distinct sharp ballistic peak via Eq.~\eqref{eq:xxzGHDformula}.

The observed discrepancy in the sharp peak amplitudes can be attributed to finite-size effects and the use of a spatial moving average in our tDMRG simulations. In particular, the tDMRG results exhibit a persistent growth in peak height with increasing time. If this trend were to persist in the thermodynamic limit, it would imply the emergence of a $\delta$-peak in the normalized scaling form of the correlation function—signaling a singular contribution from a specific velocity. However, TBA analysis reveals a smooth and broadly distributed quasi-particle spectrum, without any dominant contribution at a single velocity. This inconsistency supports the interpretation that the observed peak growth is a finite-size artifact rather than a genuine physical divergence. As we show later in Sec.~\ref{sec:random result}, for the stochastic case these peak amplitudes become controlled and are no longer anomalously enhanced after ensemble averaging.

As a comparison, in Appendix~\ref{ap:Htdmrgvsghd} we present the same benchmark of GHD results against tDMRG simulations at the level of two-point correlation function for the XXZ Hamiltonian~\eqref{eq:xxzH} at anistropy $\Delta=\cos(\pi/3)$. Again, we find excellent agreement in the bulk and the expected deviations at the fronts. 

\subsection{Isotropic point}
\begin{figure}[t]
    \centering
    \includegraphics[width=0.8\linewidth]{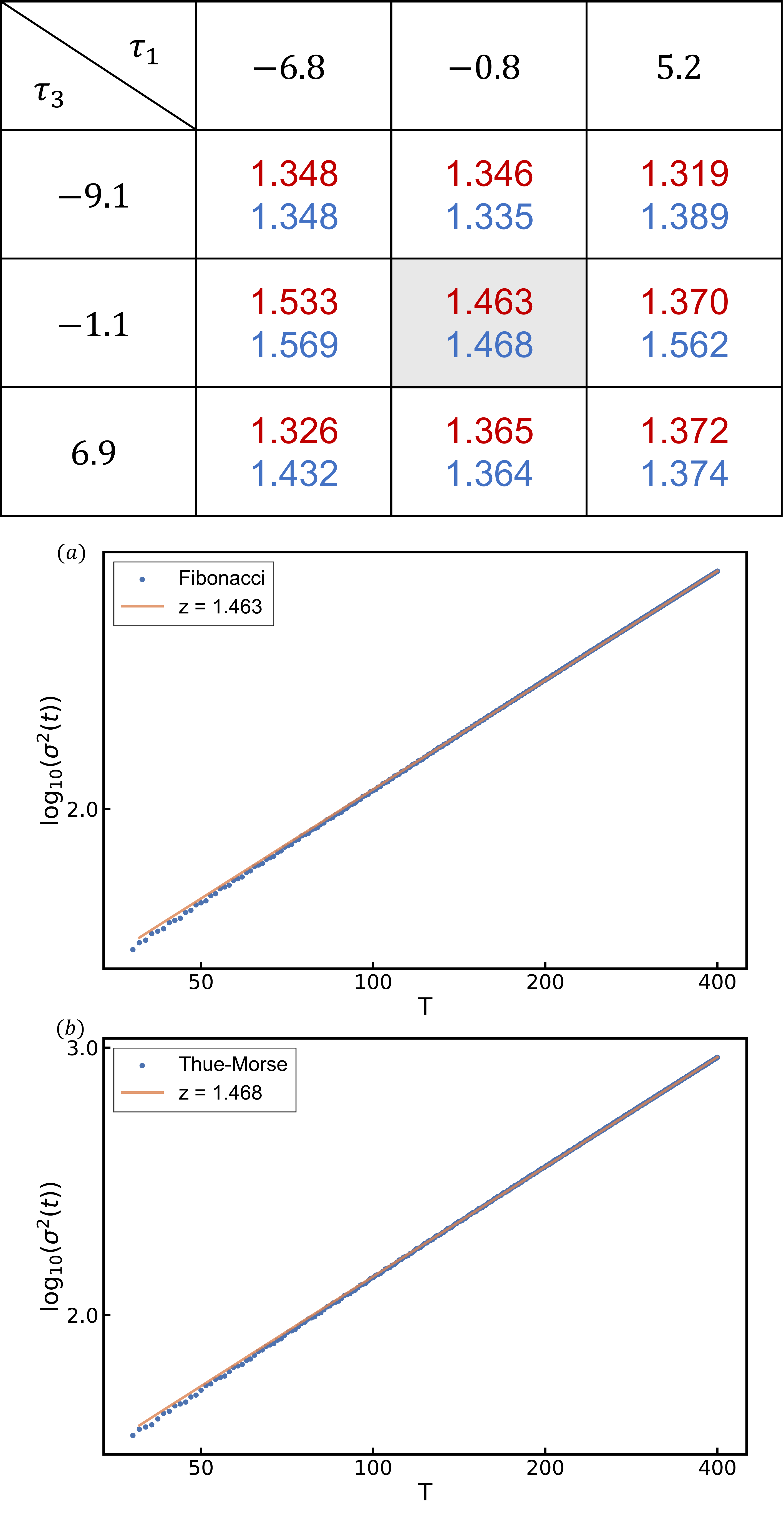}
    \caption{Estimated values of the dynamical exponent \( z \) at the isotropic point under Fibonacci (Thue--Morse) sequence, shown in red (blue), respectively. When $(\tau_1,\tau_3)$ are sufficiently large in magnitude (e.g. $|\tau_1|,|\tau_3|>5$), the dynamical exponent $z$ remains consistently less than $1.5$. This is a consequence of the coexistence of the intrinsic $z=3/2$ mode and the emergent $z=1$ mode. These results are obtained with system size $L= 1500$, bond dimension $\chi= 256$, time step $=400$ and $\mu=0.002$.}
    \label{tab:XXX z(t)}
\end{figure}

At the isotropic point, we sample the following spectral parameters for the $PR$ gate,
\begin{equation*}
    \begin{aligned}
        \tau_1 \in \{-6.8,\,-0.8,\,5.2\}, \\
        \tau_3 \in \{-9.1,\,-1.1,\,6.9\}.
    \end{aligned}
\end{equation*}
Following the same protocol as in the gapless regime, we extract the dynamical exponent $z$ for each pair $(\tau_1,\tau_3)$. The results are summarized in Fig.~\ref{tab:XXX z(t)}. A key observation is that, although the dynamics remain in the superdiffusive regime, the fitted exponents remain consistently smaller than the value $z=3/2$ expected at the isotropic point for both Hamiltonian and Floquet dynamics. As we show below, the smaller value of the fitted exponent arises from an emergent ballistic mode coexisting with the usual intrinsic superdiffusive mode with $z=3/2$, which is elucidated by the profile of the correlation function and the corresponding TBA data.

\begin{figure*} 
    \centering
    \includegraphics[width=0.9\linewidth]{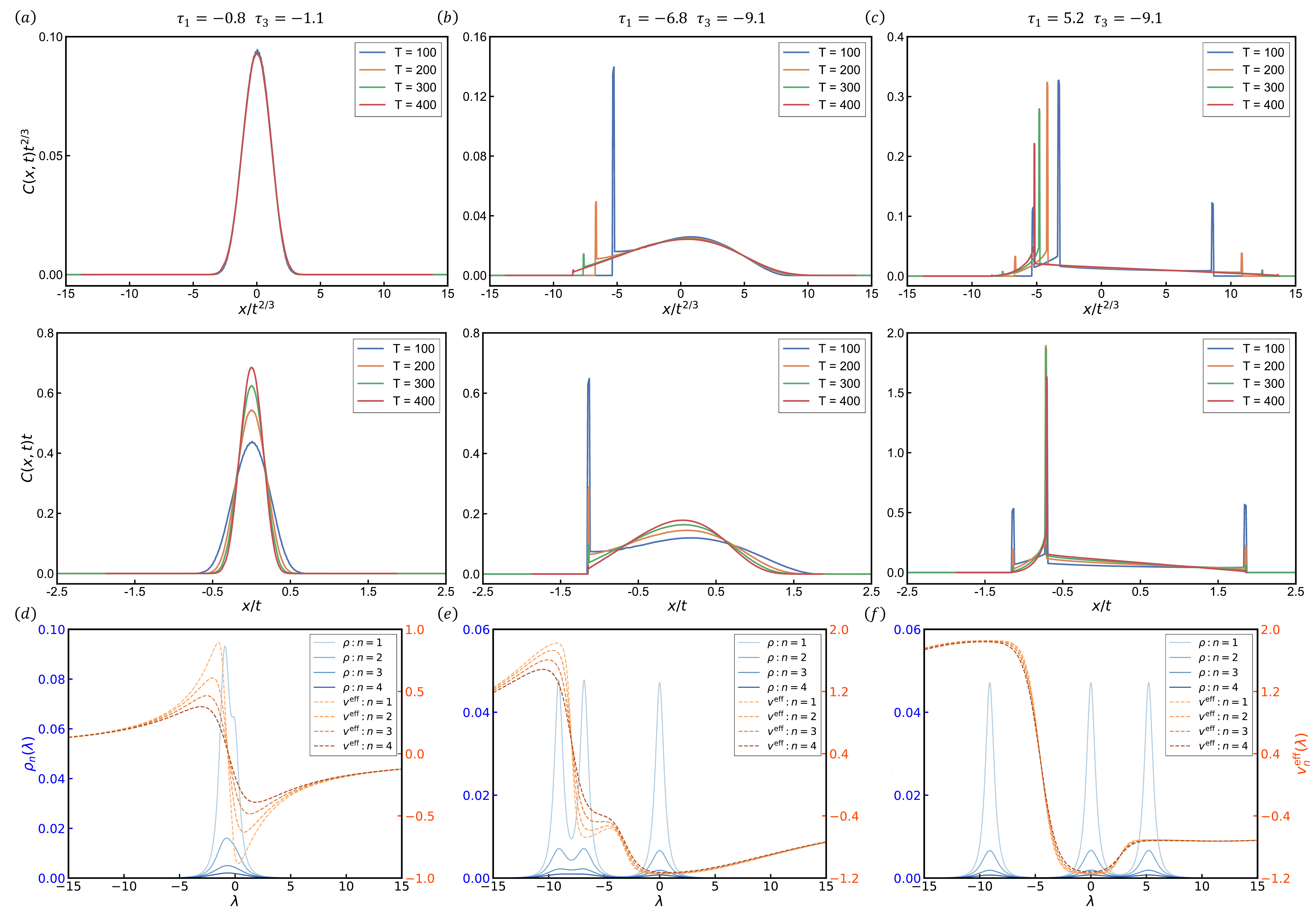}
    \caption{{\bf Isotropic point.} Panels (a), (b) and (c) show the scaled spin-spin correlation function $C(x,t)$ under Fibonacci sequence using the scaling hypothesis Eq.~(\ref{eq:collapse}) with two different choices of the dynamical exponent: $z=3/2$ (first row)  and $z=1.0$ (second row), respectively. The coexistence of superdiffusive $z=3/2$ transport and emergent $z=1$ ballistic mode due to proximity to dual unitarity is clearly visible in panels (b)\&(c). Results were obtained with system size $L = 1500$, bond dimension $\chi = 256$ and $\mu=0.002$.
    Panel (d), (e) and (f) present the corresponding TBA data. Here we only present the first four strings. }
    \label{fig:XXXFB TEBD TBA}
\end{figure*}

In Fig.~\ref{fig:XXXFB TEBD TBA}, we show three representative choices of $(\tau_1,\tau_3)$, similarly to the gapless case. In the first and second row of Fig.~\ref{fig:XXXFB TEBD TBA}, we attempt data collapse using the following scaling hypothesis with two different choices of $z$: 
\begin{equation}
    C(x,t)=\frac{1}{t^{1/z}}\,f\!\left(\frac{x}{t^{1/z}}\right),
    \label{eq:collapse}
\end{equation}
where $z=3/2$ in the top panel, and $z=1$ in the middle panel. This allows for a clear isolation of the ballistic and superdiffusive components contributing to $C(x,t)$. The bottom row shows the corresponding TBA data, namely the root densities $\rho_n(\lambda)$ and the average effective velocities $v_n^{\rm eff}(\lambda)$.

We begin with the ``regular" case $(\tau_1,\tau_3)=(-0.8,-1.1)$, shown in Fig.~\ref{fig:XXXFB TEBD TBA}(a). Here the correlation profile exhibits clean superdiffusive scaling and collapses well with $z=3/2$. The associated TBA data in Fig.~\ref{fig:XXXFB TEBD TBA}(d) show that the weights $\rho_n(\lambda)$ are strongly concentrated at $\lambda=0$ and the effective velocities $v_n^{\rm eff}(\lambda)$ exhibit a strong string dependence. The spread of velocities across different string species suppresses the formation of a single ballistic peak. Instead, it results in a slowly broadening profile, consistent with the kinetic picture of isotropic superdiffusion~\cite{sarangromainkinetic}.

For $(\tau_1,\tau_3)=(-6.8,-9.1)$, Fig.~\ref{fig:XXXFB TEBD TBA}(b) shows a pronounced sharp peak superimposed on a broad background. The scaling collapse now reveals a clear separation of dynamical components: the narrow peak region collapses under ballistic scaling ($z=1$), while the remaining smooth part continues to collapse under the superdiffusive exponent $z=3/2$. A subtle but important point is that, in the $z=1$ collapse (second row), the peak height decreases as $t$ increases, suggesting that this ballistic contribution carries vanishing weight in the infinite-time limit and is therefore likely a finite-time component rather than an asymptotic feature. The TBA data in Fig.~\ref{fig:XXXFB TEBD TBA}(e) provide a direct microscopic interpretation. The root densities $\rho_n(\lambda)$ develop a sharply localized component near $\lambda=0$, together with a broader structure at finite rapidity. The localized component corresponds to quasiparticles whose effective velocities cluster tightly around a common value (close to $v_n^{\rm eff}\simeq -1$), thereby concentrating spectral weight onto a narrow range of rays $\xi=x/t$ and generating the fast ballistic peak. In contrast, the broader component spans a wide range of effective velocities and produces the slowly spreading superdiffusive background. Overall, this points to the coexistence of a transient ballistic-like feature with the intrinsic $z=3/2$ dynamics within the time window accessible to our numerics.

Finally, for $(\tau_1,\tau_3)=(5.2,-9.1)$, Fig.~\ref{fig:XXXFB TEBD TBA}(c) shows that the correlation function develops multiple sharp peaks, which can be viewed as the natural multi-component counterpart of the single ballistic-like peak observed at
$(\tau_1,\tau_3)=(-6.8,-9.1)$. Correspondingly, the TBA data of Fig.~\ref{fig:XXXFB TEBD TBA} (f) show an even clearer separation of $\rho_n(\lambda)$ in rapidity space: different string species become localized into distinct rapidity ``packets", and within each packet the effective velocities of the contributing strings become nearly degenerate. Each packet therefore behaves as an emergent quasiparticle mode with a well-defined velocity, producing a sharp, almost ballistically propagating peak. This mechanism closely parallels the peak formation discussed in the gapless regime, now realized at the isotropic point through a reorganization of the relevant TBA weights and velocities.

The emergence of a transient ballistic mode can be understood by noting that the $PR$ gate becomes nearly dual unitary for the choices of $(\tau_1, \tau_3)$ in Fig.~\ref{fig:XXXFB TEBD TBA}(b)\&(c)~\cite{DU19prl,dualunitaryRMP}. Recall that our circuit layers contain three types of two-site gates (Fig.~\ref{fig:Our Circuits}), whose spectral parameters are \([\pm \tau_1,\, \pm \tau_3,\, \pm(\tau_1-\tau_3)]\) (with the convention $\tau_2=0$). At the isotropic point, taking $|\tau|\to\infty$ drives the gate $PR$ towards the swap gate (up to inessential phases), which is a prototypical dual-unitary gate and yields effectively non-interacting, strictly ballistic propagation. For $(\tau_1,\tau_3)=(5.2,-9.1)$, all three gate types have large spectral parameters, placing the local gates close to this swap/dual-unitary regime. Nevertheless, since dual unitarity is only approximate and is in general not compatible with $R$ matrix at the isotropic point, this ballistic mode is not a stable quasiparticle mode, and hence the ballistic component in $C(x,t)$ is only transient. This provides an intuitive explanation for the numerically observed sharp ballistic peaks on top of the intrinsic superdiffusive background.

\begin{figure}[!t] 
    \centering
    \includegraphics[width=1.0\linewidth]{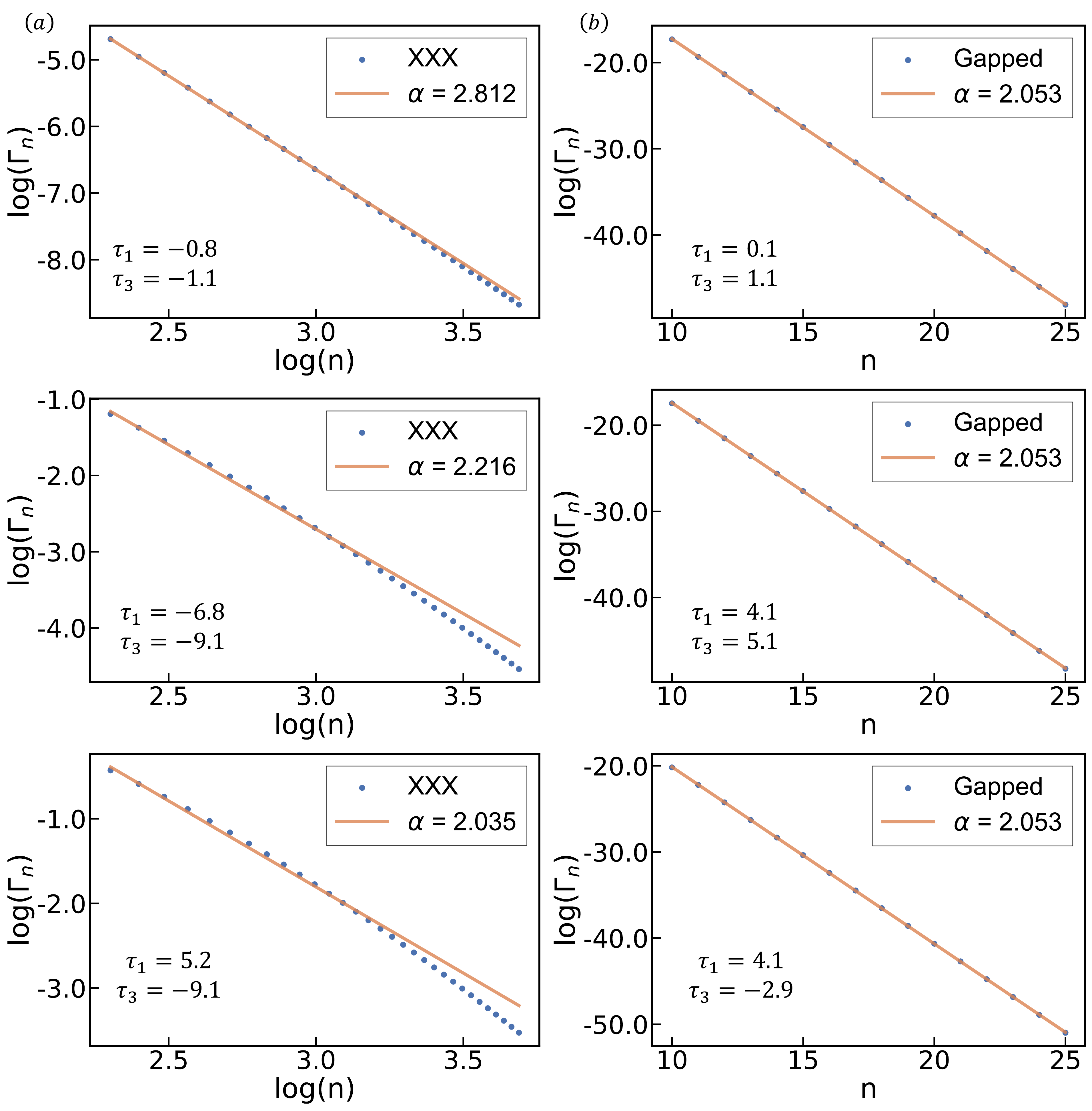}
    \caption{Scaling behavior of the rapidity integral $\Gamma(n)$ in Eq.~(\ref{eq:integralDDW}) as a function of string length. The left panels show results at the isotropic point, while the right panels correspond to the gapped phase.}
    \label{fig:ddwintscaling}
\end{figure}

Through the GHD approach we obtain an alternative route to demonstrate that
spin transport at the isotropic point is superdiffusive. It has been shown that, for
locally interacting spin chains, the spin diffusion constant $D$ is bounded
from below by the curvature of the Drude weight $\mathcal{D}$ with respect to
the magnetic field $h$ (conjugate to the total magnetization)~\cite{boundDiffc,xxxsuperdif18ilievskiprosen}. In the high-temperature regime and
in the vicinity of half filling ($h\to 0$), this bound can be written as
\begin{equation}
    D \;\ge\; \lim_{\beta \to 0} 
    \frac{2}{\beta\, v_{LR}}
    \,\partial_{h}^{2} \mathcal{D}(\beta,h)\big|_{h=0},
\end{equation}
where $v_{LR}$ is the Lieb--Robinson velocity and
$\partial_{h}^{2} \mathcal{D}(\beta,h)\big|_{h=0}$ is the curvature of the
Drude weight at half filling. Within the GHD framework, using the expression
\eqref{eq:DWghd} for the Drude weight, this curvature can be expressed as
\begin{equation}
    \frac{\beta}{2} \sum_{n\ge 1} 
    \int \frac{\mathrm{d}\lambda}{2\pi}\,
    \theta_n(\lambda)\bigl[1-\theta_n(\lambda)\bigr]\,
    p'_n(\lambda)\,
    \bigl[v_n^{\mathrm{eff}}(\lambda)\bigr]^2
    \left.\frac{\partial^2 \bigl[m_n^{\mathrm{dr}}(\lambda)\bigr]^2}{\partial h^2}\right|_{h=0}.
\end{equation}
In the high-temperature, half-filled regime that we consider, the filling
functions are known explicitly and take the simple form
\begin{equation}
    \theta_n(\lambda) = \frac{1}{(1+n)^2}.
\end{equation}
For the dressed magnetizations, in the limit $h\to 0$ and for $nh\ll 1$, one
finds~\cite{xxxsuperdif18ilievskiprosen}
\begin{equation}
    m_n^{\mathrm{dr}}(\lambda)
    \sim \frac{1}{3}(n+1)^2\, h + \mathcal{O}\bigl(h^3\bigr),
\end{equation}
so that
\begin{equation}
    \theta_n(1-\theta_n)
    \left.\frac{\partial^2 \bigl[m_n^{\mathrm{dr}}\bigr]^2}{\partial h^2}\right|_{h=0}
    \propto (n+1)^2 \sim n^2 \quad (n\gg 1).
\end{equation}
Substituting the GHD expression for the curvature into the diffusion bound and
taking the limit $\beta\to 0$, one obtains
\begin{equation}
    D\ge \frac{1}{v_{LR}} 
    \sum_{n\ge 1} \theta_n(1-\theta_n)
    \left.\frac{\partial^2 \bigl[m_n^{\mathrm{dr}}\bigr]^2}{\partial h^2}\right|_{h=0}
    \int \frac{\mathrm{d}\lambda}{2\pi}\,
         p'_n(\lambda)\,
         \bigl[v_n^{\mathrm{eff}}(\lambda)\bigr]^2.
\end{equation}
The string-length dependence of the prefactor in front of the rapidity
integral thus scales as $n^2$ for large $n$. Therefore the asymptotic scaling
of the integral
\begin{equation}
    \Gamma(n)
    = \int \mathrm{d}\lambda\,
      p'_n(\lambda)\,\bigl[v^{\mathrm{eff}}_n(\lambda)\bigr]^2
    \label{eq:integralDDW}
\end{equation}
determines whether the diffusion constant diverges. In practice, we compute
$\Gamma(n)$ by first solving the dressed-velocity equations
\eqref{eq:Linear source in EdrTBA} numerically.

For our circuit at the isotropic point, we find that the
large-$n$ behaviour of $\Gamma(n)$ at all inhomogeneity parameters considered
is well described by a power-law decay,
\begin{equation}
    \Gamma(n)
    = \int_{-\infty}^{\infty}\! \mathrm{d}\lambda\,
      p'_n(\lambda)\,\bigl[v^{\mathrm{eff}}_n(\lambda)\bigr]^2
    \simeq \frac{1}{n^{\alpha}},
\end{equation}
with exponents $\alpha<3$ [Fig.~\ref{fig:ddwintscaling}(a)]. Combining this with the $n^2$ growth of the
prefactor, the contribution of large strings to the bound behaves as
$\sum_s n^2 \Gamma(n)\sim \sum_n n^{2-\alpha}$, which diverges for
$\alpha\le 3$. Hence the diffusion constant is divergent, and spin transport
is superdiffusive at the isotropic point.

\begin{figure}[!tb] 
    \centering
    \includegraphics[width=0.9\linewidth]{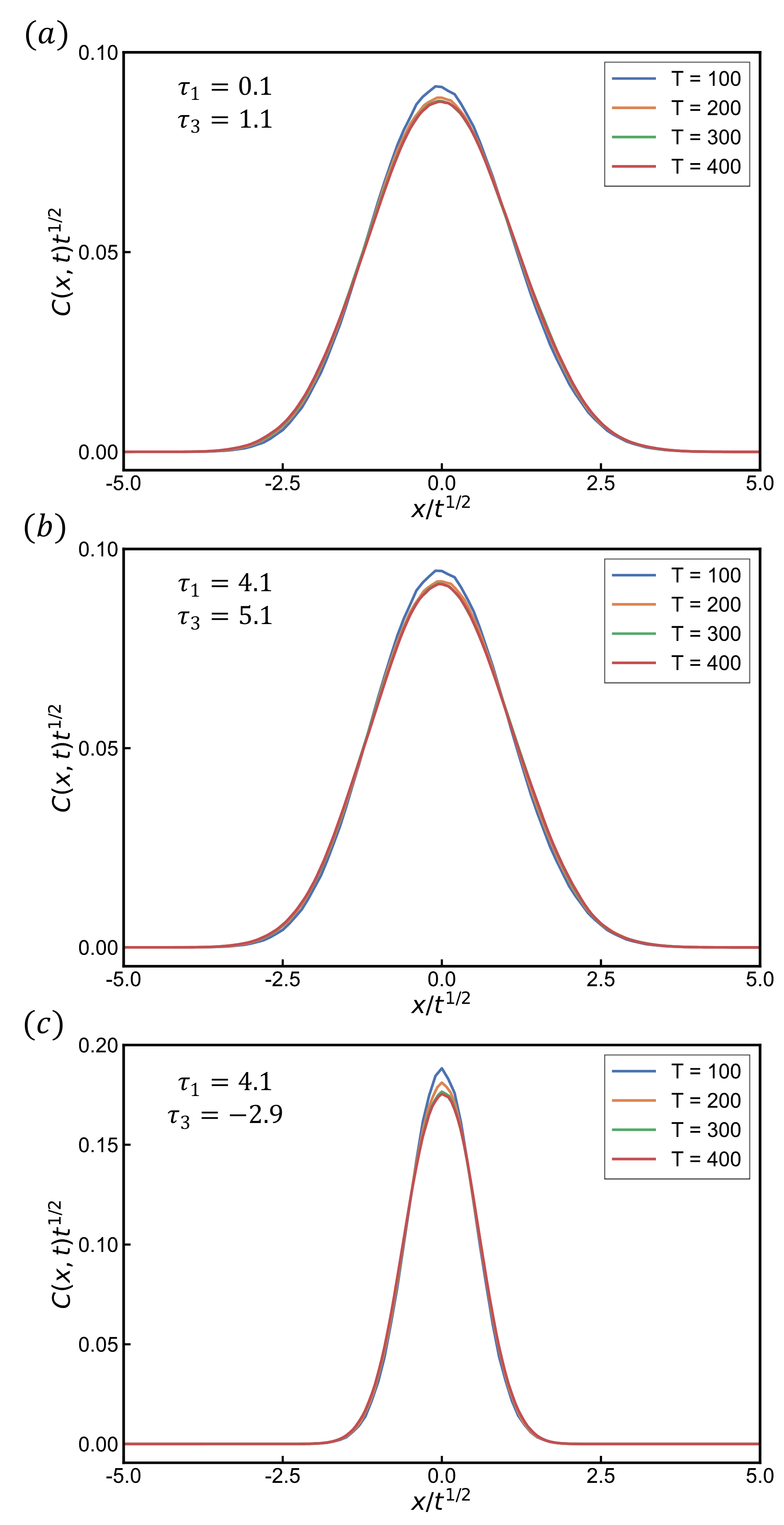}
    \caption{Scaling collapse of the correlation function $C(x, t)$ obtained via tDMRG simulations under Fibonacci sequence with dynamical exponent $z = 2.0$. While the shape and width of the scaling function depend on the spectral parameters, spin transport is diffusive throughout the entire phase. Results were obtained with system size $L = 1000$, bond dimension $\chi = 256$ and $\mu=0.002$.}
    \label{fig:3x GappedFB}
\end{figure}

\subsection{Gapped phase}
For the gapped phase, we fix $\eta = \mathrm{i} \frac{\pi}{3}$ and sample the spectral parameters from the following set:
\begin{equation*}
    \begin{aligned}
       % \eta &= \mathrm{i} \frac{\pi}{3}, \\
        \tau_1 &\in \{-3.9, 0.1, 4.1\}, \\
        \tau_3 &\in \{-2.9, 1.1, 5.1\}.
    \end{aligned}
\end{equation*}
Under Fibonacci sequence, the correlation function \(C(x,t)\) obtained from tDMRG shows an excellent collapse according to the scaling hypothesis~(\ref{eq:collapse}) with dynamical exponent \(z=2\) for all choices of $(\tau_1, \tau_3)$, as shown in Fig.~\ref{fig:3x GappedFB}. This suggests that spin transport becomes diffusive in this regime. Moreover, in contrast to the isotropic limit, in the gapped phase we find that the function \(\Gamma(n)\) decays exponentially with the string length,
\begin{equation}
   \Gamma(n)
   = \int_{-\pi/2}^{\pi/2} \frac{\mathrm{d}\lambda}{2\pi}\,
     p'_n(\lambda)\,\bigl[v^{\mathrm{eff}}_n(\lambda)\bigr]^2
   \simeq \mathrm{e}^{-\alpha n},
\end{equation}
with some positive constant \(\alpha>0\), as shown in the right panels of Fig.~\ref{fig:ddwintscaling}. In this case the sum over \(n\) is convergent, which implies a finite diffusion constant, in agreement with expectations for the gapped XXZ regime.

\begin{figure*} 
    \centering
    \includegraphics[width=0.9\linewidth]{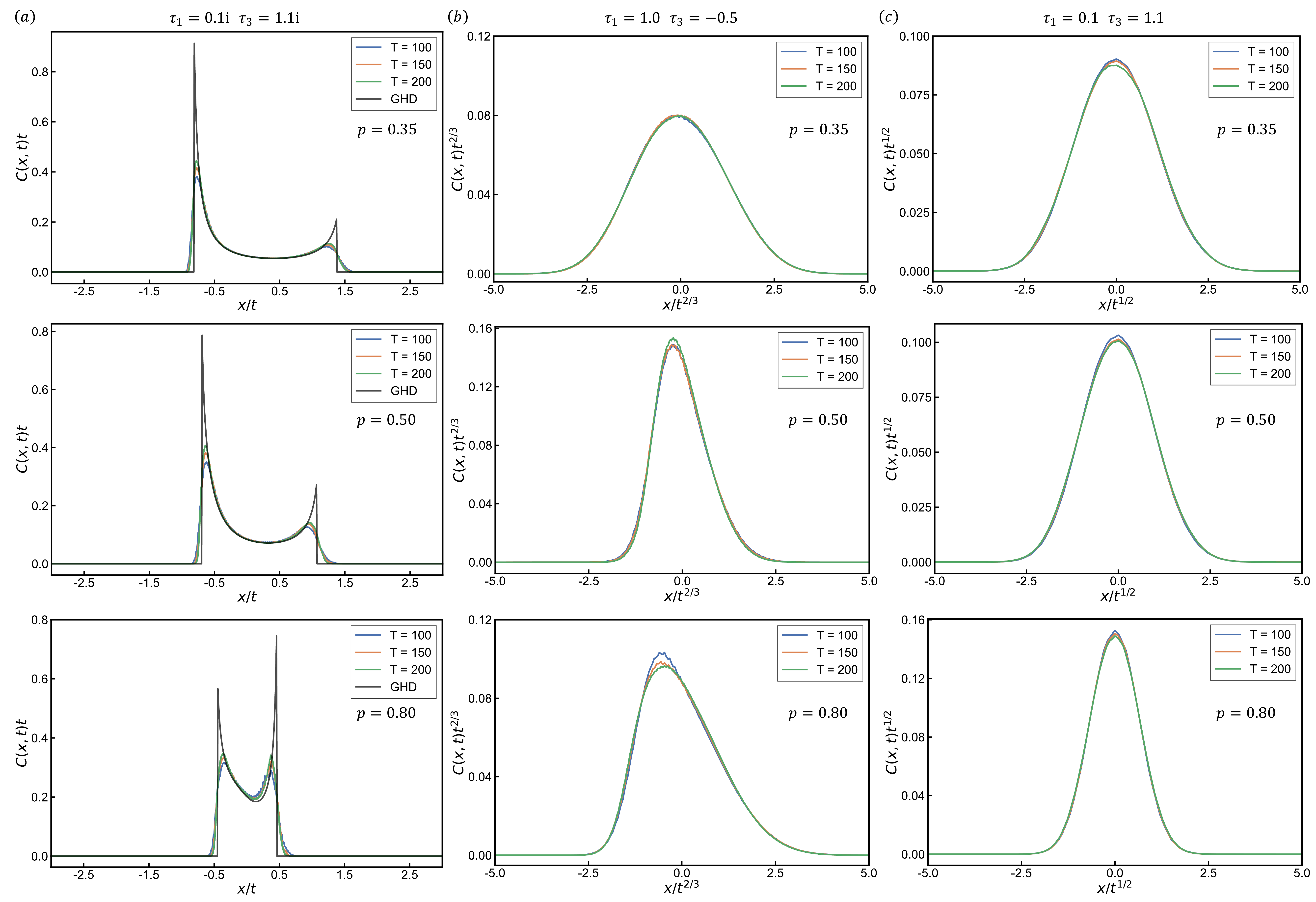}
    \caption{{\bf Integrable random circuits.} Scaling collapse of the correlation function \(C(x,t)\) under a stochastic sequence with probability \((p,1-p)\) for applying \((U_1, U_0)\). (a) Gapless phase with anisotropic parameter \(\eta=\pi/3\) and \((\tau_1,\tau_3)=(0.1\mathrm{i},1.1\mathrm{i})\), computed with system size \(L=1200\). (b) Isotropic point with spectral parameters \((\tau_1,\tau_3)=(1.0,-0.5)\), computed with system size \(L=800\). (c) Gapped phase with anisotropic parameter \(\eta=\frac{\pi}{3}\mathrm{i}\) and \((\tau_1,\tau_3)=(0.1,1.1)\), computed with system size \(L=800\). All results are obtained with bond dimension \(\chi = 128\), \(\mu = 0.002\), and an average over 100 circuit realizations.} \label{fig:Random 3x3}
\end{figure*}

\section{Numerical results for random quantum circuits} 
\label{sec:random result}
We now extend our construction to random quantum circuits by applying the circuit layers $U_0$ and $U_1$ stochastically, following the protocol described in Sec.~\ref{stochastic protocol}, and test whether the three dynamical phases controlled by the underlying spectral parameters remain robust in this case. Furthermore, it also provides a useful benchmark on the validity of GHD in describing stochastic integrable quantum dynamics.
For random quantum circuits, we consider the ensemble-averaged spin correlation function.
In the limit of many realizations, the frequencies of $U_1$ and $U_0$ converge to $p$ and $1-p$, respectively. The corresponding
ensemble-averaged correlation function is therefore expected to be described by GHD using a weight-averaged effective velocity, as discussed in Sec.~\ref{Constructing Correlations via TBA and GHD}. 

Our results are summarized in Fig.~\ref{fig:Random 3x3}. In each dynamical phase, we fix the spectral parameters $(\tau_1,\tau_3)$ and consider different values of $p$, which controls the densities of $U_1$ and $U_0$. The vertical panels (a), (b) and (c) in Fig.~~\ref{fig:Random 3x3} correspond to gapless, isotropic, and gapped regimes, respectively. Again, we find that in all cases that we have tested, the correlation function shows an excellent scaling form of Eq.~(\ref{eq:collapse}), with dynamical exponents matching those determined by the spectral parameters of the $R$ matrices. 
Changing $p$ noticeably affects the lineshape of $C(x,t)$, yet the scaling behavior within each regime remains unchanged. Moreover, in the gapless case we observe excellent agreement between tDMRG and the GHD reconstruction in the bulk. 
 
\begin{figure}[!tb] 
    \centering
    \includegraphics[width=0.9\linewidth]{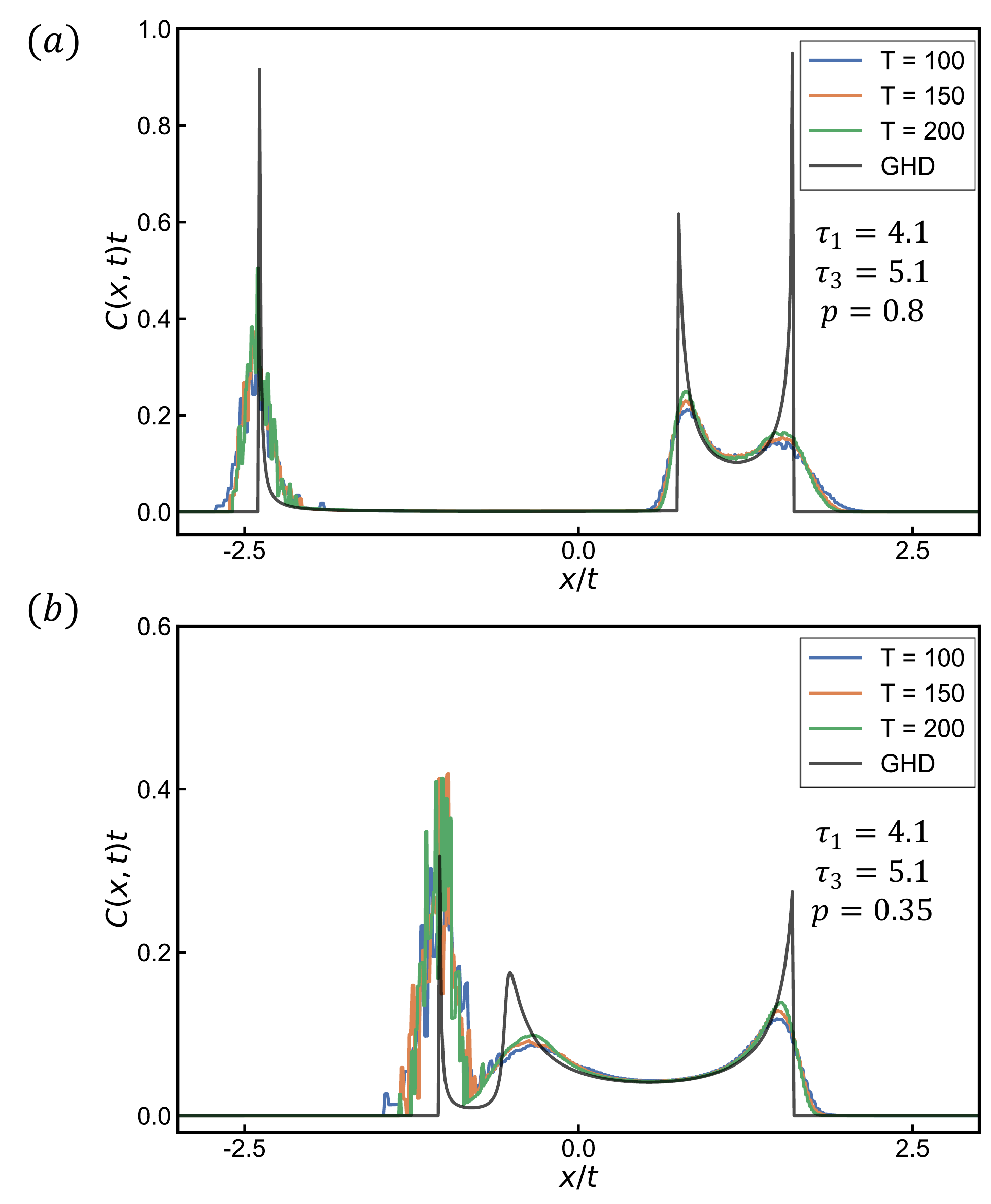}
    \caption{{\bf Integrable random circuit.} Scaling collapse of the correlation functions $C(x,t)$ and the corresponding GHD  $C_{\text{ghd}}(x,t)$  for $(\tau_1,\tau_3)=(4.1\mathrm{i},5.1\mathrm{i})$ at fixed anisotropic parameter $\eta=\frac{\pi}{3}$. The results are shown for (a) $p=0.8$ and (b) $p=0.35$. GHD gives excellent predicitions for the structure and locations of the peaks, while discrepancies in the width of the peaks are due to a combination of finite samples, finite-size effect, and beyond Euler-scale contributions. Simulations are performed with system size $L=1200$, bond dimension $\chi = 128$, $\mu = 0.002$ and an average over $200$ circuit realizations.}
    \label{fig:Random GHDvsTEBD}
\end{figure}

Finally, we consider a particular choice of parameters in the gapless phase where degenerate effective velocities leads to sharp ballistic peaks, as shown in Fig.~\ref{fig:FB TEBD GHD TBA}(b)\&(c).
Under stochastic gate sequences with varying $p$, the results are shown in Fig.~\ref{fig:Random GHDvsTEBD}. For $(\tau_1,\tau_3)=(4.1\mathrm{i},\,5.1\mathrm{i})$, the Euler-scale GHD prediction $C_{\mathrm{ghd}}(x,t)$ contains a sharp peak associated with a near-degeneracy of effective velocities. In tDMRG, finite sampling and finite-size effects broaden this feature, but in contrast to the quasi-periodic case as in Fig.\ref{fig:FB TEBD GHD TBA}(b), the peak height is no longer anomalously enhanced: the observed amplitude remains of the same order as, and indeed comparable to, the GHD prediction after ensemble averaging. At the same time, the peak positions and the propagation fronts remain in excellent agreement with GHD predictions. Overall, our results clearly demonstrate that, when appropriately taking into account weight-averaging in the effective velocities, GHD gives excellent prediction of linear-response functions under stochastic integrable quantum circuit dynamics.

\section{Conclusion and discussion} \label{sec:Outlook}
In this work, we construct a family of integrable quantum circuit models that do not have time-translation symmetry. Starting from the XXZ (six–vertex) \(R\)-matrix, we build a family of three-site inhomogeneous transfer matrices and use them to define two commuting yet unitarily inequivalent circuit layers, \(U_0\) and \(U_1\), which commute with the same transfer matrix and therefore share an identical infinite set of conserved charges. This yields a broad class of quantum circuit evolution protocols built from $\{U_0, U_1\}$ -- including quasiperiodic Fibonacci and Thue-Morse sequences as well as completely random sequences — in which time-translation symmetry is broken while Yang–Baxter integrability is preserved at all times. Using large-scale tDMRG simulations at infinite temperature and half filling, we map out the spin transport phase diagram of our integrable circuits, which is closely analogous to that of the Hamiltonian and Floquet counterpart:
ballistic spin transport in the easy–plane regime, superdiffusion at the isotropic point, and normal diffusion in the easy–axis regime. Within each phase we further show that the spatiotemporal profile of spin–spin correlation depends sensitively on the inhomogeneity parameters \(\{\tau_i\}\), including the coexistence of emergent ballistic modes and superdiffusive wavepacket due to proximity to dual unitarity. To interpret these phenomena, we develope a thermodynamic Bethe–ansatz and generalized-hydrodynamics description adapted to our circuit, providing Euler-scale predictions for dressed quasiparticle
content, Drude weights, diffusion bounds, and correlation profiles. The comparison between GHD and tDMRG highlights both the predictive power and the limitations of GHD for these random integrable circuits beyond static Hamiltonian and time-periodic Floquet settings.

Our results prompt several directions for future work.
(i) \emph{GHD vs tDMRG correlations and front scaling.}
While Euler-scale GHD reproduces the bulk of the spin--spin correlator remarkably well throughout this work, it systematically misses the physics at the propagation
fronts~\cite{freefront,xxzfront19,xxzfrontrev}: tDMRG shows broadened fronts and fine structure beyond the strictly ballistic GHD prediction. An important direction is therefore to quantify the front scaling in our time-dependent yet integrable circuits and to test its robustness across regimes,
inhomogeneity parameters, and driving protocols. In particular, it is natural to ask whether stochastic driving produces genuinely new front universality classes or instead renormalizes known sub-ballistic broadening mechanisms in integrable dynamics.
Answering this will require extracting front widths and their time scaling for different initial states and relating them to the TBA velocity landscape (e.g., near-degeneracies
that generate sharp ballistic components), as well as developing controlled subleading hydrodynamic corrections to the present protocol-averaged GHD framework. 

(ii) \emph{Controlled integrability breaking.} What is the fate of spin transport and of the GHD description under weak perturbations that violate the Yang--Baxter structure? In view of existing GHD frameworks for weak spatial inhomogeneities and slowly varying drives \cite{GHDstinh}, it would be natural to develop a controlled perturbative extension that predicts nontrivial crossover behaviour in transport. 

(iii) \emph{Beyond linear response.}
Recent progress on higher-order fluctuations \cite{fullcountingTEBD} and on anomalous generalized Gibbs ensembles in integrable systems \cite{SQGGE} motivates a systematic study of higher current cumulants and anomalous
thermalization in our time-dependent integrable circuits. A closely related and more fundamental question is whether one can construct an explicit discrete GGE~\cite{dGGE} for our quantum-circuit model. 

(iv) \emph{Gate noncommutativity and new protocols.} In the present construction, \(U_0\) and \(U_1\) commute as a consequence of their Yang--Baxter origin. An open challenge is to realize genuinely interacting, time-translation–symmetry–breaking integrable circuits in which the elementary gates need not commute, while still preserving an extensive set of conserved charges beyond the framework explored here.

\section*{Acknowledgment}
We are grateful to Bal\'{a}zs Pozsgay and Jie Ren for useful discussions. This work is supported by Grant No. 12375027 from the National Natural Science Foundation of China (S.W., C.L., and Z.-C.Y.). H.Z. is supported by Quantum Science and Technology-National Science and Technology Major Project
(No. 2024ZD0301800) and by the National Natural Science
Foundation of China (Grant No. 12474214).
Numerical simulations based on tensor networks use the ITensor package~\cite{iTensor} and were performed on the High-performance Computing Platform of Peking University. 

\section*{Data Availability}
The numerical data and simulation code used in this work are openly available at Zenodo, DOI: 10.5281/zenodo.18503174.

\appendix
\section{Classification of R-matrix and parameter choice}
\label{Appendix A}
Within the Yang-Baxter framework, the quantum phase of the XXZ chain is directly encoded in the functional form of its $R$-matrix: the gapped phase ($\Delta>1$) is described by a \emph{hyperbolic} R-matrix \eqref{RGapped}, the gapless phase ($\Delta<1$) by a \emph{trigonometric} one \eqref{RGapless}, and the isotropic point ($\Delta=1$) by a \emph{rational} one \eqref{RXXX}. All three cases originate from the same general $R$-matrix expression given in Eq.~\eqref{Rmatrix} by extending the spectral parameter $u$ to the complex plane and evaluating it in the appropriate domain.

\begin{figure*}
    \centering
    \includegraphics[width=0.95\linewidth]{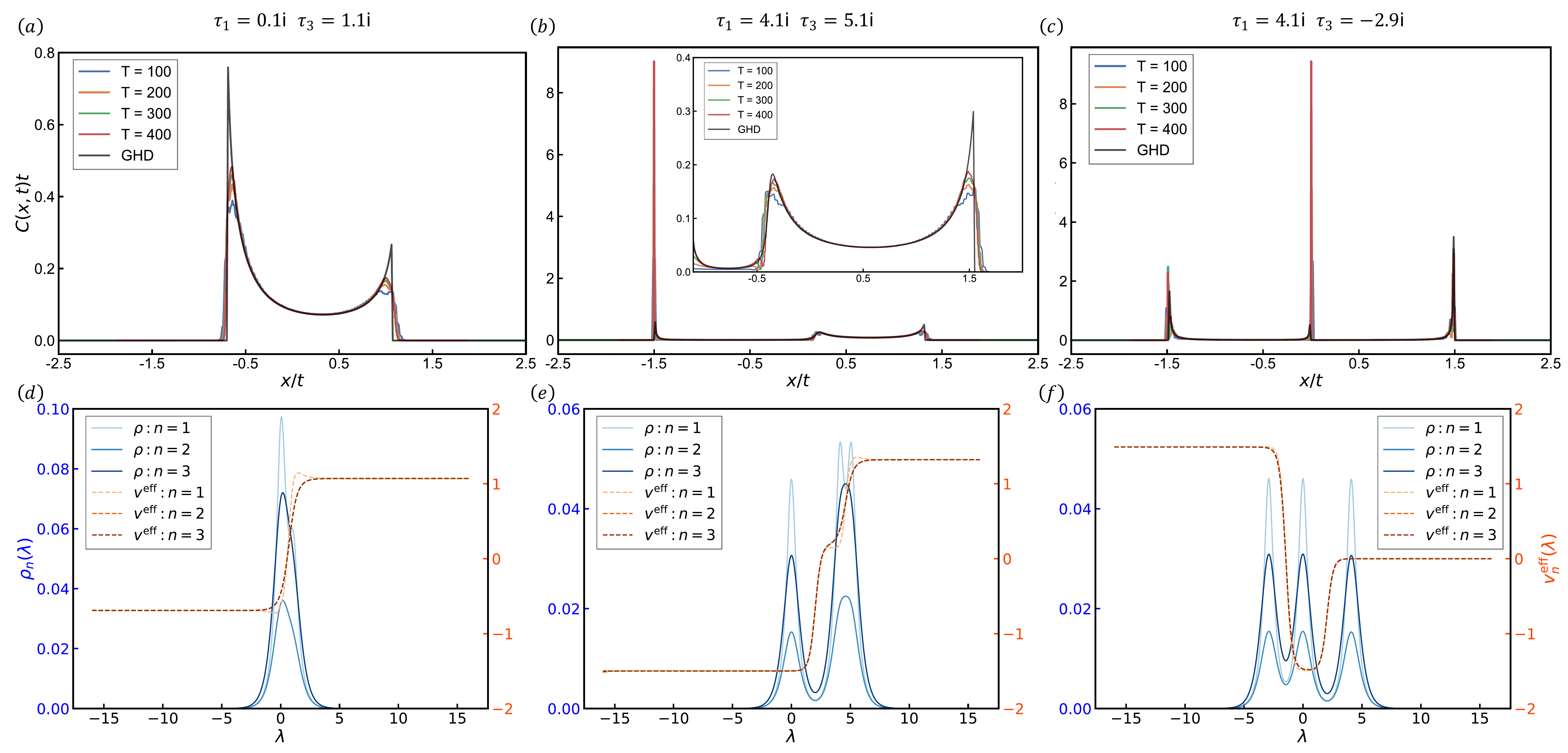}
    \caption{(a), (b), and (c): Scaling collapse of the correlation function $C(x,t)$ obtained from tDMRG simulations compared with the GHD prediction $C_{\text{ghd}}(x,t)$ under Thue--Morse sequence. Simulations are performed with system size $L = 1500$, bond dimension $\chi = 256$ and $\mu=0.002$. The inset of (b) is a zoom-in plot of the low-amplitude region of the profile, which shows excellent agreement between tDMRG and GHD there.
    (d), (e), and (f): Corresponding TBA results for the effective velocity $v_{\mathrm{eff}}(\lambda)$ and quasi-particle density $\rho_n(\lambda)$ used as inputs for GHD calculation.
    This set of pictures serves as the Thue--Morse counterpart to the results under Fibonacci sequence presented in the main text (see Fig.~\ref{fig:FB TEBD GHD TBA}).}
    \label{fig:TM TEBD GHD TBA}
\end{figure*}

To ensure that the overall quantum evolution remains unitary, the local gate $PR$ must itself be unitary. Since the swap operator $P$ is unitary, this requirement reduces to imposing unitarity on the $R$-matrix alone. To derive the corresponding parameter constraints explicitly, we note that the $R$-matrix (Eq.~\eqref{Rmatrix}) is related to the XXZ Hamiltonian via an exponential map \cite{XXZcircuitprosen}:

\begin{equation}
    \begin{aligned}
         \operatorname{e}^{\mathrm{i} \Delta \tau} \operatorname{e}^{-\mathrm{i}\tau (\sigma^x \sigma^x + \sigma^y \sigma^y + \Delta \sigma^z \sigma^z)} &=
    \begin{pmatrix}
        1 & 0 & 0 & 0 \\
        0 & a & b & 0 \\
        0 & b & a & 0 \\
        0 & 0 & 0 & 1
    \end{pmatrix} ,\\
        a = \frac{1}{2}\operatorname{e}^{2\mathrm{i}(\Delta-1)\tau}(1+\operatorname{e}^{4\mathrm{i}\tau}) &= \frac{\sin(u)}{\sin(u+\eta)} ,\\
        b = \frac{1}{2}\operatorname{e}^{2\mathrm{i}(\Delta-1)\tau}(1-\operatorname{e}^{4\mathrm{i}\tau}) &= \frac{\sin(\eta)}{\sin(u+\eta)}.
    \end{aligned}
    \label{expHxxz}
\end{equation}

From these relations, we obtain the explicit parametric mapping between the physical parameters $(\tau,\Delta)$ and the Yang-Baxter parameters $(u,\eta)$:

\begin{equation}
     e^{2i(\Delta \pm 1)\tau} = \frac{\sin u \mp \sin \eta}{\sin(u + \eta)}.
\end{equation}
$\Delta$ and $\tau$ must be real to ensure unitarity and lead to the constriant

\begin{equation}
    \begin{aligned}
         \left| \frac{\sin u \mp \sin \eta}{\sin(u + \eta)} \right| &= 1.
    \end{aligned}
    \label{constraint}
\end{equation}

Equation~\eqref{constraint} is satisfied in two mutually exclusive regimes:
\begin{equation}
    \begin{aligned}
        &u \in \mathbb{R} ~ \text{and} ~\eta \in \mathrm{i} \mathbb{R}, \
        &\text{or} \quad \
        &u \in \mathrm{i} \mathbb{R} ~ \text{and} ~ \eta \in \mathbb{R}.
    \end{aligned}
\end{equation}

These two regions correspond precisely to the two main thermodynamic phases of the XXZ chain: the regime with $u \in R$ describes the gapless phase, whereas $u \in \mathrm{i} R$ corresponds to the gapped phase. Notice that
\begin{equation}
    PR(u;\eta) = R(\eta;u).    
\end{equation}
Thus, for our unitary circuit constructed from $PR$, the parameters are constrained as follows:
\begin{itemize}
\item \textbf{Gapped phase (hyperbolic)}: $u \in \mathbb{R}$ and $\eta \in \mathrm{i} \mathbb{R}$.
\item \textbf{Gapless phase (trigonometric)}: $u \in \mathrm{i} \mathbb{R}$ and $\eta \in \mathbb{R}$
\item \textbf{Isotropic point (rational)}: This is obtained as a scaling limit of the above phases, where the spectral parameter $u$ in Eq.~\eqref{RXXX} is real.
\end{itemize}

\begin{figure*}
    \centering
    \includegraphics[width=0.95\linewidth]{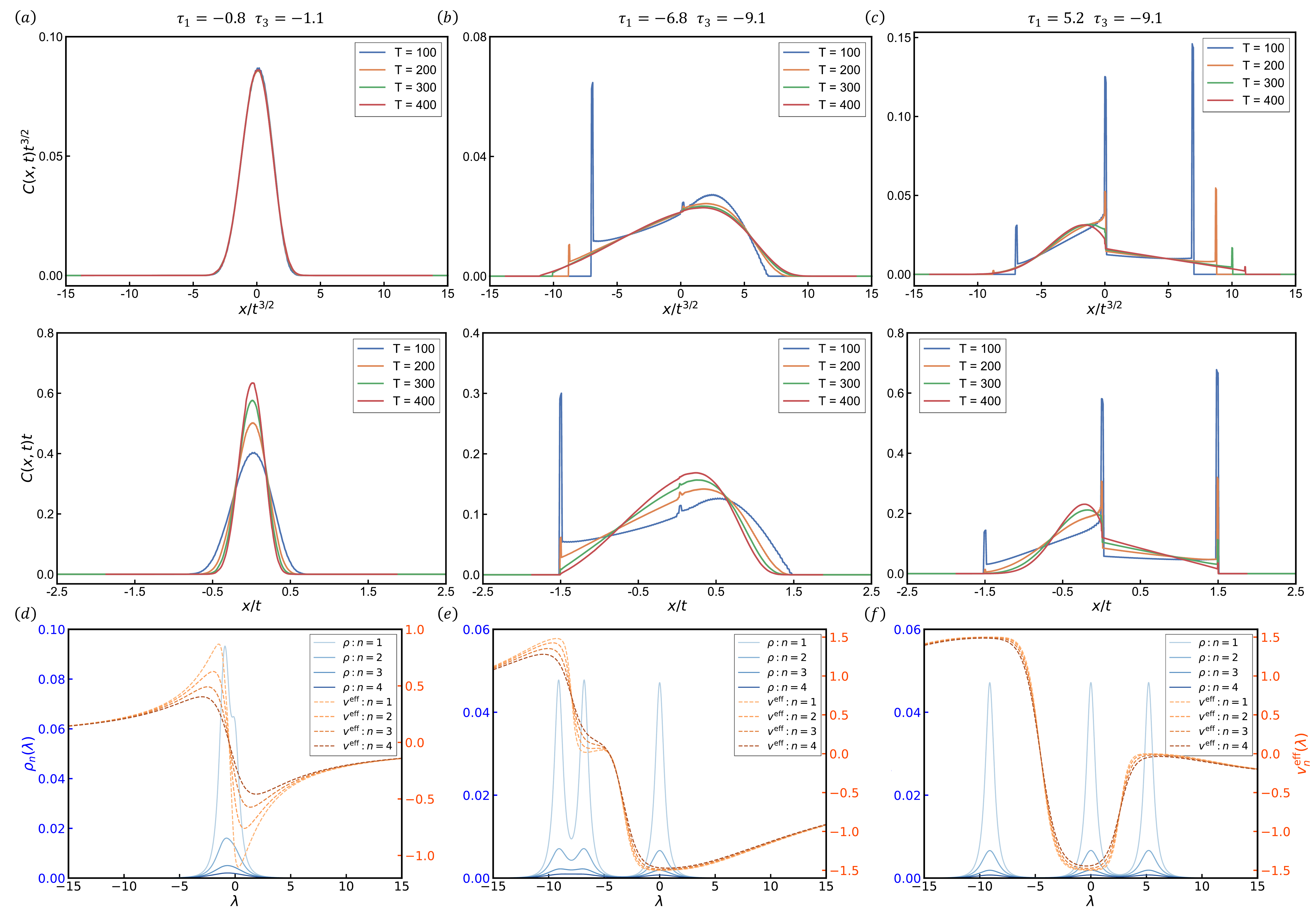}
    \caption{Panels (a), (b) and (c) show the scaled spin-spin correlation function $C(x,t)$ under Thue-Morse sequence using the scaling hypothesis Eq.~(\ref{eq:collapse}) with two different choices of the dynamical exponent: $z=3/2$ (first row)  and $z=1.0$ (second row), respectively. The coexistence of superdiffusive $z=3/2$ transport and emergent $z=1$ ballistic mode due to proximity to dual unitarity is clearly visible in panels (b)\&(c). Results were obtained with system size $L = 1500$, bond dimension $\chi = 256$ and $\mu=0.002$.
    Panel (d), (e) and (f) present the corresponding TBA data. Here we only present the first four strings. This set of pictures serves as the Thue--Morse counterpart to the results under Fibonacci sequence presented in the main text (see Fig.~\ref{fig:XXXFB TEBD TBA}.)}
    \label{fig:XXXTM TEBD TBA}
\end{figure*}

\section{Quasiperiodic quantum circuits from Thue--Morse sequence} 
\label{Appendix B}
This appendix presents the scaling collapse of the correlation function $C(x,t)$ under
Thue--Morse (TM) driving. Since $U_0$ and $U_1$ commute, the dynamics at even time steps
are governed by the effective Floquet unitary $U_F=U_0U_1$, so TM driving can be viewed as
a Floquet evolution sampled at even times. For a direct comparison with the main text, we
use the same model parameters as in the Fibonacci (FB) analysis. The corresponding results
in the gapless regime and at the isotropic point are shown in
Fig.~\ref{fig:TM TEBD GHD TBA} and Fig.~\ref{fig:XXXTM TEBD TBA}, respectively.

Overall, the TM sequence yields the same dynamical scaling exponents as FB driving and
produces qualitatively similar structures (smooth backgrounds coexisting with sharp
ballistic-like peaks), but the detailed \emph{lineshapes} are sequence dependent. This is
illustrated by explicit examples. In the gapless regime at
$(\tau_1,\tau_3)=(4.1\mathrm{i},-2.9\mathrm{i})$, the correlation function develops the same
distinctive three-peak structure as in the Fibonacci case, see
Fig.~\ref{fig:TM TEBD GHD TBA}(c) (cf.~Fig.~\ref{fig:FB TEBD GHD TBA}). At the isotropic
point, TM driving likewise exhibits a clear evolution of profile shapes across parameter
choices: as shown in Fig.~\ref{fig:XXXTM TEBD TBA}(a)--(c), the correlations cross over
from a broadly spreading profile to one with a single sharp peak and finally to a
three-peak structure.

From the GHD perspective, the stationary quasiparticle root densities \(\rho_n(\lambda)\) are uniquely determined by the inhomogeneous transfer matrix and remain independent of the specific circuit sequence employed. Therefore, the choice of circuit sequence only affects the dressed dynamical quantities—in particular, the effective velocities \(v^\mathrm{eff}_n(\lambda)\). As shown in Fig.~\ref{fig:FB TEBD GHD TBA}(c) and Fig.~\ref{fig:TM TEBD GHD TBA}(c), for fixed parameters \((\tau_1,\tau_3)\), the qualitative structure of the effective velocities as functions of rapidity remains largely unchanged across the two protocols. Specifically, rapidity regions where the velocities of different string species are nearly degenerate under one sequence remain nearly degenerate under the other and vice versa. This persistence of the degeneracy and splitting patterns implies that the corresponding spin–spin correlation functions exhibit the same qualitative space–time structure under both Fibonacci and Thue--Morse sequences. The choice of protocol primarily influences quantitative features—such as front velocities and peak amplitudes—rather than altering the overall profile.

\begin{figure}[!t] 
    \centering 
    \includegraphics[width=0.95\linewidth]{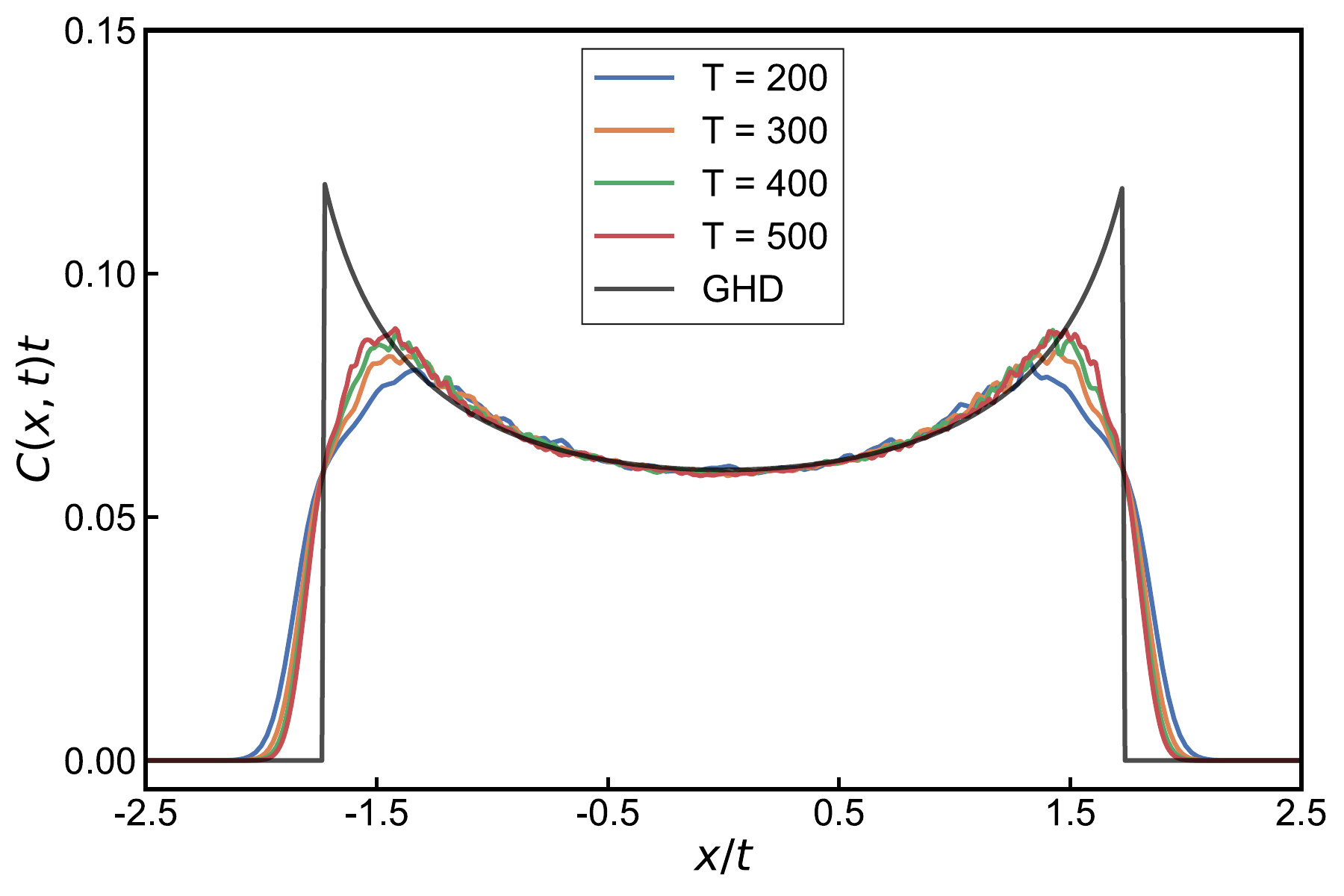}
    \caption{Comparison of the spin--spin correlation function \(C(x,t)\) for the static XXZ Hamiltonian computed using tDMRG and the GHD formula. Dashed curves show the GHD prediction, while solid curves (with scaling collapse) correspond to the tDMRG data. The tDMRG results are obtained for a chain of length \(L=500\) with time step \(dt=0.1\) and bond dimension \(\chi=400\), which we find to be sufficient for convergence.}
    \label{fig:ghdcheck}
\end{figure}

\section{Comparison of GHD and tDMRG in the Hamiltonian case}
\label{ap:Htdmrgvsghd}
In this appendix we benchmark the GHD formula for spin correlations in the
simpler setting of a homogeneous XXZ Hamiltonian, in order to disentangle
genuine circuit effects from possible limitations of the hydrodynamic
description. We again focus on infinite temperature and the gapless regime,
choosing \(\Delta = \cos\frac{\pi}{3} = 0.5\). At this anisotropy the string
content is the simplest one and coincides with the gapless string structure
used in the main text for the circuit; however, the closed integral equations
for the dressed quantities differ because the transfer matrix is now
homogeneous.

In the Hamiltonian case the bare derivative of the quasiparticle momentum is
related to the total density \(\rho^t_j(\lambda)\) via
\(p'^{\rm dr}_j(\lambda) = 2\pi \sigma_j \rho^t_j(\lambda)\), and can equivalently be
written as
\(p'_j(\lambda) = 2\pi a_j(\lambda)\).
The dressed momentum derivatives satisfy the linear integral equation
\begin{equation}
    p'^{\rm dr}_j(\lambda)
    = 2\pi a_j(\lambda)
      - \sum_{k=1}^{P} \sigma_k \int_{-\infty}^{\infty}
        \mathrm{d}\lambda'\, A_{jk}(\lambda-\lambda')\,
        \theta_k(\lambda')\, p'^{\rm dr}_k(\lambda'),
\end{equation}
where \(A_{jk}\) and \(n_k(\lambda)\) are the usual scattering kernel and
filling functions of the gapless XXZ TBA.

To obtain the dressed velocities of the quasiparticles we also need the
dressed quasiparticle energies. For the homogeneous Hamiltonian the bare
energy of the \(j\)-th string is
\begin{align}
    E_j(\lambda)
    &= \frac{2\sin\gamma\,\sin(n_j\gamma)}
            {v_j\cosh(2\lambda)-\cos(n_j\gamma)}, \\
    E'_j(\lambda)
    &= \frac{-4 v_j \sin\gamma\,\sin(n_j\gamma)\,\sinh(2\lambda)}
            {\bigl[v_j\cosh(2\lambda)-\cos(n_j\gamma)\bigr]^2},
\end{align}
so that the dressed energy derivatives are determined by
\begin{equation}
    E_j'^{\rm dr}(\lambda)
    = E'_j(\lambda)
      - \sum_{k=1}^{P} \sigma_k \int_{-\infty}^{\infty}
        \mathrm{d}\lambda'\, A_{jk}(\lambda-\lambda')\,
        \theta_k(\lambda')\, E_k'^{\rm dr}(\lambda').
\end{equation}
The dressed velocities then follow as
\(v_j^{\rm eff}(\lambda) = E_j'^{\rm dr}(\lambda)/p_j'^{\rm dr}(\lambda)\),
which we insert into the GHD correlation formula
Eq.~\eqref{eq:GHDformula}.
As shown in Fig.~\ref{fig:ghdcheck}, the tDMRG and GHD results for the
spin–spin correlation function at \(\Delta = 0.5\) agree very well over the
bulk of the light cone, Small discrepancies remain at the edges because of the front scaling. So the GHD formula provides an accurate description of the spin correlations at Euler scaling.

\begin{figure}[!t]
    \centering 
    \includegraphics[width=0.95\linewidth]{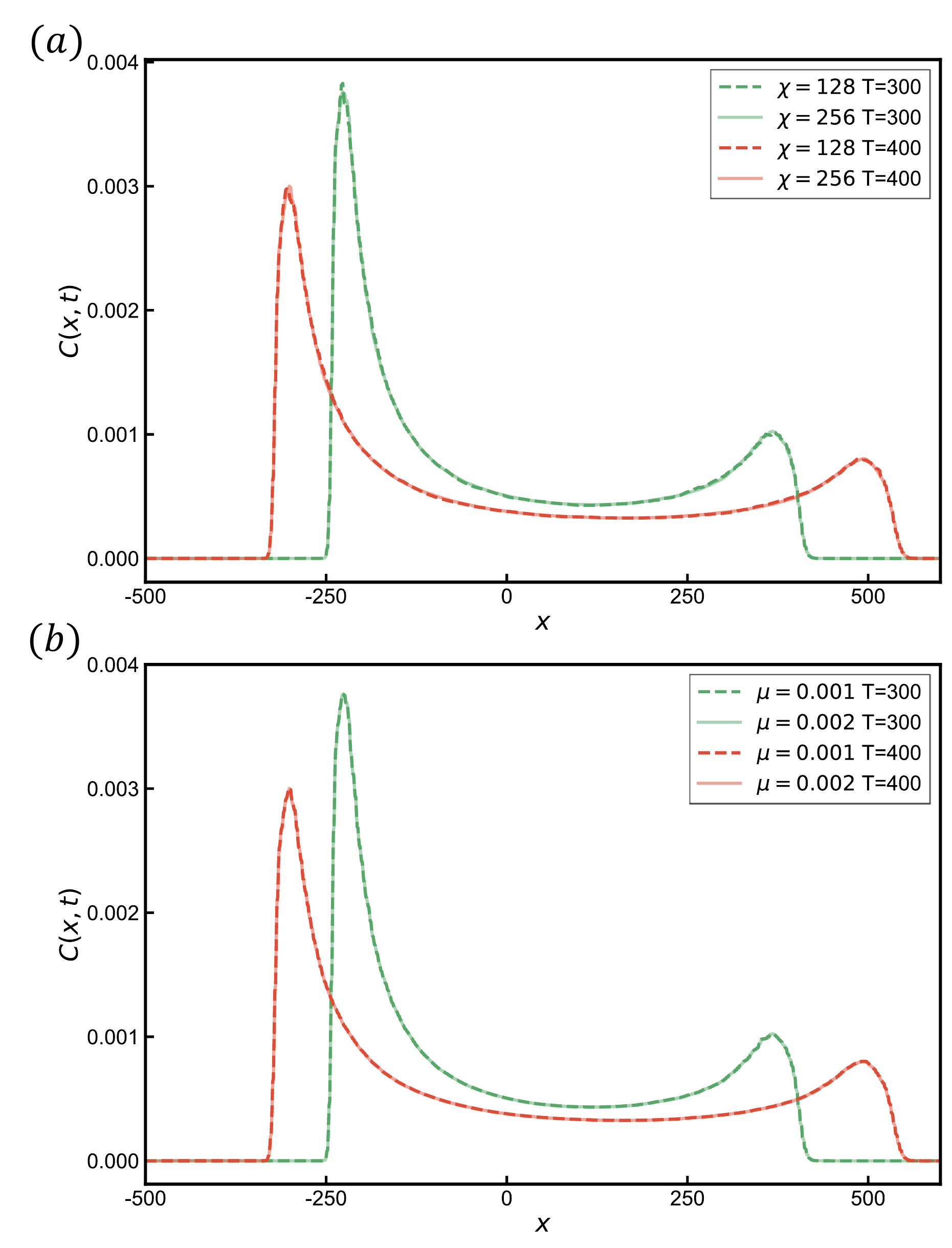}
    \caption{Numerical convergence analysis of the spin-spin correlation function $C(x,t)$ under Fibonacci sequence at the representative parameter point $(\tau_1,\tau_3) = (0.1i,1.1i)$. (a) Comparison between bond dimensions $\chi = 128$ and $\chi = 256$  at times $T=300$ (green) and $T=400$ (red). (b) Comparison between bias parameters
     $\mu=0.001$ and $\mu=0.002$ at the same times. In both cases, the curves exhibit excellent agreement, confirming that the numerical results converge with respect to both $\chi$ and $\mu$.}
    \label{fig:Check}
\end{figure}

\section{Analysis of numerical convergence}
\label{appendix:numerical-check} 
This appendix provides a detailed analysis of the numerical accuracy of the results presented throughout the main text. To ensure the reliability of our conclusions, we have performed comprehensive checks, including convergence tests with respect to bond dimension $\chi$ and bias parameter $\mu$ in the initial state. These analyses confirm that numerical uncertainties are well controlled and do not affect the quantitative features of the results. Here, we present a representative example to demonstrate the robustness of our results. Specifically, we compare the correlation function $C(x,t)$
under Fibonacci sequence, for two values of the bias parameter $\mu=0.001$ and $\mu=0.002$, as well as two bond dimensions $\chi=128$ and $\chi=256$, at a typical parameter point $(\tau_1,\tau_3) = (0.1i,1.1i)$. As shown in Fig.~\ref{fig:Check}, the results exhibit excellent agreement.

\clearpage
%\newpage
\bibliographystyle{apsrev4-2}
\bibliography{reference}

@article{kadanoff1963,
title = {Hydrodynamic equations and correlation functions},
journal = {Annals of Physics},
volume = {24},
pages = {419-469},
year = {1963},
issn = {0003-4916},
doi = {https://doi.org/10.1016/0003-4916(63)90078-2},
url = {https://www.sciencedirect.com/science/article/pii/0003491663900782},
author = {Leo P Kadanoff and Paul C Martin}
}

@book{spohnbook,
  title={Large scale dynamics of interacting particles},
  author={Spohn, Herbert},
  year={2012},
  publisher={Springer Science \& Business Media}
}

@book{forsterbook,
  title={Hydrodynamic fluctuations, broken symmetry, and correlation functions},
  author={Forster, Dieter},
  year={2018},
  publisher={CRC Press}
}

@book{chaikinbook,
  title={Principles of condensed matter physics},
  author={Chaikin, Paul M and Lubensky, Tom C and Witten, Thomas A},
  volume={10},
  year={1995},
  publisher={Cambridge university press Cambridge}
}

@Article{iTensor,
	title={{The ITensor Software Library for Tensor Network Calculations}},
	author={Matthew Fishman and Steven R. White and E. Miles Stoudenmire},
	journal={SciPost Phys. Codebases},
	pages={4},
	year={2022},
	publisher={SciPost},
	doi={10.21468/SciPostPhysCodeb.4},
	url={https://scipost.org/10.21468/SciPostPhysCodeb.4},
}

@article{cmekpz,
  title={Detection of Kardar--Parisi--Zhang hydrodynamics in a quantum Heisenberg spin-1/2 chain},
  author={Scheie, Allen and Sherman, NE and Dupont, M and Nagler, SE and Stone, MB and Granroth, GE and Moore, JE and Tennant, DA},
  journal={Nature Physics},
  volume={17},
  number={6},
  pages={726--730},
  year={2021},
  doi={https://doi.org/10.1038/s41567-021-01191-6},
  URL={https://www.nature.com/articles/s41567-021-01191-6},
  publisher={Nature Publishing Group UK London}
}

@article{coldatomkpz,
author = {David Wei  and Antonio Rubio-Abadal  and Bingtian Ye  and Francisco Machado  and Jack Kemp  and Kritsana Srakaew  and Simon Hollerith  and Jun Rui  and Sarang Gopalakrishnan  and Norman Y. Yao  and Immanuel Bloch  and Johannes Zeiher },
title = {Quantum gas microscopy of Kardar-Parisi-Zhang superdiffusion},
journal = {Science},
volume = {376},
number = {6594},
pages = {716-720},
year = {2022},
doi = {10.1126/science.abk2397},
URL = {https://www.science.org/doi/abs/10.1126/science.abk2397},
eprint = {https://www.science.org/doi/pdf/10.1126/science.abk2397}}

@article{googlekpz,
author = {E. Rosenberg et al.},
title = {Dynamics of magnetization at infinite temperature in a Heisenberg spin chain},
journal = {Science},
volume = {384},
number = {6691},
pages = {48-53},
year = {2024},
doi = {10.1126/science.adi7877},
URL = {https://www.science.org/doi/abs/10.1126/science.adi7877},
eprint = {https://www.science.org/doi/pdf/10.1126/science.adi7877}}

@book{classicalint,
  title={Introduction to classical integrable systems},
  author={Babelon, Olivier and Bernard, Denis and Talon, Michel},
  year={2003},
  publisher={Cambridge University Press}
}

@book{arutyunov2019elements,
  title={Elements of Classical and Quantum Integrable Systems},
  author={Arutyunov, Gleb},
  year={2019},
  publisher={Springer},
  doi = {https://doi.org/10.1007/978-3-030-24198-8}
}

@book{baxterexactly,
  title={Exactly solved models in statistical mechanics},
  author={Baxter, Rodney J},
  year={2016},
  publisher={Elsevier}
}

@book{korepin1997,
  title={Quantum inverse scattering method and correlation functions},
  author={Korepin, Vladimir E and Korepin, Vladimir E and Bogoliubov, NM and Izergin, AG},
  volume={3},
  year={1997},
  publisher={Cambridge university press}
}

@article{CNYang1967,
  title = {Some Exact Results for the Many-Body Problem in one Dimension with Repulsive Delta-Function Interaction},
  author = {Yang, C. N.},
  journal = {Phys. Rev. Lett.},
  volume = {19},
  issue = {23},
  pages = {1312--1315},
  numpages = {0},
  year = {1967},
  month = {Dec},
  publisher = {American Physical Society},
  doi = {10.1103/PhysRevLett.19.1312},
  url = {https://link.aps.org/doi/10.1103/PhysRevLett.19.1312}
}

@article{BAXTER1972,
title = {Partition function of the Eight-Vertex lattice model},
journal = {Annals of Physics},
volume = {70},
number = {1},
pages = {193-228},
year = {1972},
issn = {0003-4916},
doi = {https://doi.org/10.1016/0003-4916(72)90335-1},
url = {https://www.sciencedirect.com/science/article/pii/0003491672903351},
author = {Rodney J Baxter}
}

@article{yangyangtba,
    author = {Yang, C. N. and Yang, C. P.},
    title = {Thermodynamics of a One‐Dimensional System of Bosons with Repulsive Delta‐Function Interaction},
    journal = {Journal of Mathematical Physics},
    volume = {10},
    number = {7},
    pages = {1115-1122},
    year = {1969},
    month = {07},
    issn = {0022-2488},
    doi = {10.1063/1.1664947},
    url = {https://doi.org/10.1063/1.1664947},
}

@article{Ilievski_2016,
doi = {10.1088/1742-5468/2016/06/064008},
url = {https://dx.doi.org/10.1088/1742-5468/2016/06/064008},
year = {2016},
month = {jun},
publisher = {IOP Publishing and SISSA},
volume = {2016},
number = {6},
pages = {064008},
author = {Enej Ilievski and Marko Medenjak and Tomaz Prosen and Lenart Zadnik},
title = {Quasilocal charges in integrable lattice systems},
journal = {Journal of Statistical Mechanics: Theory and Experiment},
}

@article{Essler_2016,
doi = {10.1088/1742-5468/2016/06/064002},
url = {https://dx.doi.org/10.1088/1742-5468/2016/06/064002},
year = {2016},
month = {jun},
publisher = {IOP Publishing and SISSA},
volume = {2016},
number = {6},
pages = {064002},
author = {Fabian H L Essler and Maurizio Fagotti},
title = {Quench dynamics and relaxation in isolated integrable quantum spin chains},
journal = {Journal of Statistical Mechanics: Theory and Experiment}
}

@article{Vidmar_2016,
doi = {10.1088/1742-5468/2016/06/064007},
url = {https://dx.doi.org/10.1088/1742-5468/2016/06/064007},
year = {2016},
month = {jun},
publisher = {IOP Publishing and SISSA},
volume = {2016},
number = {6},
pages = {064007},
author = {Lev Vidmar and Marcos Rigol},
title = {Generalized Gibbs ensemble in integrable lattice models},
journal = {Journal of Statistical Mechanics: Theory and Experiment}
}

@article{prosenRMPtransport,
  title = {Finite-temperature transport in one-dimensional quantum lattice models},
  author = {Bertini, B. and Heidrich-Meisner, F. and Karrasch, C. and Prosen, T. and Steinigeweg, R. and \ifmmode \check{Z}\else \v{Z}\fi{}nidari\ifmmode \check{c}\else \v{c}\fi{}, M.},
  journal = {Rev. Mod. Phys.},
  volume = {93},
  issue = {2},
  pages = {025003},
  numpages = {71},
  year = {2021},
  month = {May},
  publisher = {American Physical Society},
  doi = {10.1103/RevModPhys.93.025003},
  url = {https://link.aps.org/doi/10.1103/RevModPhys.93.025003}
}

@article{prosen17NC,
  title={Spin diffusion from an inhomogeneous quench in an integrable system},
  author={Ljubotina, Marko and {\v{Z}}nidari{\v{c}}, Marko and Prosen, Toma{\v{z}}},
  journal={Nature communications},
  volume={8},
  number={1},
  pages={16117},
  year={2017},
  publisher={Nature Publishing Group UK London},
  doi = {10.1038/ncomms16117},
  url = {https://doi.org/10.1038/ncomms16117}
}

@article{GHD2016doyon,
  title = {Emergent Hydrodynamics in Integrable Quantum Systems Out of Equilibrium},
  author = {Castro-Alvaredo, Olalla A. and Doyon, Benjamin and Yoshimura, Takato},
  journal = {Phys. Rev. X},
  volume = {6},
  issue = {4},
  pages = {041065},
  numpages = {17},
  year = {2016},
  month = {Dec},
  publisher = {American Physical Society},
  doi = {10.1103/PhysRevX.6.041065},
  url = {https://link.aps.org/doi/10.1103/PhysRevX.6.041065}
}

@article{GHDatomchip,
  title = {Generalized Hydrodynamics on an Atom Chip},
  author = {Schemmer, M. and Bouchoule, I. and Doyon, B. and Dubail, J.},
  journal = {Phys. Rev. Lett.},
  volume = {122},
  issue = {9},
  pages = {090601},
  numpages = {7},
  year = {2019},
  month = {Mar},
  publisher = {American Physical Society},
  doi = {10.1103/PhysRevLett.122.090601},
  url = {https://link.aps.org/doi/10.1103/PhysRevLett.122.090601}
}

@article{GHDstinh,
  title = {Generalized Hydrodynamics with Space-Time Inhomogeneous Interactions},
  author = {Bastianello, Alvise and Alba, Vincenzo and Caux, Jean-S\'ebastien},
  journal = {Phys. Rev. Lett.},
  volume = {123},
  issue = {13},
  pages = {130602},
  numpages = {7},
  year = {2019},
  month = {Sep},
  publisher = {American Physical Society},
  doi = {10.1103/PhysRevLett.123.130602},
  url = {https://link.aps.org/doi/10.1103/PhysRevLett.123.130602}
}

@article{GHDperspective,
  title = {Generalized Hydrodynamics: A Perspective},
  author = {Doyon, Benjamin and Gopalakrishnan, Sarang and M\o{}ller, Frederik and Schmiedmayer, J\"org and Vasseur, Romain},
  journal = {Phys. Rev. X},
  volume = {15},
  issue = {1},
  pages = {010501},
  numpages = {28},
  year = {2025},
  month = {Jan},
  publisher = {American Physical Society},
  doi = {10.1103/PhysRevX.15.010501},
  url = {https://link.aps.org/doi/10.1103/PhysRevX.15.010501}
}

@article{SQGGE,
  title = {Squeezed Ensembles and Anomalous Dynamic Roughening in Interacting Integrable Chains},
  author = {Cecile, Guillaume and De Nardis, Jacopo and Ilievski, Enej},
  journal = {Phys. Rev. Lett.},
  volume = {132},
  issue = {13},
  pages = {130401},
  numpages = {6},
  year = {2024},
  month = {Mar},
  publisher = {American Physical Society},
  doi = {10.1103/PhysRevLett.132.130401},
  url = {https://link.aps.org/doi/10.1103/PhysRevLett.132.130401}
}

@article{GHDrevintro,
doi = {10.1088/1742-5468/ac3e6a},
url = {https://dx.doi.org/10.1088/1742-5468/ac3e6a},
year = {2022},
month = {jan},
publisher = {IOP Publishing and SISSA},
volume = {2022},
number = {1},
pages = {014001},
author = {Bastianello, Alvise and Bertini, Bruno and Doyon, Benjamin and Vasseur, Romain},
title = {Introduction to the Special Issue on Emergent Hydrodynamics in Integrable Many-Body Systems},
journal = {Journal of Statistical Mechanics: Theory and Experiment}
}

@article{Alba_2021,
doi = {10.1088/1742-5468/ac257d},
url = {https://dx.doi.org/10.1088/1742-5468/ac257d},
year = {2021},
month = {nov},
publisher = {IOP Publishing and SISSA},
volume = {2021},
number = {11},
pages = {114004},
author = {Alba, Vincenzo and Bertini, Bruno and Fagotti, Maurizio and Piroli, Lorenzo and Ruggiero, Paola},
title = {Generalized-hydrodynamic approach to inhomogeneous quenches: correlations, entanglement and quantum effects},
journal = {Journal of Statistical Mechanics: Theory and Experiment}
}

@article{DeNardis_2022,
doi = {10.1088/1742-5468/ac3658},
url = {https://dx.doi.org/10.1088/1742-5468/ac3658},
year = {2022},
month = {jan},
publisher = {IOP Publishing and SISSA},
volume = {2022},
number = {1},
pages = {014002},
author = {De Nardis, Jacopo and Doyon, Benjamin and Medenjak, Marko and Panfil, Miłosz},
title = {Correlation functions and transport coefficients in generalised hydrodynamics},
journal = {Journal of Statistical Mechanics: Theory and Experiment}
}

@article{Bulchandani_superdiffusion,
doi = {10.1088/1742-5468/ac12c7},
url = {https://dx.doi.org/10.1088/1742-5468/ac12c7},
year = {2021},
month = {aug},
publisher = {IOP Publishing and SISSA},
volume = {2021},
number = {8},
pages = {084001},
author = {Bulchandani, Vir B and Gopalakrishnan, Sarang and Ilievski, Enej},
title = {Superdiffusion in spin chains},
journal = {Journal of Statistical Mechanics: Theory and Experiment}
}

@article{sarangromain_rev2023,
doi = {10.1088/1361-6633/acb36e},
url = {https://dx.doi.org/10.1088/1361-6633/acb36e},
year = {2023},
month = {feb},
publisher = {IOP Publishing},
volume = {86},
number = {3},
pages = {036502},
author = {Gopalakrishnan, Sarang and Vasseur, Romain},
title = {Anomalous transport from hot quasiparticles in interacting spin chains},
journal = {Reports on Progress in Physics}
}

@article{sarangromain_rev2024,
   author = "Gopalakrishnan, Sarang and Vasseur, Romain",
   title = "Superdiffusion from Nonabelian Symmetries in Nearly Integrable Systems", 
   journal= "Annual Review of Condensed Matter Physics",
   year = "2024",
   volume = "15",
   number = "Volume 15, 2024",
   pages = "159-176",
   doi = "https://doi.org/10.1146/annurev-conmatphys-032922-110710",
   url = "https://www.annualreviews.org/content/journals/10.1146/annurev-conmatphys-032922-110710",
   publisher = "Annual Reviews",
   issn = "1947-5462",
   type = "Journal Article",
  }

@book{Vsamaj—book,
  title={Introduction to the statistical physics of integrable many-body systems},
  author={{\v{S}}amaj, Ladislav and Bajnok, Zolt{\'a}n},
  year={2013},
  publisher={Cambridge University Press}
}

@Article{intfloquet2017,
	title={{Integrable Floquet dynamics}},
	author={Vladimir Gritsev and Anatoli Polkovnikov},
	journal={SciPost Phys.},
	volume={2},
	pages={021},
	year={2017},
	publisher={SciPost},
	doi={10.21468/SciPostPhys.2.3.021},
	url={https://scipost.org/10.21468/SciPostPhys.2.3.021},
}

@article{Paletta2025IntegrabilityAC,
  title={Integrability and charge transport in asymmetric quantum-circuit geometries},
  author={Chiara Paletta and Urban Duh and Bal{\'a}zs Pozsgay and Lenart Zadnik},
  journal={Journal of Physics A: Mathematical and Theoretical},
  year={2025},
  url={https://api.semanticscholar.org/CorpusID:276812874}
}

@article{KPZ2019,
   title={Kardar-Parisi-Zhang Physics in the Quantum Heisenberg Magnet},
   volume={122},
   ISSN={1079-7114},
   url={http://dx.doi.org/10.1103/PhysRevLett.122.210602},
   DOI={10.1103/physrevlett.122.210602},
   number={21},
   journal={Physical Review Letters},
   publisher={American Physical Society (APS)},
   author={Ljubotina, Marko and Žnidarič, Marko and Prosen, Tomaž},
   year={2019},
   month=may }

@article{XXZcircuitprosen,
   title={Ballistic Spin Transport in a Periodically Driven Integrable Quantum System},
   volume={122},
   ISSN={1079-7114},
   url={http://dx.doi.org/10.1103/PhysRevLett.122.150605},
   DOI={10.1103/physrevlett.122.150605},
   number={15},
   journal={Physical Review Letters},
   publisher={American Physical Society (APS)},
   author={Ljubotina, Marko and Zadnik, Lenart and Prosen, Tomaž},
   year={2019},
   month=apr }

@article{XXXcircuit,
  title = {Integrable Trotterization: Local Conservation Laws and Boundary Driving},
  author = {Vanicat, Matthieu and Zadnik, Lenart and Prosen, Toma\ifmmode \check{z}\else \v{z}\fi{}},
  journal = {Phys. Rev. Lett.},
  volume = {121},
  issue = {3},
  pages = {030606},
  numpages = {6},
  year = {2018},
  month = {Jul},
  publisher = {American Physical Society},
  doi = {10.1103/PhysRevLett.121.030606},
  url = {https://link.aps.org/doi/10.1103/PhysRevLett.121.030606}
}

@article{ALEINERcircuit,
title = {Bethe Ansatz solutions for certain Periodic Quantum Circuits},
journal = {Annals of Physics},
volume = {433},
pages = {168593},
year = {2021},
issn = {0003-4916},
doi = {https://doi.org/10.1016/j.aop.2021.168593},
url = {https://www.sciencedirect.com/science/article/pii/S0003491621001998},
author = {Igor L. Aleiner}
}

@misc{faddeev1996ABA,
      title={How Algebraic Bethe Ansatz works for integrable model}, 
      author={L. D. Faddeev},
      year={1996},
      eprint={hep-th/9605187},
      archivePrefix={arXiv},
      primaryClass={hep-th},
      url={https://arxiv.org/abs/hep-th/9605187}, 
}

@book{takahashiTBA,
  title={Thermodynamics of one-dimensional solvable models},
  author={Takahashi, Minoru},
  year={1999},
  publisher={Cambridge university press Cambridge}
}

@article{takahashi1972,
    author = {Takahashi, Minoru and Suzuki, Masuo},
    title = {One-Dimensional Anisotropic Heisenberg Model at Finite Temperatures},
    journal = {Progress of Theoretical Physics},
    volume = {48},
    number = {6},
    pages = {2187-2209},
    year = {1972},
    month = {12},
    issn = {0033-068X},
    doi = {10.1143/PTP.48.2187},
    url = {https://doi.org/10.1143/PTP.48.2187},
    eprint = {https://academic.oup.com/ptp/article-pdf/48/6/2187/5255323/48-6-2187.pdf},
}

@article{intoutreview2016,
doi = {10.1088/1742-5468/2016/06/064001},
url = {https://doi.org/10.1088/1742-5468/2016/06/064001},
year = {2016},
month = {jun},
publisher = {IOP Publishing and SISSA},
volume = {2016},
number = {6},
pages = {064001},
author = {Calabrese, Pasquale and Essler, Fabian H L and Mussardo, Giuseppe},
title = {Introduction to ‘Quantum Integrability in Out of Equilibrium Systems’},
journal = {Journal of Statistical Mechanics: Theory and Experiment}
}

@article{xxxsuperdif18ilievskiprosen,
  title = {Superdiffusion in One-Dimensional Quantum Lattice Models},
  author = {Ilievski, Enej and De Nardis, Jacopo and Medenjak, Marko and Prosen, Toma\ifmmode \check{z}\else \v{z}\fi{}},
  journal = {Phys. Rev. Lett.},
  volume = {121},
  issue = {23},
  pages = {230602},
  numpages = {6},
  year = {2018},
  month = {Dec},
  publisher = {American Physical Society},
  doi = {10.1103/PhysRevLett.121.230602},
  url = {https://link.aps.org/doi/10.1103/PhysRevLett.121.230602}
}

@article{gapghd2017,
  title = {Transport in out-of-equilibrium XXZ chains: Nonballistic behavior and correlation functions},
  author = {Piroli, Lorenzo and De Nardis, Jacopo and Collura, Mario and Bertini, Bruno and Fagotti, Maurizio},
  journal = {Phys. Rev. B},
  volume = {96},
  issue = {11},
  pages = {115124},
  numpages = {12},
  year = {2017},
  month = {Sep},
  publisher = {American Physical Society},
  doi = {10.1103/PhysRevB.96.115124},
  url = {https://link.aps.org/doi/10.1103/PhysRevB.96.115124}
}

@Article{ghdnewton,
	title={{Hydrodynamics of the interacting Bose gas in the Quantum Newton Cradle setup}},
	author={Jean-Sébastien Caux and Benjamin Doyon and Jérôme Dubail and Robert Konik and Takato Yoshimura},
	journal={SciPost Phys.},
	volume={6},
	pages={070},
	year={2019},
	publisher={SciPost},
	doi={10.21468/SciPostPhys.6.6.070},
	url={https://scipost.org/10.21468/SciPostPhys.6.6.070},
}

@article{sarangromainkinetic,
  title = {Kinetic Theory of Spin Diffusion and Superdiffusion in $XXZ$ Spin Chains},
  author = {Gopalakrishnan, Sarang and Vasseur, Romain},
  journal = {Phys. Rev. Lett.},
  volume = {122},
  issue = {12},
  pages = {127202},
  numpages = {6},
  year = {2019},
  month = {Mar},
  publisher = {American Physical Society},
  doi = {10.1103/PhysRevLett.122.127202},
  url = {https://link.aps.org/doi/10.1103/PhysRevLett.122.127202}
}

@Article{diffusionGHD,
	title={{Diffusion in generalized hydrodynamics and quasiparticle scattering}},
	author={Jacopo De Nardis and Denis Bernard and Benjamin Doyon},
	journal={SciPost Phys.},
	volume={6},
	pages={049},
	year={2019},
	publisher={SciPost},
	doi={10.21468/SciPostPhys.6.4.049},
	url={https://scipost.org/10.21468/SciPostPhys.6.4.049},
}

@Article{XXZCircuitGHD,
	title={{Generalized hydrodynamics of integrable quantum circuits}},
	author={Friedrich Hübner and Eric Vernier and Lorenzo Piroli},
	journal={SciPost Phys.},
	volume={18},
	pages={135},
	year={2025},
	publisher={SciPost},
	doi={10.21468/SciPostPhys.18.4.135},
	url={https://scipost.org/10.21468/SciPostPhys.18.4.135},
}

@article{XXZCircuitGGE,
  title = {Integrable Digital Quantum Simulation: Generalized Gibbs Ensembles and Trotter Transitions},
  author = {Vernier, Eric and Bertini, Bruno and Giudici, Giuliano and Piroli, Lorenzo},
  journal = {Phys. Rev. Lett.},
  volume = {130},
  issue = {26},
  pages = {260401},
  numpages = {7},
  year = {2023},
  month = {Jun},
  publisher = {American Physical Society},
  doi = {10.1103/PhysRevLett.130.260401},
  url = {https://link.aps.org/doi/10.1103/PhysRevLett.130.260401}
}

@Article{doyonGHDnote,
	title={{Lecture notes on Generalised Hydrodynamics}},
	author={Benjamin Doyon},
	journal={SciPost Phys. Lect. Notes},
	pages={18},
	year={2020},
	publisher={SciPost},
	doi={10.21468/SciPostPhysLectNotes.18},
	url={https://scipost.org/10.21468/SciPostPhysLectNotes.18},
}

@Article{doyonspohnDrude,
	title={{Drude Weight for the Lieb-Liniger Bose Gas}},
	author={Benjamin Doyon and Herbert Spohn},
	journal={SciPost Phys.},
	volume={3},
	pages={039},
	year={2017},
	publisher={SciPost},
	doi={10.21468/SciPostPhys.3.6.039},
	url={https://scipost.org/10.21468/SciPostPhys.3.6.039},
}

@article{DenardisIlievskiDrude,
  title = {Ballistic transport in the one-dimensional Hubbard model: The hydrodynamic approach},
  author = {Ilievski, Enej and De Nardis, Jacopo},
  journal = {Phys. Rev. B},
  volume = {96},
  issue = {8},
  pages = {081118},
  numpages = {6},
  year = {2017},
  month = {Aug},
  publisher = {American Physical Society},
  doi = {10.1103/PhysRevB.96.081118},
  url = {https://link.aps.org/doi/10.1103/PhysRevB.96.081118}
}

@article{GHD2016prl,
  title = {Transport in Out-of-Equilibrium $XXZ$ Chains: Exact Profiles of Charges and Currents},
  author = {Bertini, Bruno and Collura, Mario and De Nardis, Jacopo and Fagotti, Maurizio},
  journal = {Phys. Rev. Lett.},
  volume = {117},
  issue = {20},
  pages = {207201},
  numpages = {8},
  year = {2016},
  month = {Nov},
  publisher = {American Physical Society},
  doi = {10.1103/PhysRevLett.117.207201},
  url = {https://link.aps.org/doi/10.1103/PhysRevLett.117.207201}
}

@article{EsslerGHDnote,
title = {A short introduction to Generalized Hydrodynamics},
journal = {Physica A: Statistical Mechanics and its Applications},
volume = {631},
pages = {127572},
year = {2023},
note = {Lecture Notes of the 15th International Summer School of Fundamental Problems in Statistical Physics},
issn = {0378-4371},
doi = {https://doi.org/10.1016/j.physa.2022.127572},
url = {https://www.sciencedirect.com/science/article/pii/S0378437122003971},
author = {Fabian H.L. Essler}
}

@article{integrablecircuitratchets,
  title = {Quantum Many-Body Spin Ratchets},
  author = {Zadnik, Lenart and Ljubotina, Marko and Krajnik, \ifmmode \check{Z}\else \v{Z}\fi{}iga and Ilievski, Enej and Prosen, Toma\ifmmode \check{z}\else \v{z}\fi{}},
  journal = {PRX Quantum},
  volume = {5},
  issue = {3},
  pages = {030356},
  numpages = {26},
  year = {2024},
  month = {Sep},
  publisher = {American Physical Society},
  doi = {10.1103/PRXQuantum.5.030356},
  url = {https://link.aps.org/doi/10.1103/PRXQuantum.5.030356}
}

@Article{floquetbaxter,
	title={{The Floquet Baxterisation}},
	author={Yuan Miao and Vladimir Gritsev and Denis V. Kurlov},
	journal={SciPost Phys.},
	volume={16},
	pages={078},
	year={2024},
	publisher={SciPost},
	doi={10.21468/SciPostPhys.16.3.078},
	url={https://scipost.org/10.21468/SciPostPhys.16.3.078},
}

@article{randominteEssler,
  title = {Integrable spin chains with random interactions},
  author = {Essler, Fabian H. L. and van den Berg, Rianne and Gritsev, Vladimir},
  journal = {Phys. Rev. B},
  volume = {98},
  issue = {2},
  pages = {024203},
  numpages = {10},
  year = {2018},
  month = {Jul},
  publisher = {American Physical Society},
  doi = {10.1103/PhysRevB.98.024203},
  url = {https://link.aps.org/doi/10.1103/PhysRevB.98.024203}
}

@article{randomintsr,
  title = {Generalized hydrodynamics, quasiparticle diffusion, and anomalous local relaxation in random integrable spin chains},
  author = {Agrawal, Utkarsh and Gopalakrishnan, Sarang and Vasseur, Romain},
  journal = {Phys. Rev. B},
  volume = {99},
  issue = {17},
  pages = {174203},
  numpages = {12},
  year = {2019},
  month = {May},
  publisher = {American Physical Society},
  doi = {10.1103/PhysRevB.99.174203},
  url = {https://link.aps.org/doi/10.1103/PhysRevB.99.174203}
}

@Article{randintegrablekrajnik2025,
	title={{Integrable fishnet circuits and Brownian solitons}},
	author={Žiga Krajnik and Enej Ilievski and Tomaz Prosen and Benjamin J. A. Héry and Vincent Pasquier},
	journal={SciPost Phys.},
	volume={19},
	pages={027},
	year={2025},
	publisher={SciPost},
	doi={10.21468/SciPostPhys.19.1.027},
	url={https://scipost.org/10.21468/SciPostPhys.19.1.027},
}

@article{znidaricinthaar,
  title = {Integrability is generic in homogeneous U(1)-invariant nearest-neighbor qubit circuits},
  author = {\ifmmode \check{Z}\else \v{Z}\fi{}nidari\ifmmode \check{c}\else \v{c}\fi{}, Marko and Duh, Urban and Zadnik, Lenart},
  journal = {Phys. Rev. B},
  volume = {112},
  issue = {2},
  pages = {L020302},
  numpages = {6},
  year = {2025},
  month = {Jul},
  publisher = {American Physical Society},
  doi = {10.1103/tqy8-ynpd},
  url = {https://link.aps.org/doi/10.1103/tqy8-ynpd}
}

@article{vznidarivcnonlocalsu2,
  title={Inhomogeneous SU (2) symmetries in homogeneous integrable U (1) circuits and transport},
  author={{\v{Z}}nidari{\v{c}}, Marko},
  journal={Nature communications},
  volume={16},
  number={1},
  pages={4336},
  year={2025},
  publisher={Nature Publishing Group UK London},
  doi={ https://doi.org/10.1038/s41467-025-59705-2}
}

@article{fullcountingTEBD,
  title = {Efficient Computation of Cumulant Evolution and Full Counting Statistics: Application to Infinite Temperature Quantum Spin Chains},
  author = {Valli, Angelo and Moca, C\ifmmode \u{a}\else \u{a}\fi{}t\ifmmode \u{a}\else \u{a}\fi{}lin Pa\ifmmode \mbox{\c{s}}\else \c{s}\fi{}cu and Werner, Mikl\'os Antal and Kormos, M\'arton and Krajnik, \ifmmode \check{Z}\else \v{Z}\fi{}iga and Prosen, Toma\ifmmode \check{z}\else \v{z}\fi{} and Zar\'and, Gergely},
  journal = {Phys. Rev. Lett.},
  volume = {135},
  issue = {10},
  pages = {100401},
  numpages = {7},
  year = {2025},
  month = {Sep},
  publisher = {American Physical Society},
  doi = {10.1103/f3c4-n21z},
  url = {https://link.aps.org/doi/10.1103/f3c4-n21z}
}

@article{boundDiffc,
  title = {Lower Bounding Diffusion Constant by the Curvature of Drude Weight},
  author = {Medenjak, Marko and Karrasch, Christoph and Prosen, Toma\ifmmode \check{z}\else \v{z}\fi{}},
  journal = {Phys. Rev. Lett.},
  volume = {119},
  issue = {8},
  pages = {080602},
  numpages = {5},
  year = {2017},
  month = {Aug},
  publisher = {American Physical Society},
  doi = {10.1103/PhysRevLett.119.080602},
  url = {https://link.aps.org/doi/10.1103/PhysRevLett.119.080602}
}

@article{KPZ1986,
  title = {Dynamic Scaling of Growing Interfaces},
  author = {Kardar, Mehran and Parisi, Giorgio and Zhang, Yi-Cheng},
  journal = {Phys. Rev. Lett.},
  volume = {56},
  issue = {9},
  pages = {889--892},
  numpages = {0},
  year = {1986},
  month = {Mar},
  publisher = {American Physical Society},
  doi = {10.1103/PhysRevLett.56.889},
  url = {https://link.aps.org/doi/10.1103/PhysRevLett.56.889}
}

@article{znidaricxxx2011,
  title = {Spin Transport in a One-Dimensional Anisotropic Heisenberg Model},
  author = {\ifmmode \check{Z}\else \v{Z}\fi{}nidari\ifmmode \check{c}\else \v{c}\fi{}, Marko},
  journal = {Phys. Rev. Lett.},
  volume = {106},
  issue = {22},
  pages = {220601},
  numpages = {4},
  year = {2011},
  month = {May},
  publisher = {American Physical Society},
  doi = {10.1103/PhysRevLett.106.220601},
  url = {https://link.aps.org/doi/10.1103/PhysRevLett.106.220601}
}

@article{xxzqlc1,
  title = {Open $XXZ$ Spin Chain: Nonequilibrium Steady State and a Strict Bound on Ballistic Transport},
  author = {Prosen, Toma\ifmmode \check{z}\else \v{z}\fi{}},
  journal = {Phys. Rev. Lett.},
  volume = {106},
  issue = {21},
  pages = {217206},
  numpages = {4},
  year = {2011},
  month = {May},
  publisher = {American Physical Society},
  doi = {10.1103/PhysRevLett.106.217206},
  url = {https://link.aps.org/doi/10.1103/PhysRevLett.106.217206}
}

@article{xxzqlc2,
  title = {Families of Quasilocal Conservation Laws and Quantum Spin Transport},
  author = {Prosen, Toma\ifmmode \check{z}\else \v{z}\fi{} and Ilievski, Enej},
  journal = {Phys. Rev. Lett.},
  volume = {111},
  issue = {5},
  pages = {057203},
  numpages = {5},
  year = {2013},
  month = {Aug},
  publisher = {American Physical Society},
  doi = {10.1103/PhysRevLett.111.057203},
  url = {https://link.aps.org/doi/10.1103/PhysRevLett.111.057203}
}

@article{xxzqlc3,
  title = {Exact Nonequilibrium Steady State of a Strongly Driven Open $XXZ$ Chain},
  author = {Prosen, Toma\ifmmode \check{z}\else \v{z}\fi{}},
  journal = {Phys. Rev. Lett.},
  volume = {107},
  issue = {13},
  pages = {137201},
  numpages = {5},
  year = {2011},
  month = {Sep},
  publisher = {American Physical Society},
  doi = {10.1103/PhysRevLett.107.137201},
  url = {https://link.aps.org/doi/10.1103/PhysRevLett.107.137201}
}

@article{affleckxxz1,
  title = {Diffusion and Ballistic Transport in One-Dimensional Quantum Systems},
  author = {Sirker, J. and Pereira, R. G. and Affleck, I.},
  journal = {Phys. Rev. Lett.},
  volume = {103},
  issue = {21},
  pages = {216602},
  numpages = {4},
  year = {2009},
  month = {Nov},
  publisher = {American Physical Society},
  doi = {10.1103/PhysRevLett.103.216602},
  url = {https://link.aps.org/doi/10.1103/PhysRevLett.103.216602}
}

@article{affleckxxz2,
  title = {Conservation laws, integrability, and transport in one-dimensional quantum systems},
  author = {Sirker, J. and Pereira, R. G. and Affleck, I.},
  journal = {Phys. Rev. B},
  volume = {83},
  issue = {3},
  pages = {035115},
  numpages = {20},
  year = {2011},
  month = {Jan},
  publisher = {American Physical Society},
  doi = {10.1103/PhysRevB.83.035115},
  url = {https://link.aps.org/doi/10.1103/PhysRevB.83.035115}
}

@article{sachdevdiffusion,
  title = {Spin dynamics and transport in gapped one-dimensional Heisenberg antiferromagnets at nonzero temperatures},
  author = {Damle, Kedar and Sachdev, Subir},
  journal = {Phys. Rev. B},
  volume = {57},
  issue = {14},
  pages = {8307--8339},
  numpages = {0},
  year = {1998},
  month = {Apr},
  publisher = {American Physical Society},
  doi = {10.1103/PhysRevB.57.8307},
  url = {https://link.aps.org/doi/10.1103/PhysRevB.57.8307}
}

@article{super-superdiffusion,
  title = {Superuniversality of Superdiffusion},
  author = {Ilievski, Enej and De Nardis, Jacopo and Gopalakrishnan, Sarang and Vasseur, Romain and Ware, Brayden},
  journal = {Phys. Rev. X},
  volume = {11},
  issue = {3},
  pages = {031023},
  numpages = {31},
  year = {2021},
  month = {Jul},
  publisher = {American Physical Society},
  doi = {10.1103/PhysRevX.11.031023},
  url = {https://link.aps.org/doi/10.1103/PhysRevX.11.031023}
}

@article{KPZdmt,
  title = {Universal Kardar-Parisi-Zhang Dynamics in Integrable Quantum Systems},
  author = {Ye, Bingtian and Machado, Francisco and Kemp, Jack and Hutson, Ross B. and Yao, Norman Y.},
  journal = {Phys. Rev. Lett.},
  volume = {129},
  issue = {23},
  pages = {230602},
  numpages = {7},
  year = {2022},
  month = {Nov},
  publisher = {American Physical Society},
  doi = {10.1103/PhysRevLett.129.230602},
  url = {https://link.aps.org/doi/10.1103/PhysRevLett.129.230602}
}

@article{coupleKPZ,
  title = {Nonlinear Fluctuating Hydrodynamics for Kardar-Parisi-Zhang Scaling in Isotropic Spin Chains},
  author = {De Nardis, Jacopo and Gopalakrishnan, Sarang and Vasseur, Romain},
  journal = {Phys. Rev. Lett.},
  volume = {131},
  issue = {19},
  pages = {197102},
  numpages = {6},
  year = {2023},
  month = {Nov},
  publisher = {American Physical Society},
  doi = {10.1103/PhysRevLett.131.197102},
  url = {https://link.aps.org/doi/10.1103/PhysRevLett.131.197102}
}

@article{KPZfullcheck2025,
  title = {Partial Yet Definite Emergence of the Kardar-Parisi-Zhang Class in Isotropic Spin Chains},
  author = {Takeuchi, Kazumasa A. and Takasan, Kazuaki and Busani, Ofer and Ferrari, Patrik L. and Vasseur, Romain and De Nardis, Jacopo},
  journal = {Phys. Rev. Lett.},
  volume = {134},
  issue = {9},
  pages = {097104},
  numpages = {8},
  year = {2025},
  month = {Mar},
  publisher = {American Physical Society},
  doi = {10.1103/PhysRevLett.134.097104},
  url = {https://link.aps.org/doi/10.1103/PhysRevLett.134.097104}
}

@article{KPZgaugemode,
  title = {Kardar-Parisi-Zhang universality from soft gauge modes},
  author = {Bulchandani, Vir B.},
  journal = {Phys. Rev. B},
  volume = {101},
  issue = {4},
  pages = {041411},
  numpages = {6},
  year = {2020},
  month = {Jan},
  publisher = {American Physical Society},
  doi = {10.1103/PhysRevB.101.041411},
  url = {https://link.aps.org/doi/10.1103/PhysRevB.101.041411}
}

@article{XXXmagicformula,
  title = {Anomalous Spin Diffusion in One-Dimensional Antiferromagnets},
  author = {De Nardis, Jacopo and Medenjak, Marko and Karrasch, Christoph and Ilievski, Enej},
  journal = {Phys. Rev. Lett.},
  volume = {123},
  issue = {18},
  pages = {186601},
  numpages = {6},
  year = {2019},
  month = {Oct},
  publisher = {American Physical Society},
  doi = {10.1103/PhysRevLett.123.186601},
  url = {https://link.aps.org/doi/10.1103/PhysRevLett.123.186601}
}

@article{KPZsoliton,
  title = {Superdiffusion from Emergent Classical Solitons in Quantum Spin Chains},
  author = {De Nardis, Jacopo and Gopalakrishnan, Sarang and Ilievski, Enej and Vasseur, Romain},
  journal = {Phys. Rev. Lett.},
  volume = {125},
  issue = {7},
  pages = {070601},
  numpages = {7},
  year = {2020},
  month = {Aug},
  publisher = {American Physical Society},
  doi = {10.1103/PhysRevLett.125.070601},
  url = {https://link.aps.org/doi/10.1103/PhysRevLett.125.070601}
}

@article{otocint,
  title = {Hydrodynamics of operator spreading and quasiparticle diffusion in interacting integrable systems},
  author = {Gopalakrishnan, Sarang and Huse, David A. and Khemani, Vedika and Vasseur, Romain},
  journal = {Phys. Rev. B},
  volume = {98},
  issue = {22},
  pages = {220303},
  numpages = {6},
  year = {2018},
  month = {Dec},
  publisher = {American Physical Society},
  doi = {10.1103/PhysRevB.98.220303},
  url = {https://link.aps.org/doi/10.1103/PhysRevB.98.220303}
}

@article{inteautomata2017,
  title = {Diffusion in Deterministic Interacting Lattice Systems},
  author = {Medenjak, Marko and Klobas, Katja and Prosen, Toma\ifmmode \check{z}\else \v{z}\fi{}},
  journal = {Phys. Rev. Lett.},
  volume = {119},
  issue = {11},
  pages = {110603},
  numpages = {5},
  year = {2017},
  month = {Sep},
  publisher = {American Physical Society},
  doi = {10.1103/PhysRevLett.119.110603},
  url = {https://link.aps.org/doi/10.1103/PhysRevLett.119.110603}
}

@misc{yoshimura2506,
      title={Hydrodynamic fluctuations of stochastic charged cellular automata}, 
      author={Takato Yoshimura and Žiga Krajnik},
      year={2025},
      eprint={2506.05247},
      archivePrefix={arXiv},
      primaryClass={cond-mat.stat-mech},
      url={https://arxiv.org/abs/2506.05247}, 
}

@article{PhysRevE.111.024141,
  title = {Anomalous current fluctuations from Euler hydrodynamics},
  author = {Yoshimura, Takato and Krajnik, \ifmmode \check{Z}\else \v{Z}\fi{}iga},
  journal = {Phys. Rev. E},
  volume = {111},
  issue = {2},
  pages = {024141},
  numpages = {9},
  year = {2025},
  month = {Feb},
  publisher = {American Physical Society},
  doi = {10.1103/PhysRevE.111.024141},
  url = {https://link.aps.org/doi/10.1103/PhysRevE.111.024141}
}

@article{Klobas_2018,
doi = {10.1088/1742-5468/aae853},
url = {https://doi.org/10.1088/1742-5468/aae853},
year = {2018},
month = {dec},
publisher = {IOP Publishing and SISSA},
volume = {2018},
number = {12},
pages = {123202},
author = {Klobas, Katja and Medenjak, Marko and Prosen, Tomaz},
title = {Exactly solvable deterministic lattice model of crossover between ballistic and diffusive transport},
journal = {Journal of Statistical Mechanics: Theory and Experiment}
}

@article{Krajnik_2025,
doi = {10.1088/1742-5468/add513},
url = {https://doi.org/10.1088/1742-5468/add513},
year = {2025},
month = {may},
publisher = {IOP Publishing},
volume = {2025},
number = {5},
pages = {053209},
author = {Krajnik, Ziga and Klobas, Katja and Bertini, Bruno and Prosen, Tomaz},
title = {Fluctuations of stochastic charged cellular automata},
journal = {Journal of Statistical Mechanics: Theory and Experiment}
}

@article{Pereira_2014,
doi = {10.1088/1742-5468/2014/09/P09037},
url = {https://doi.org/10.1088/1742-5468/2014/09/P09037},
year = {2014},
month = {sep},
publisher = {IOP Publishing and SISSA},
volume = {2014},
number = {9},
pages = {P09037},
author = {Pereira, R G and Pasquier, V and Sirker, J and Affleck, I},
title = {Exactly conserved quasilocal operators for the XXZ spin chain},
journal = {Journal of Statistical Mechanics: Theory and Experiment}
}

@article{PROSENnpb,
title = {Quasilocal conservation laws in XXZ spin-1/2 chains: Open, periodic and twisted boundary conditions},
journal = {Nuclear Physics B},
volume = {886},
pages = {1177-1198},
year = {2014},
issn = {0550-3213},
doi = {https://doi.org/10.1016/j.nuclphysb.2014.07.024},
url = {https://www.sciencedirect.com/science/article/pii/S0550321314002430},
author = {Tomaz Prosen}
}

@Article{ghdentanglement,
	title={{Entanglement evolution and generalised hydrodynamics: interacting integrable systems}},
	author={Vincenzo Alba and Bruno Bertini and Maurizio Fagotti},
	journal={SciPost Phys.},
	volume={7},
	pages={005},
	year={2019},
	publisher={SciPost},
	doi={10.21468/SciPostPhys.7.1.005},
	url={https://scipost.org/10.21468/SciPostPhys.7.1.005},
}

@article{qppentanglement,
author = {Vincenzo Alba  and Pasquale Calabrese },
title = {Entanglement and thermodynamics after a quantum quench in integrable systems},
journal = {Proceedings of the National Academy of Sciences},
volume = {114},
number = {30},
pages = {7947-7951},
year = {2017},
doi = {10.1073/pnas.1703516114},
URL = {https://www.pnas.org/doi/abs/10.1073/pnas.1703516114},
eprint = {https://www.pnas.org/doi/pdf/10.1073/pnas.1703516114}}

@article{ions1,
author = {B. P. Lanyon  and C. Hempel  and D. Nigg  and M. Müller  and R. Gerritsma  and F. Zähringer  and P. Schindler  and J. T. Barreiro  and M. Rambach  and G. Kirchmair  and M. Hennrich  and P. Zoller  and R. Blatt  and C. F. Roos },
title = {Universal Digital Quantum Simulation with Trapped Ions},
journal = {Science},
volume = {334},
number = {6052},
pages = {57-61},
year = {2011},
doi = {10.1126/science.1208001},
URL = {https://www.science.org/doi/abs/10.1126/science.1208001},
eprint = {https://www.science.org/doi/pdf/10.1126/science.1208001}}

@article{ions3,
  title = {Programmable quantum simulations of spin systems with trapped ions},
  author = {Monroe, C. and Campbell, W. C. and Duan, L.-M. and Gong, Z.-X. and Gorshkov, A. V. and Hess, P. W. and Islam, R. and Kim, K. and Linke, N. M. and Pagano, G. and Richerme, P. and Senko, C. and Yao, N. Y.},
  journal = {Rev. Mod. Phys.},
  volume = {93},
  issue = {2},
  pages = {025001},
  numpages = {57},
  year = {2021},
  month = {Apr},
  publisher = {American Physical Society},
  doi = {10.1103/RevModPhys.93.025001},
  url = {https://link.aps.org/doi/10.1103/RevModPhys.93.025001}
}

@article{ions4,
author = {M. K. Joshi  and F. Kranzl  and A. Schuckert  and I. Lovas  and C. Maier  and R. Blatt  and M. Knap  and C. F. Roos },
title = {Observing emergent hydrodynamics in a long-range quantum magnet},
journal = {Science},
volume = {376},
number = {6594},
pages = {720-724},
year = {2022},
doi = {10.1126/science.abk2400},
URL = {https://www.science.org/doi/abs/10.1126/science.abk2400},
eprint = {https://www.science.org/doi/pdf/10.1126/science.abk2400}}

@article{endres2016,
  title={Atom-by-atom assembly of defect-free one-dimensional cold atom arrays},
  author={Endres, Manuel and Bernien, Hannes and Keesling, Alexander and Levine, Harry and Anschuetz, Eric R and Krajenbrink, Alexandre and Senko, Crystal and Vuletic, Vladan and Greiner, Markus and Lukin, Mikhail D},
  journal={Science},
  volume={354},
  number={6315},
  pages={1024--1027},
  year={2016},
  doi={10.1126/science.aah3752},
  publisher={American Association for the Advancement of Science}
}

@article{Rydberg,
	Author = {Bernien, Hannes and Schwartz, Sylvain and Keesling, Alexander and Levine, Harry and Omran, Ahmed and Pichler, Hannes and Choi, Soonwon and Zibrov, Alexander S. and Endres, Manuel and Greiner, Markus and Vuleti{\'c}, Vladan and Lukin, Mikhail D.},
	Da = {2017/11/01},
	Doi = {10.1038/nature24622},
	Id = {Bernien2017},
	Isbn = {1476-4687},
	Journal = {Nature},
	Number = {7682},
	Pages = {579--584},
	Title = {Probing many-body dynamics on a 51-atom quantum simulator},
	Ty = {JOUR},
	Url = {https://doi.org/10.1038/nature24622},
	Volume = {551},
	Year = {2017}
}

@article{LRSYKtransport,
  title = {Hydrodynamic Modes and Operator Spreading in a Long-Range Center-of-Mass-Conserving Brownian Sachdev-Ye-Kitaev Model},
  author = {Cheng, Bai-Lin and Jian, Shao-Kai and Yang, Zhi-Cheng},
  journal = {Phys. Rev. Lett.},
  volume = {134},
  issue = {15},
  pages = {156301},
  numpages = {7},
  year = {2025},
  month = {Apr},
  publisher = {American Physical Society},
  doi = {10.1103/PhysRevLett.134.156301},
  url = {https://link.aps.org/doi/10.1103/PhysRevLett.134.156301}
}

@article{constraintsubdiff,
  title = {Subdiffusion and Many-Body Quantum Chaos with Kinetic Constraints},
  author = {Singh, Hansveer and Ware, Brayden A. and Vasseur, Romain and Friedman, Aaron J.},
  journal = {Phys. Rev. Lett.},
  volume = {127},
  issue = {23},
  pages = {230602},
  numpages = {6},
  year = {2021},
  month = {Dec},
  publisher = {American Physical Society},
  doi = {10.1103/PhysRevLett.127.230602},
  url = {https://link.aps.org/doi/10.1103/PhysRevLett.127.230602}
}

@article{longrangeconstraints,
  title = {Hydrodynamics in long-range interacting systems with center-of-mass conservation},
  author = {Morningstar, Alan and O'Dea, Nicholas and Richter, Jonas},
  journal = {Phys. Rev. B},
  volume = {108},
  issue = {2},
  pages = {L020304},
  numpages = {8},
  year = {2023},
  month = {Jul},
  publisher = {American Physical Society},
  doi = {10.1103/PhysRevB.108.L020304},
  url = {https://link.aps.org/doi/10.1103/PhysRevB.108.L020304}
}

@article{fractontransport,
  title = {Fracton hydrodynamics},
  author = {Gromov, Andrey and Lucas, Andrew and Nandkishore, Rahul M.},
  journal = {Phys. Rev. Res.},
  volume = {2},
  issue = {3},
  pages = {033124},
  numpages = {11},
  year = {2020},
  month = {Jul},
  publisher = {American Physical Society},
  doi = {10.1103/PhysRevResearch.2.033124},
  url = {https://link.aps.org/doi/10.1103/PhysRevResearch.2.033124}
}

@article{PhysRevE.103.022142,
  title = {Multipole conservation laws and subdiffusion in any dimension},
  author = {Iaconis, Jason and Lucas, Andrew and Nandkishore, Rahul},
  journal = {Phys. Rev. E},
  volume = {103},
  issue = {2},
  pages = {022142},
  numpages = {7},
  year = {2021},
  month = {Feb},
  publisher = {American Physical Society},
  doi = {10.1103/PhysRevE.103.022142},
  url = {https://link.aps.org/doi/10.1103/PhysRevE.103.022142}
}

@article{PhysRevB.100.214301,
  title = {Anomalous subdiffusion from subsystem symmetries},
  author = {Iaconis, Jason and Vijay, Sagar and Nandkishore, Rahul},
  journal = {Phys. Rev. B},
  volume = {100},
  issue = {21},
  pages = {214301},
  numpages = {16},
  year = {2019},
  month = {Dec},
  publisher = {American Physical Society},
  doi = {10.1103/PhysRevB.100.214301},
  url = {https://link.aps.org/doi/10.1103/PhysRevB.100.214301}
}

@article{disordertransport4,
  title = {Superdiffusive transport on lattices with nodal impurities},
  author = {Wang, Yu-Peng and Ren, Jie and Fang, Chen},
  journal = {Phys. Rev. B},
  volume = {110},
  issue = {14},
  pages = {144201},
  numpages = {6},
  year = {2024},
  month = {Oct},
  publisher = {American Physical Society},
  doi = {10.1103/PhysRevB.110.144201},
  url = {https://link.aps.org/doi/10.1103/PhysRevB.110.144201}
}

@article{disordertransport1,
  title = {Absence of localization in a random-dimer model},
  author = {Dunlap, David H. and Wu, H-L. and Phillips, Philip W.},
  journal = {Phys. Rev. Lett.},
  volume = {65},
  issue = {1},
  pages = {88--91},
  numpages = {0},
  year = {1990},
  month = {Jul},
  publisher = {American Physical Society},
  doi = {10.1103/PhysRevLett.65.88},
  url = {https://link.aps.org/doi/10.1103/PhysRevLett.65.88}
}

@article{disordertransport2,
title = {Quantum intermittency in almost-periodic lattice systems derived from their spectral properties},
journal = {Physica D: Nonlinear Phenomena},
volume = {103},
number = {1},
pages = {576-589},
year = {1997},
note = {Lattice Dynamics},
issn = {0167-2789},
doi = {https://doi.org/10.1016/S0167-2789(96)00287-4},
url = {https://www.sciencedirect.com/science/article/pii/S0167278996002874},
author = {Giorgio Mantica}
}

@article{disordertransport3,
  title={Dynamics of an Electron in Quasiperiodic Systems. I. Fibonacci Model},
  author={Hisashi Hiramoto and Shuji Abe},
  journal={Journal of the Physical Society of Japan},
  volume={57},
  number={1},
  pages={230-240},
  year={1988},
  doi={10.1143/JPSJ.57.230}
}

@article{nahumprx7,
  title = {Quantum Entanglement Growth under Random Unitary Dynamics},
  author = {Nahum, Adam and Ruhman, Jonathan and Vijay, Sagar and Haah, Jeongwan},
  journal = {Phys. Rev. X},
  volume = {7},
  issue = {3},
  pages = {031016},
  numpages = {30},
  year = {2017},
  month = {Jul},
  publisher = {American Physical Society},
  doi = {10.1103/PhysRevX.7.031016},
  url = {https://link.aps.org/doi/10.1103/PhysRevX.7.031016}
}

@article{nahumprx8,
  title = {Operator Spreading in Random Unitary Circuits},
  author = {Nahum, Adam and Vijay, Sagar and Haah, Jeongwan},
  journal = {Phys. Rev. X},
  volume = {8},
  issue = {2},
  pages = {021014},
  numpages = {30},
  year = {2018},
  month = {Apr},
  publisher = {American Physical Society},
  doi = {10.1103/PhysRevX.8.021014},
  url = {https://link.aps.org/doi/10.1103/PhysRevX.8.021014}
}

@article{vonrcnocharge,
  title = {Operator Hydrodynamics, OTOCs, and Entanglement Growth in Systems without Conservation Laws},
  author = {von Keyserlingk, C. W. and Rakovszky, Tibor and Pollmann, Frank and Sondhi, S. L.},
  journal = {Phys. Rev. X},
  volume = {8},
  issue = {2},
  pages = {021013},
  numpages = {19},
  year = {2018},
  month = {Apr},
  publisher = {American Physical Society},
  doi = {10.1103/PhysRevX.8.021013},
  url = {https://link.aps.org/doi/10.1103/PhysRevX.8.021013}
}

@article{vonrcwithcharge,
  title = {Diffusive Hydrodynamics of Out-of-Time-Ordered Correlators with Charge Conservation},
  author = {Rakovszky, Tibor and Pollmann, Frank and von Keyserlingk, C. W.},
  journal = {Phys. Rev. X},
  volume = {8},
  issue = {3},
  pages = {031058},
  numpages = {28},
  year = {2018},
  month = {Sep},
  publisher = {American Physical Society},
  doi = {10.1103/PhysRevX.8.031058},
  url = {https://link.aps.org/doi/10.1103/PhysRevX.8.031058}
}

@article{circuitreivewnahum,
   author = "Fisher, Matthew P.A. and Khemani, Vedika and Nahum, Adam and Vijay, Sagar",
   title = "Random Quantum Circuits", 
   journal= "Annual Review of Condensed Matter Physics",
   year = "2023",
   volume = "14",
   number = "Volume 14, 2023",
   pages = "335-379",
   doi = "https://doi.org/10.1146/annurev-conmatphys-031720-030658",
   publisher = "Annual Reviews",
   issn = "1947-5462",
   type = "Journal Article",
  }

@article{else,
  title = {Long-Lived Interacting Phases of Matter Protected by Multiple Time-Translation Symmetries in Quasiperiodically Driven Systems},
  author = {Else, Dominic V. and Ho, Wen Wei and Dumitrescu, Philipp T.},
  journal = {Phys. Rev. X},
  volume = {10},
  issue = {2},
  pages = {021032},
  numpages = {40},
  year = {2020},
  month = {May},
  publisher = {American Physical Society},
  doi = {10.1103/PhysRevX.10.021032},
  url = {https://link.aps.org/doi/10.1103/PhysRevX.10.021032}
}

@article{vasseurqpdr,
  title = {Logarithmically Slow Relaxation in Quasiperiodically Driven Random Spin Chains},
  author = {Dumitrescu, Philipp T. and Vasseur, Romain and Potter, Andrew C.},
  journal = {Phys. Rev. Lett.},
  volume = {120},
  issue = {7},
  pages = {070602},
  numpages = {6},
  year = {2018},
  month = {Feb},
  publisher = {American Physical Society},
  doi = {10.1103/PhysRevLett.120.070602},
  url = {https://link.aps.org/doi/10.1103/PhysRevLett.120.070602}
}

@article{mori,
  title = {Rigorous Bounds on the Heating Rate in Thue-Morse Quasiperiodically and Randomly Driven Quantum Many-Body Systems},
  author = {Mori, Takashi and Zhao, Hongzheng and Mintert, Florian and Knolle, Johannes and Moessner, Roderich},
  journal = {Phys. Rev. Lett.},
  volume = {127},
  issue = {5},
  pages = {050602},
  numpages = {6},
  year = {2021},
  month = {Jul},
  publisher = {American Physical Society},
  doi = {10.1103/PhysRevLett.127.050602},
  url = {https://link.aps.org/doi/10.1103/PhysRevLett.127.050602}
}

@article{CHSE,
  title = {Complete Hilbert-Space Ergodicity in Quantum Dynamics of Generalized Fibonacci Drives},
  author = {Pilatowsky-Cameo, Sa\'ul and Dag, Ceren B. and Ho, Wen Wei and Choi, Soonwon},
  journal = {Phys. Rev. Lett.},
  volume = {131},
  issue = {25},
  pages = {250401},
  numpages = {7},
  year = {2023},
  month = {Dec},
  publisher = {American Physical Society},
  doi = {10.1103/PhysRevLett.131.250401},
  url = {https://link.aps.org/doi/10.1103/PhysRevLett.131.250401}
}

@article{GGE17ilievski,
  title = {From interacting particles to equilibrium statistical ensembles},
  author = {Ilievski, Enej and Quinn, Eoin and Caux, Jean-S\'ebastien},
  journal = {Phys. Rev. B},
  volume = {95},
  issue = {11},
  pages = {115128},
  numpages = {9},
  year = {2017},
  month = {Mar},
  publisher = {American Physical Society},
  doi = {10.1103/PhysRevB.95.115128},
  url = {https://link.aps.org/doi/10.1103/PhysRevB.95.115128}
}

@article{fibdrivenfree,
  title = {Self-Similar Phase Diagram of the Fibonacci-Driven Quantum Ising Model},
  author = {Schmid, Harald and Peng, Yang and Refael, Gil and von Oppen, Felix},
  journal = {Phys. Rev. Lett.},
  volume = {134},
  issue = {24},
  pages = {240404},
  numpages = {6},
  year = {2025},
  month = {Jun},
  publisher = {American Physical Society},
  doi = {10.1103/hn66-j8pt},
  url = {https://link.aps.org/doi/10.1103/hn66-j8pt}
}

@article{prx7aperiodic,
  title = {Aperiodically Driven Integrable Systems and Their Emergent Steady States},
  author = {Nandy, Sourav and Sen, Arnab and Sen, Diptiman},
  journal = {Phys. Rev. X},
  volume = {7},
  issue = {3},
  pages = {031034},
  numpages = {17},
  year = {2017},
  month = {Aug},
  publisher = {American Physical Society},
  doi = {10.1103/PhysRevX.7.031034},
  url = {https://link.aps.org/doi/10.1103/PhysRevX.7.031034}
}

@misc{dualunitaryRMP,
      title={Exactly solvable many-body dynamics from space-time duality}, 
      author={Bruno Bertini and Pieter W. Claeys and Tomaz Prosen},
      year={2025},
      eprint={2505.11489},
      archivePrefix={arXiv},
      primaryClass={cond-mat.stat-mech},
      url={https://arxiv.org/abs/2505.11489}, 
}

@article{DU19prl,
  title = {Exact Correlation Functions for Dual-Unitary Lattice Models in $1+1$ Dimensions},
  author = {Bertini, Bruno and Kos, Pavel and Prosen, Toma\ifmmode \check{z}\else \v{z}\fi{}},
  journal = {Phys. Rev. Lett.},
  volume = {123},
  issue = {21},
  pages = {210601},
  numpages = {6},
  year = {2019},
  month = {Nov},
  publisher = {American Physical Society},
  doi = {10.1103/PhysRevLett.123.210601},
  url = {https://link.aps.org/doi/10.1103/PhysRevLett.123.210601}
}

@article{lfepXXZ,
  title = {Anomalous low-frequency conductivity in easy-plane XXZ spin chains},
  author = {Agrawal, Utkarsh and Gopalakrishnan, Sarang and Vasseur, Romain and Ware, Brayden},
  journal = {Phys. Rev. B},
  volume = {101},
  issue = {22},
  pages = {224415},
  numpages = {6},
  year = {2020},
  month = {Jun},
  publisher = {American Physical Society},
  doi = {10.1103/PhysRevB.101.224415},
  url = {https://link.aps.org/doi/10.1103/PhysRevB.101.224415}
}

@article{prlide2017,
  title = {Microscopic Origin of Ideal Conductivity in Integrable Quantum Models},
  author = {Ilievski, Enej and De Nardis, Jacopo},
  journal = {Phys. Rev. Lett.},
  volume = {119},
  issue = {2},
  pages = {020602},
  numpages = {6},
  year = {2017},
  month = {Jul},
  publisher = {American Physical Society},
  doi = {10.1103/PhysRevLett.119.020602},
  url = {https://link.aps.org/doi/10.1103/PhysRevLett.119.020602}
}

@article{xxzfront19,
  title = {Subdiffusive front scaling in interacting integrable models},
  author = {Bulchandani, Vir B. and Karrasch, Christoph},
  journal = {Phys. Rev. B},
  volume = {99},
  issue = {12},
  pages = {121410},
  numpages = {7},
  year = {2019},
  month = {Mar},
  publisher = {American Physical Society},
  doi = {10.1103/PhysRevB.99.121410},
  url = {https://link.aps.org/doi/10.1103/PhysRevB.99.121410}
}

@article{freefront,
  title = {Full Counting Statistics in a Propagating Quantum Front and Random Matrix Spectra},
  author = {Eisler, Viktor and R\'acz, Zolt\'an},
  journal = {Phys. Rev. Lett.},
  volume = {110},
  issue = {6},
  pages = {060602},
  numpages = {5},
  year = {2013},
  month = {Feb},
  publisher = {American Physical Society},
  doi = {10.1103/PhysRevLett.110.060602},
  url = {https://link.aps.org/doi/10.1103/PhysRevLett.110.060602}
}

@article{tdkondo,
  title = {Integrability of the Kondo model with time-dependent interaction strength},
  author = {Pasnoori, Parameshwar R.},
  journal = {Phys. Rev. B},
  volume = {112},
  issue = {6},
  pages = {L060409},
  numpages = {6},
  year = {2025},
  month = {Aug},
  publisher = {American Physical Society},
  doi = {10.1103/78xb-5lmw},
  url = {https://link.aps.org/doi/10.1103/78xb-5lmw}
}

@article{GHDmoore,
  title = {Bethe-Boltzmann hydrodynamics and spin transport in the XXZ chain},
  author = {Bulchandani, Vir B. and Vasseur, Romain and Karrasch, Christoph and Moore, Joel E.},
  journal = {Phys. Rev. B},
  volume = {97},
  issue = {4},
  pages = {045407},
  numpages = {13},
  year = {2018},
  month = {Jan},
  publisher = {American Physical Society},
  doi = {10.1103/PhysRevB.97.045407},
  url = {https://link.aps.org/doi/10.1103/PhysRevB.97.045407}
}

@article{xxzfrontrev,
  title={Scaling of fronts and entanglement spreading during a domain wall melting},
  author={Scopa, Stefano and Karevski, Dragi},
  journal={The European Physical Journal Special Topics},
  volume={232},
  number={11},
  pages={1763--1781},
  year={2023},
  publisher={Springer},
  doi = {10.1140/epjs/s11734-023-00845-1},
  url = {https://doi.org/10.1140/epjs/s11734-023-00845-1}
}

@article{pozsgaymedint,
  title = {Integrable spin chains and cellular automata with medium-range interaction},
  author = {Gombor, Tam\'as and Pozsgay, Bal\'azs},
  journal = {Phys. Rev. E},
  volume = {104},
  issue = {5},
  pages = {054123},
  numpages = {30},
  year = {2021},
  month = {Nov},
  publisher = {American Physical Society},
  doi = {10.1103/PhysRevE.104.054123},
  url = {https://link.aps.org/doi/10.1103/PhysRevE.104.054123}
}

@article{XXZgaplessdwGHD,
  title = {Analytic solution of the domain-wall nonequilibrium stationary state},
  author = {Collura, Mario and De Luca, Andrea and Viti, Jacopo},
  journal = {Phys. Rev. B},
  volume = {97},
  issue = {8},
  pages = {081111},
  numpages = {6},
  year = {2018},
  month = {Feb},
  publisher = {American Physical Society},
  doi = {10.1103/PhysRevB.97.081111},
  url = {https://link.aps.org/doi/10.1103/PhysRevB.97.081111}
}

@article{dGGE,
  title = {Integrable Digital Quantum Simulation: Generalized Gibbs Ensembles and Trotter Transitions},
  author = {Vernier, Eric and Bertini, Bruno and Giudici, Giuliano and Piroli, Lorenzo},
  journal = {Phys. Rev. Lett.},
  volume = {130},
  issue = {26},
  pages = {260401},
  numpages = {7},
  year = {2023},
  month = {Jun},
  publisher = {American Physical Society},
  doi = {10.1103/PhysRevLett.130.260401},
  url = {https://link.aps.org/doi/10.1103/PhysRevLett.130.260401}
}

\end{document}